\documentclass[aps,pra,superscriptaddress,nofootinbib,twocolumn]{revtex4-2}
\usepackage{amsmath,amsthm,amsfonts,amssymb,amscd} 
\usepackage[utf8]{inputenc}
\usepackage{indentfirst}
\usepackage{graphicx}
\usepackage[T1]{fontenc}
\usepackage[american,ngerman,greek,brazilian,british]{babel}
\usepackage{float}
\usepackage[dvipsnames]{xcolor}
\usepackage[colorlinks=true,citecolor=Mulberry,linkcolor=prettygreen,urlcolor=MidnightBlue,hyperindex,breaklinks]{hyperref}
\usepackage{cleveref}
\usepackage{mathtools}
\usepackage{bm}
\usepackage{ulem}
\usepackage{enumerate}
\usepackage{sidecap}
\usepackage{microtype}
\usepackage{enumitem}
\usepackage{bbold}
\usepackage{listings}
\usepackage{algorithm}
\usepackage{algpseudocode}
\usepackage{multirow}
\usepackage{setspace}
\usepackage{csquotes}
\usepackage{soul}
\usepackage{tikz}
\usetikzlibrary{quantikz2}
\usepackage{dsfont}
\usepackage{physics}
\usepackage{soul}
\setstcolor{red}
\usepackage{mathrsfs}

\definecolor{prettygreen}{RGB}{5,125,143}
\definecolor{JessicaColour}{rgb}{0,0.502,0.502}

\newtheorem{theorem}{Theorem}%[chapter]
\newtheorem*{theorem*}{Theorem}
\newtheorem{lemma}{Lemma}%[chapter]
\newtheorem{conjecture}{Conjecture}% [chapter]
\def\d{{\rm d}}
\newcommand{\1}{\mathbb{1}}

\renewcommand{\tr}{\text{\normalfont tr}}
\newcommand{\id}{\mathbb{1}}
\newcommand{\map}[1]{\mathcal{#1}}
\newcommand{\supermap}[1]{\map{#1}}
\newcommand{\ope}[1]{\mathsf{#1}}

\usepackage{listings}
\newcommand{\dket}[1]{\vert#1\rangle\!\rangle}
\newcommand{\dbra}[1]{\langle\!\langle #1\vert}

\newcommand{\dketbra}[2]{\left.\left\vert #1 \right\rangle \hspace{-.8mm} \right\rangle \hspace{-1mm} \left\langle\hspace{-.8mm}\left\langle #2 \right\vert\right.}

\makeatletter
\def\dbraket#1{%
    \@ifnextchar\bgroup{%
        \dbraket@{#1}%
    }{%
        \langle\!\langle {#1} \vert {#1} \rangle\!\rangle%
    }%
}
\def\dbraket@#1#2{%
    \langle\!\langle {#1} \vert {#2} \rangle\!\rangle%
}
\makeatother

\makeatletter
\def\dketbra#1{%
    \@ifnextchar\bgroup{%
        \dketbra@{#1}%
    }{%
        \vert {#1} \rangle\!\rangle\!\langle\!\langle {#1} \vert%
    }%
}
\def\dketbra@#1#2{%
    \vert {#1} \rangle\!\rangle\!\langle\!\langle {#2} \vert%
}
\makeatother
\newcommand{\todai}{Department of Physics, Graduate School of Science, The University of Tokyo, Hongo 7-3-1, Bunkyo-ku, Tokyo 113-0033, Japan}
\newcommand{\transscale} {Trans-scale Quantum Science Institute, The University of Tokyo, Hongo 7-3-1, Bunkyo-ku, Tokyo 113-0033, Japan}
\newcommand{\perimeter}{Perimeter Institute for Theoretical Physics, 31 Caroline Street North, Waterloo, Ontario, N2L 2Y5, Canada}
\newcommand{\iqc}{Institute for Quantum Computing, University of Waterloo, 200 University Avenue West, Waterloo, Ontario, N2L 3G1, Canada}
\newcommand{\unige}{Department of Applied Physics, University of Geneva, 1205 Geneva, Switzerland}
\newcommand{\sorbonne}{Sorbonne Universit\'{e}, CNRS, LIP6, F-75005 Paris, France}
\newcommand{\manchester}{Department of Physics \& Astronomy, University of Manchester, Manchester M13 9PL, United Kingdom}
\newcommand{\umontreal}{Department of Computer Science and Operations Research, Universit{\' e} de Montr{\' e}al, Montr{\' e}al, Qu{\' e}bec, H3T 1J4, Canada}
\begin{document}

\author{Jessica Bavaresco}
\affiliation{\unige}
\affiliation{\sorbonne}

\author{Hl\'{e}r Kristj\'{a}nsson}
\affiliation{\perimeter}
\affiliation{\iqc}
\affiliation{\todai}
\affiliation{\umontreal}

\author{Mio Murao}
\affiliation{\todai}
\affiliation{\transscale}

\author{\mbox{Tatsuki Odake}}
\affiliation{\todai}

\author{Marco T\'{u}lio Quintino}
\email{Marco.Quintino@lip6.fr}
\affiliation{\sorbonne}

\author{Philip Taranto}
\affiliation{\manchester}
\affiliation{\todai}

\author{Satoshi Yoshida}
\affiliation{\todai}

\date{\today}

\title{Simulating the quantum switch with quantum circuits is computationally hard} 

\begin{abstract}
Higher-order transformations acting on input quantum channels in an indefinite causal order---such as the quantum switch---cannot be described by quantum circuits using the same number of calls to the input channels. A natural question is whether they can be simulated, i.e., whether their action can be exactly and deterministically reproduced by a quantum circuit with more calls to the input channels. Here, we prove that the quantum switch acting on two $n$-qubit channels cannot be simulated by any quantum circuit using $k$ calls to one channel and one to the other, if $k<2^n$. This establishes an exponential separation in quantum query complexity between processes with indefinite causal order and quantum circuits. Moreover, even with one extra call to both input channels, such a simulation remains impossible. We further demonstrate the robustness of this separation by extending the result to probabilistic and approximate simulations scenarios.
\end{abstract}

\maketitle

%%%%%%%%%%%%%%%%%%%%%%%%%%%%%%%%%%%%%%%%%%%%
\section{Introduction}\label{sec::intro}
%%%%%%%%%%%%%%%%%%%%%%%%%%%%%%%%%%%%%%%%%%%%

Indefinite causal order is a fundamental property that emerges in the exploration of higher-order transformations within quantum theory~\cite{chiribella2008quantum,chiribella2008transforming,chiribella2009theoretical,bisio2019theoretical,milz2024characterising}. Higher-order quantum transformations are operations whose inputs and outputs are not quantum states, but rather transformations that themselves act on quantum states, such as quantum channels.
An example of such a transformation is a quantum circuit with open slots where arbitrary quantum channels can be inserted and acted upon in a sequential, temporally ordered manner
~\cite{chiribella2008quantum,chiribella2009theoretical}. 
Remarkably, there also exist well-defined higher-order transformations that act on their input quantum channels in an indefinite causal order~\cite{chiribella2013quantum,oreshkov2012quantum,araujo2015witnessing}. Such transformations cannot be described by any quantum circuit that uses the same number of calls, or queries, of the input quantum channels, challenging the standard notion of computation in which operations are performed on a system in a fixed order. 

The ability to perform operations in such an indefinite order has been shown to provide advantages in a variety of information-processing settings, such as quantum channel discrimination~\cite{chiribella2012perfect,bavaresco2021strict,bavaresco2022unitary}, quantum metrology~\cite{zhao2020quantum,mothe2024reassessing,liu2023optimal,yin2023experimental}, quantum computational complexity~\cite{araujo2014computational,renner2022computational}, quantum query complexity~\cite{abbott2024quantum}, transformations of black-box unitaries and isometries~\cite{quintino2019reversing,quintino2019probabilistic,quintino2022deterministic,yoshida2023universal,yoshida2025universal}, among others. 

A prominent example of indefinite causal order that is responsible for several of these theoretically predicted advantages is the quantum switch~\cite{chiribella2013quantum}, a higher-order transformation that coherently controls the causal order of two quantum channels. The information-processing advantages of the quantum switch have mostly been shown via comparison with higher-order transformations that act in a fixed order on a single call of each input channel~\cite{chiribella2012perfect,araujo2015witnessing,zhao2020quantum,bavaresco2021strict}. However, their true practical significance hinges on the extent to which these advantages would still hold when comparing the quantum switch with quantum circuits that use a larger number of calls to its inputs.

In the context of quantum computation, whether the quantum switch exhibits a true complexity-theoretic advantage depends upon whether its action can be efficiently reproduced by using quantum circuits, given extra queries to the input channels. Until now, no exponential separation has been demonstrated between the query complexity of computations using indefinite causal order versus quantum circuits. In fact, for unitary channels, the action of the quantum switch can be reproduced by a quantum circuit with just one extra query~\cite{chiribella2013quantum}, significantly limiting the computational power of the quantum switch in this case. However, a crucial open question is whether this limitation extends to general quantum channels.

More concretely, this question can be phrased in terms of the simulability of the quantum switch, or more generally, of any higher-order transformation with indefinite causal order:

\begin{displayquote}
    Can a higher-order transformation with indefinite causal order that acts on arbitrary quantum channels be simulated by quantum circuits that have access to more calls of the input quantum channels?
\end{displayquote}

In the case where the quantum circuit performing the simulation has access to an infinite number of calls of the input channels, the answer is: yes. In this case, a process tomography protocol~\cite{chiribella2009theoretical,poyatos1997complete,chuang1997prescription} can completely characterize the input channels and simply prepare the output channel expected from the higher-order transformation with indefinite causal order. However, such a simulation requires infinite resources. Another possible way that the quantum switch can be simulated is using post-selection~\cite{chiribella2013quantum,milz2018entanglement}. However, such a simulation does not work deterministically. Additionally, in the particular case where the input channels are restricted to being unitary channels, it has been shown that a simulation of the quantum switch exists~\cite{chiribella2013quantum}. However, such a simulation is not universal, inasmuch as it only works for a restricted set of input channels.

We hence ask the question, which has remained largely unexplored so far: does any finite number of extra calls of the input channels suffice to deterministically and universally simulate the action of the quantum switch with a quantum circuit?

In this work, we prove that the quantum switch acting on arbitrary channels $A$ and $B$ cannot be deterministically and universally simulated by a quantum circuit (or even a quantum circuit with classical control of the causal order~\cite{wechs2021quantum}) that has access to $k_A<2^n$ calls of $A$ and $k_B=1$ call of $B$, where $A$ and $B$ are arbitrary $n$-qubit quantum channels. Our theorem demonstrates an exponential separation in quantum query complexity for computational tasks using quantum processes with indefinite causal order versus quantum circuits with fixed or classically-controlled causal order, in terms of the number of qubits. We moreover prove that when one extra call of each quantum channel is available ($k_A=k_B=2$), it remains impossible to simulate the action of the quantum switch, even for single-qubit channels. We demonstrate the extent of the robustness of these results in two different ways: with probabilistic and approximate simulations. We show that even when approximate simulations are considered, the probability of simulating a higher-order operation which is $\epsilon$-close to the quantum switch is significantly below one for an $\epsilon$ significantly above zero. We furthermore thoroughly analyze the problem of simulating the quantum switch when it acts only on part of its input quantum channels or on quantum instruments. Finally, we show some new particular (non-universal) cases in which simulations of the quantum switch are possible. In light of these results, we conjecture that a deterministic simulation of the quantum switch, if possible, would require a number of queries to both quantum channels that grows at least exponentially with the size of the system they act upon. If our conjecture holds true, it would imply that processes with indefinite causal order cannot be efficiently simulated even by quantum circuits with classically-controlled causal order.

%%%%%%%%%%%%%%%%%%%%%%%%%%%%%%%%%%%%%%%%%%%%
\section{Results}\label{sec::results}
%%%%%%%%%%%%%%%%%%%%%%%%%%%%%%%%%%%%%%%%%%%%

%%%%%%%%%%%%%%%%%
\subsection*{Quantum switch simulation}\label{subsec::task}
%%%%%%%%%%%%%%%%%

\begin{figure}%[h!]
\begin{center}
	\includegraphics[width=\columnwidth]{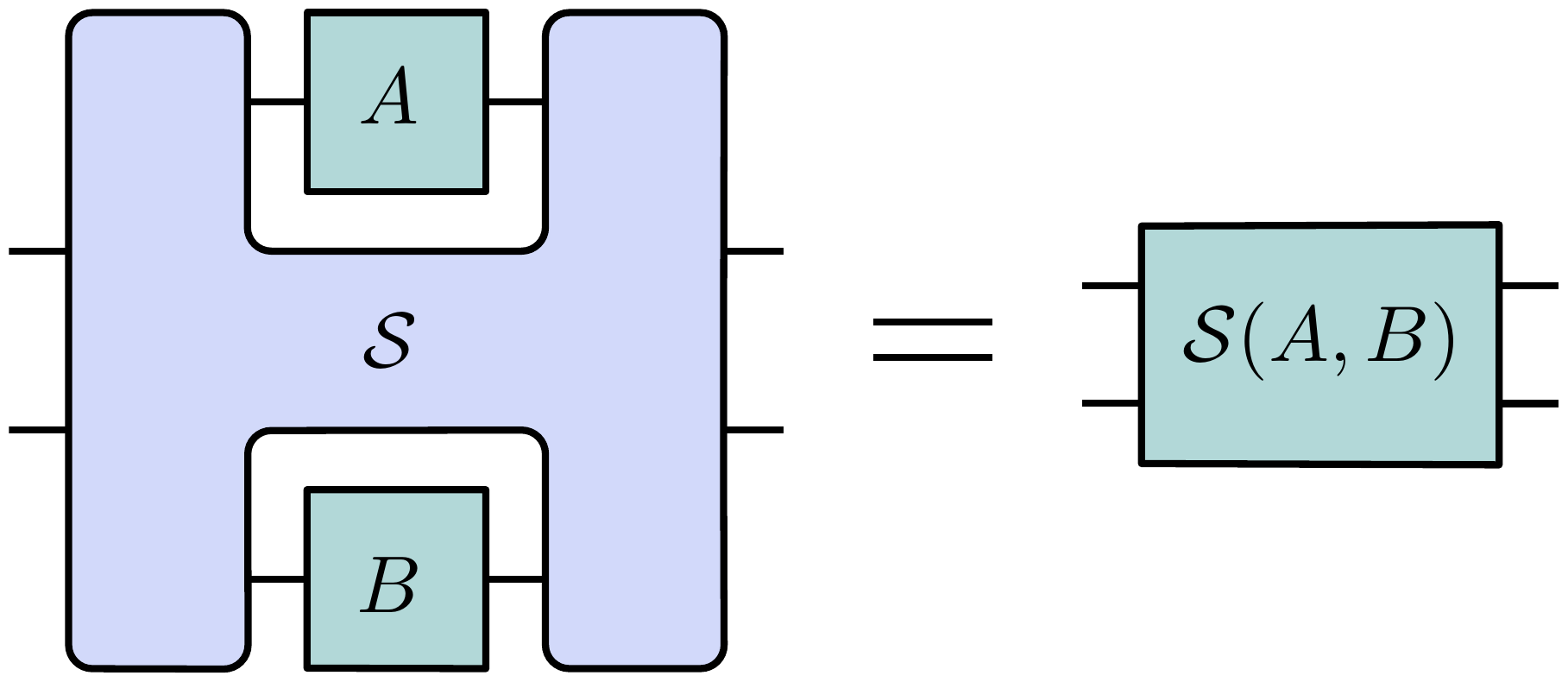}
	\caption{\textbf{The quantum switch transformation.} The quantum switch {$\map{S}$} is a higher-order transformation that takes as input {any} two quantum channels $A$ and $B$ and transforms them into a {another} quantum channel $\map{S}(A,B)$. The resulting channel $\map{S}(A,B)$ acts on a qubit control system and a qudit target system.}
\label{fig::output_switch}
\end{center}
\end{figure}

The quantum switch $\map{S}$ is a higher-order transformation that takes two arbitrary quantum channels (i.e. completely positive, trace-preserving maps) $A$ and $B$ as input, where $A: \map{L}(\map{H}^{A_I}) \to \map{L}(\map{H}^{A_O})$ and $B: \map{L}(\map{H}^{B_I}) \to \map{L}(\map{H}^{B_O})$ are channels that act on qudit systems, and transforms them into a channel $\map{S}(A,B): \map{L}(\map{H}^{c_I}\otimes\map{H}^{t_I}) \to \map{L}(\map{H}^{c_O}\otimes\map{H}^{t_O})$ that acts on a qubit control system and a qudit target system. The output channel resulting from the action of the quantum switch~\cite{chiribella2013quantum} on its input channels is defined as
\begin{equation}\label{eq::switch}
    \map{S}(A,B)[\,\sigma_{c}\otimes\rho_{t}\,] \coloneqq  \sum_{i,j} \ope{S}_{ij} (\sigma_{c}\otimes\rho_{t}) \ope{S}_{ij}^\dagger,
\end{equation}
where $\sigma_{c}\in\map{L}(\map{H}^{c_I})$ is the state of the input qubit control system, $\rho_t\in\map{L}(\map{H}^{t_I})$ is that of the qudit target system, and $\ope{S}_{ij}$ is given by
\begin{equation}\label{eq::switchkraus}
    \ope{S}_{ij} \coloneqq  \ketbra{0}{0}\otimes \ope{B}_{j} \ope{A}_{i} + \ketbra{1}{1}\otimes \ope{A}_{i} \ope{B}_{j},
\end{equation} 
where $\ope{A}_{i}:\map{H}^{A_I}\to\map{H}^{A_O}$ and $\ope{B}_{j}:\map{H}^{B_I}\to\map{H}^{B_O}$ are Kraus operators~\cite{watrous2018theory} of the channels $A$ and $B$, respectively; i.e., $A[\rho] = \sum_i \ope{A}_i\,\rho\, \ope{A}_i^\dagger$ and $B[\rho] = \sum_i \ope{B}_i\,\rho\, \ope{B}_i^\dagger$. This transformation is depicted in Fig.~\ref{fig::output_switch}. The quantum switch acts on its input channels in an order that is conditioned on the state of a quantum control system. Since the quantum control system may be initiated in a superposition state, such as $\ket{+}\coloneqq \frac{1}{\sqrt{2}}(\ket{0}+\ket{1})$, the overall quantum switch transformation may be understood as a conditioned superposition of two different circuits, one with the control system in state $\ket{0}$ and the quantum channels being applied in the order $A$ before $B$, and another with the control system in state $\ket{1}$ and the quantum channels being applied in the order $B$ before $A$.

%%%%%%%%%%%%%%%%%%%%%%%%%%%%%%%%%%%%%%%%%%%
%%%%%%%%%%%%%%%%%%%%%%%%%%%%%%%%%%%%%%%%%%%

\begin{figure*}%[h!]
\begin{center}
	\includegraphics[width=\textwidth]{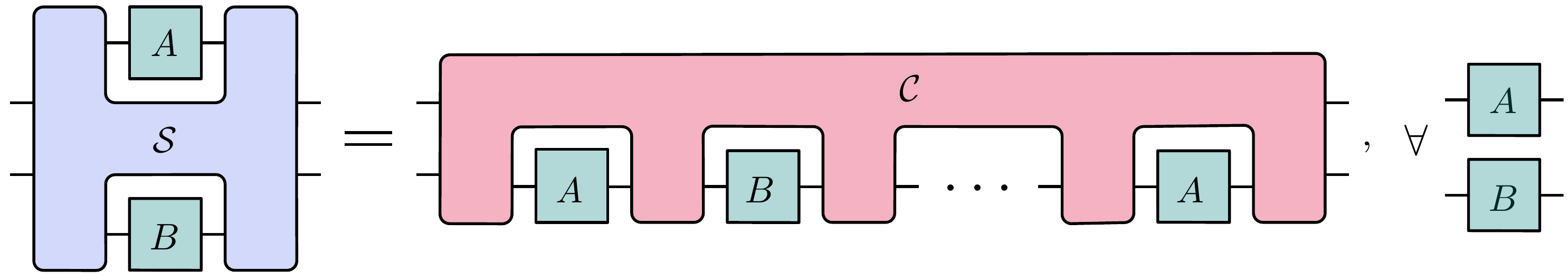}
	\caption{\textbf{Simulating the quantum switch.} A higher-order transformation $\map{C}$ corresponding to an open-slot quantum circuit with fixed or classically-controlled causal order, which acts on several copies of the input quantum channels $A$ and $B$, is a simulation of the quantum switch $\map{S}$ if it reproduces the action of the quantum switch on all arbitrary channels $A$ and $B$.}
\label{fig::simulation}
\end{center}
\end{figure*}

A deterministic and exact simulation of the quantum switch is a higher-order transformation $\map{C}$ that obeys causal constraints and that acts on a finite number $k_A$ and $k_B$ of calls to the channels $A$ and $B$, respectively, in such a way that the channel $\map{C}(A^{\otimes k_A},B^{\otimes k_B}): \map{L}(\map{H}^{c_I}\otimes\map{H}^{t_I}) \to \map{L}(\map{H}^{c_O}\otimes\map{H}^{t_O})$ resulting from this transformation satisfies
\begin{equation}\label{eq::simulation}
    \map{C}(A^{\otimes k_A},B^{\otimes k_B}) = \map{S}(A,B) \ \ \ \forall \ A, B,
\end{equation}
where $A, B$ are arbitrary quantum channels. Natural causal constraints that one might impose on the simulation is to require that $\map{C}$ be described by a fixed-order quantum circuit with open slots, called a quantum comb~\cite{chiribella2008quantum,chiribella2009theoretical} (see App.~\ref{subapp::combs} for a formal definition). 
A more general strategy for the simulation---which could nevertheless be interpreted as having a definite causal order---would be to impose that $\map{C}$ is described by open-slot quantum circuits that have classical control of causal order. This class of higher-order transformations, proposed in Ref.~\cite{wechs2021quantum} and called ``QC-CCs'', is larger than the set of quantum combs, allowing for classical mixtures of open-slot quantum circuits as well as for classically-controlled dynamical causal orders.
Hence, permitting this class gives more power to the simulation as compared to quantum combs, while still allowing for a causally ordered interpretation of the simulation (see App.~\ref{subapp::qcccs} for a formal definition). We furthermore require the simulation to be universal: the same simulation $\map{C}$ must work for all input pairs of quantum channels. See Fig.~\ref{fig::simulation} for a graphical representation of Eq.~\eqref{eq::simulation}. The question of simulability then boils down to whether there exists, for some finite number of calls $k_A$ and $k_B$, a simulation $\map{C}$ that obeys causal constraints and that satisfies Eq.~\eqref{eq::simulation}. 

This notion of computation, where the inputs and outputs are quantum rather than classical, is suitable to treat inherently quantum problems and has been employed beyond the higher-order quantum computing paradigm explored here in problems ranging from the SWAP test~\cite{barenco1997swap,buhrman2001fingerprint} to Hamiltonian simulations~\cite{berry2009blackbox,berry2015hamiltonian} and quantum property testing~\cite{montanaro2013survey}.

%%%%%%%%%%%%%%%%%%%%%%%%%%%%%%%%%%%%%%%%%%%
%%%%%%%%%%%%%%%%%%%%%%%%%%%%%%%%%%%%%%%%%%%

%%%%%%%%%%%%%%%%%
\subsection*{Go theorem: An explicit non-universal simulation}\label{subsec::particularcase}
%%%%%%%%%%%%%%%%%

For a particular case of input channels, it is known that a non-universal simulation of the quantum switch exists. As first shown in the paper that originally defined the quantum switch~\cite{chiribella2013quantum}, in the particular case where the input channels are unitary (i.e.\ reversible), a simulation of the quantum switch by a quantum circuit is possible, requiring only an extra use of one of the input channels.

The simulation presented in Ref.~\cite{chiribella2013quantum} is given by the quantum circuit
\begin{equation}\label{cc::chiribella}
	%\begin{tikzpicture}	\node[scale=1.1]{
	\begin{quantikz}[column sep=6.5mm]
	& \ctrl{2} &   		  & \ctrl{2} & 	& \ctrl[open]{2} &          &	\ctrl[open]{2} &	\\
	& \targX{} &          & \targX{} &          & \targX{} &          & \targX{} &	\\
	& \targX{} & \gate[style={fill=JessicaColour!30},label style=black]{A} & \targX{} & \gate[style={fill=JessicaColour!30},label style=black]{B} & \targX{} & \gate[style={fill=JessicaColour!30},label style=black]{A} &	\targX{} &		
	\end{quantikz}
	%};
	%\end{tikzpicture}
\end{equation}	
where
$\begin{quantikz}[column sep=1mm] &\ctrl[open]{0}& \end{quantikz}$ $\coloneqq$ $\begin{quantikz}[column sep=1mm] &\gate{X}& &\ctrl{0}& &\gate{X}& \end{quantikz}$ 
and
$\begin{quantikz}[column sep=1mm] &\gate{X}& \end{quantikz}$ is the NOT gate. Here, the first circuit line corresponds to the control qubit system, the second to an auxiliary system, and the third to the target qudit system. This circuit can equivalently be represented by the Kraus operators $\{\ope{C}_{iji'}\}_{iji'}$, where
\begin{equation}\label{eq::chiribellakraus}
	\ope{C}_{iji'} \coloneqq  \ketbra{0}{0}\otimes \ope{A}_{i'}\otimes \ope{B}_{j}\ope{A}_{i} + \ketbra{1}{1}\otimes \ope{A}_{i}\otimes \ope{A}_{i'}\ope{B}_{j}.
\end{equation} 
Since in general $i\neq i'$, the transformation acting on the auxiliary system does not typically factor out from the transformation acting on the qubit and control systems. By comparison with Eq.~\eqref{eq::switchkraus}, it is straightforward to see that the resulting transformation does not simulate the quantum switch for arbitrary quantum channels. However, in the case where $A$ and $B$ are unitary channels, and hence are described by a single Kraus operator (i.e. $i=i'=j=0$), the transformation on the auxiliary system factorizes from that on the control and target systems. In this particular case, it is straightforward to see that, for any input state of the auxiliary system, when the output auxiliary system is discarded, the quantum circuit in Eq.~\eqref{cc::chiribella} performs a (non-universal) simulation of the quantum switch. 

A crucial point to note is that the circuit in Eq.~\eqref{cc::chiribella} acts on the input channels in the order ``ABA''. If one were to change the order of the input channels to either ``AAB'' or ``BAA'', this circuit no longer simulates the quantum switch, even if the input channels are unitary. In fact, for these different orders, we prove that there does not exist any quantum circuit that can simulate the action of the quantum switch, even when acting only on unitary channels. We present more details in App.~\ref{app::nogounitary}.

Here, we show that an extension of the circuit in Eq.~\eqref{cc::chiribella} allows one to perform a simulation of the quantum switch in a more general---albeit not fully general---scenario. This is the case where the quantum switch acts only on part of a quantum channel $A$ that is unitary and part of a quantum channel $B$ that is general. 

\begin{figure}%[h!]
\begin{center}
	\includegraphics[width=\columnwidth]{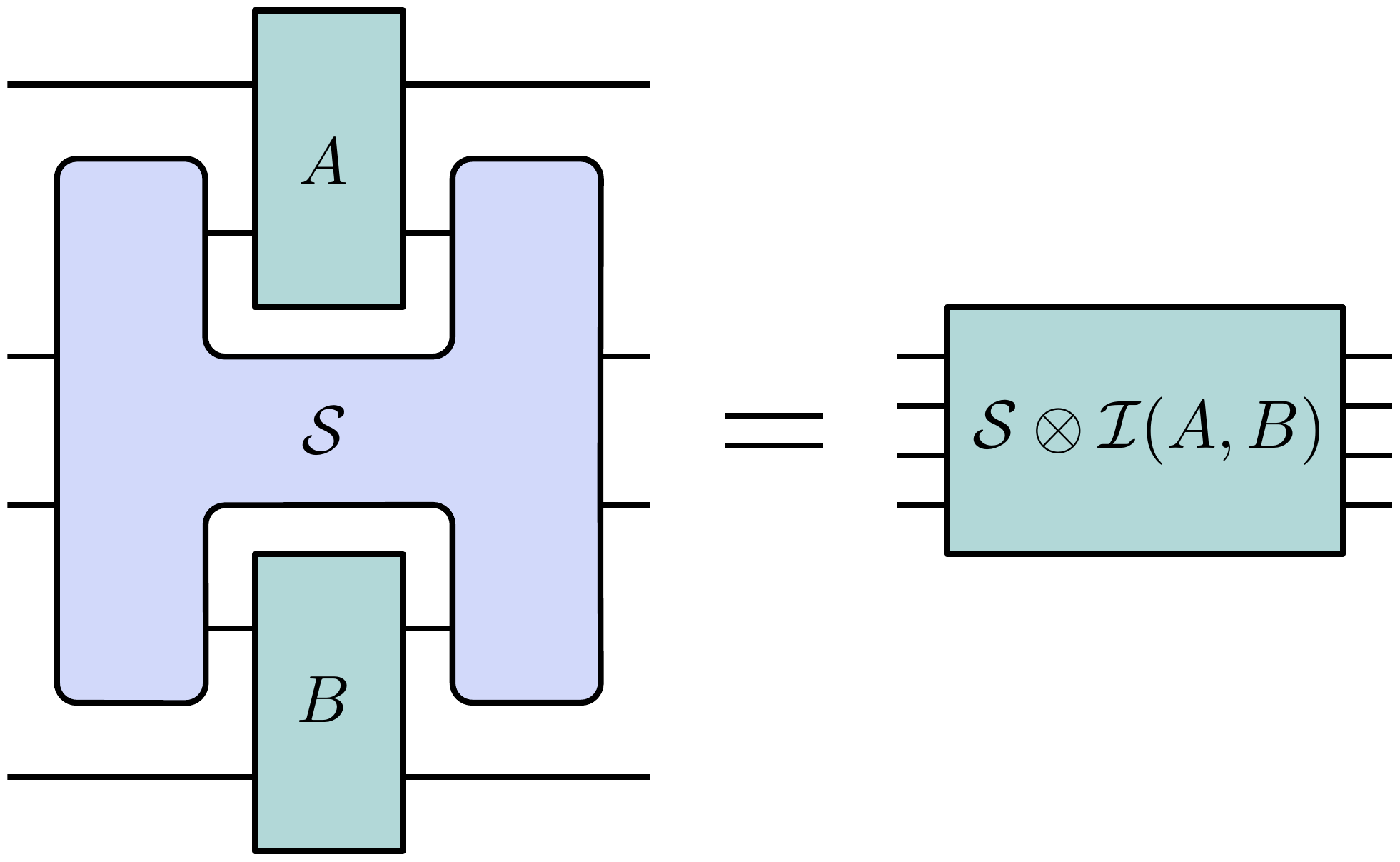}
	\caption{\textbf{The extended quantum switch transformation.} The quantum switch {$\map{S}$} is a higher-order transformation that takes as input {any two} quantum channels $A$ and $B$ and transforms them into {another} quantum channel, $\map{S}\otimes\map{I}(A,B)$, even when acting only on part of these channels.}
\label{fig::output_switch_general}
\end{center}
\end{figure}

In this simulation scenario, the quantum switch acts only on part of bipartite channels $A: \map{L}(\map{H}^{A_I}\otimes\map{H}^{A'_I}) \to \map{L}(\map{H}^{A_O}\otimes\map{H}^{A'_O})$ 
and $B: \map{L}(\map{H}^{B_I}\otimes\map{H}^{B'_I}) \to \map{L}(\map{H}^{B_O}\otimes\map{H}^{B'_O})$, both of which have two input and two output spaces. The extra input/output systems of these bipartite channels can be interpreted as environments that are not accessible to the local parties Alice and Bob. The output channel from the quantum switch transformation is then given by $\map{S}\otimes\map{I}(A,B): 
\map{L}(\map{H}^{A'_I}\otimes\map{H}^{c_I}\otimes\map{H}^{t_I}\otimes\map{H}^{B'_I}) \to
\map{L}(\map{H}^{A'_O}\otimes\map{H}^{c_O}\otimes\map{H}^{t_O}\otimes\map{H}^{B'_O})$, where $\map{I}$ is the identity higher-order transformation acting on the primed spaces. This transformation is depicted in Fig.~\ref{fig::output_switch_general}. Since the quantum switch higher-order transformation is uniquely defined by its action on single-party unitary channels~\cite{dong2023quantum}, there is no risk of ambiguity when considering such extended quantum switch transformations. 

When the input channels $A$ and $B$ are general, i.e., not restricted to being unitary, this scenario is the strongest possible simulation scenario. In other words, a simulation $\map{C}$ that is able to prepare, with some finite number of calls $k_A$ and $k_B$, a channel $\map{C}(A^{\otimes k_A},B^{\otimes k_B})$ such that 
\begin{equation}\label{eq::simulation_general}
    \map{C}(A^{\otimes k_A},B^{\otimes k_B}) = \map{S}\otimes\map{I}(A,B) \ \ \ \forall \ A,B,
\end{equation}
where $A,B$ are arbitrary quantum channels, is a higher-order transformation that can simulate the action of the quantum switch in its most general form. We discuss further details concerning general simulations in App.~\ref{subapp::switchonextendedchannels}, as well as simulations of the action of the quantum switch on quantum instruments in App.~\ref{subapp::switchoninstruments}.

In this general simulation scenario, we prove the following theorem, regarding the existence of a non-universal simulation of the action of the quantum switch on part of bipartite channels---in the particular case where $A$ is restricted to being a unitary channel and $B$ is a fully general quantum channel:
\begin{theorem}\label{thm::go}
    The action of the quantum switch on part of bipartite quantum channels can be deterministically simulated by a quantum circuit that has access to an extra call to one the input channels, as long as that channel is restricted to being unitary.

    In other words, if $A$ is a bipartite unitary channel and $B$ is a bipartite general channel, there exists a quantum circuit described by a higher-order transformation $\map{C}$ that satisfies Eq.~\eqref{eq::simulation_general} for $k_A=2$ and $k_B=1$.
\end{theorem}

We prove this result by explicitly constructing the quantum circuit that performs this simulation, which is given by
\begin{equation}\label{cc::ourcircuit}
	%\begin{tikzpicture}	\node[scale=0.8]{
	\begin{quantikz}[column sep=3.8mm]
	&\ctrl{3}&\ctrl{4}&           &\ctrl{3}&\ctrl{4}&   &\ctrl[open]{3}&\ctrl[open]{4}&        &\ctrl[open]{3}&\ctrl[open]{4}& \\
	&\targX{}&		  &			  &\targX{}&		&	        &\targX{}&		   &	   &\targX{}&	     & \\
	& 		 &\targX{}&		      &	   	   &\targX{}&	        &		 &\targX{}&		   &		&\targX{}& \\
	&\targX{}&        &\gate[2,style={fill=JessicaColour!30},label style=black]{A}&\targX{}&        &		    &\targX{}&        &\gate[2,style={fill=JessicaColour!30},label style=black]{A}&\targX{}&	 & \\
	& 		 &\targX{}& 	      &	       &\targX{}&\gate[2,style={fill=JessicaColour!30},label style=black]{B}&        &\targX{}&		   &		&\targX{}& \\
	& 		 &		  & 	      &	       &	    &		    &        &		  &		   &		&	     & 
	\end{quantikz}.
	%};
	%\end{tikzpicture}
\end{equation}
Here, the first circuit line represents the qubit control, the second and third lines represent auxiliary systems, the fourth line represents Alice's primed system, the fifth line represents the target system, and the sixth line represents Bob's primed system. In App.~\ref{subapp::proofoursimulation}, we prove Thm.~\ref{thm::go} explicitly, showing how it recovers the action of the quantum switch when $A$ is a unitary channel, and how it fails when $A$ is a general channel. 

Given that every quantum channel can be seen as the marginal channel of a unitary channel acting on a higher-dimensional space, implemented with the use of an auxiliary system---via the Stinespring dilation theorem---it may seem counter-intuitive that the circuit in Eq.~\eqref{cc::ourcircuit} fails to simulate the quantum switch for general channels. However, this is due to the fact that unitary channels acting on higher-dimensional spaces only correspond to general, non-unitary channels mapping the input to the output systems if the auxiliary system is not accessible. One way to enforce this is by requiring that the auxiliary systems be discarded. If the auxiliary system were discarded, and hence not available to be exploited during the simulation, it is straightforward to see that the circuit in Eq.~\eqref{cc::ourcircuit} would fail. In App.~\ref{subapp::unitarydilation}, we discuss in detail how Thm.~\ref{thm::go} relates to the Stinespring dilation of non-unitary quantum channels.

%%%%%%%%%%%%%%%%%%%%%%%%%%%%%%%%%%%%%%%%%%%
%%%%%%%%%%%%%%%%%%%%%%%%%%%%%%%%%%%%%%%%%%%

%%%%%%%%%%%%%%%%%
\subsection*{No-go theorems: Deterministic exact simulations}\label{subsubsec::nogo_exp}
%%%%%%%%%%%%%%%%%

We now show that the go theorem from the previous section does not generalize: not only does the circuit in Eq.~\eqref{cc::ourcircuit} not work for arbitrary quantum channels $A$, but there does not exist any other quantum circuit that can simulate the quantum switch, even when a higher number of calls of the input channel $A$ are available. In order to do so, we prove an even stronger statement: one which demonstrates an exponential separation in quantum query complexity between a higher-order transformation with indefinite causal order and quantum circuits that have a fixed or classically-controlled causal order.

\begin{theorem}\label{thm::nogo_exp}
    There is no $(k_A+1)$-slot higher-order transformation $\mathcal{C}$, described by a quantum circuit with fixed or classically-controlled causal order, that can simulate the quantum switch, i.e., that satisfies
    \begin{align}
            \mathcal{C}(A^{\otimes k_A}, B) = \mathcal{S}(A,B)\label{eq:switch_simulation_1_maintext}
    \end{align}
    for all $n$-qubit mixed unitary channels $A$ and unitary channels $B$, if
    \begin{equation}
        k_A \leq  \max(2, 2^{n}-1).
    \end{equation}

    Therefore, such a simulation also does not exist for all $n$-qubit quantum channels $A$ and $B$.
\end{theorem}
We provide a sketch of the proof in Sec.~\ref{sec::methods}, with the full proof presented in App.~\ref{app::fullproof_nogoexp}. While Thm.~\ref{thm::nogo_exp} forbids a deterministic exact simulation of the quantum switch, it also prevents arbitrarily good probabilistic approximate simulations of it. That is because it guarantees that any probabilistic approximate simulation will necessary have a maximal probability of success $p$ strictly less than one or minimum approximation error $\epsilon$ strictly greater than zero, for any notion of metric distance.

This result can be interpreted as an exponential advantage of indefinite causal order over definite causal order for a specific task, namely that of simulating the action of the quantum switch on arbitrary quantum channels $A$ and $B$ when only one call of $B$ is available. In order to formalize this result in the context of computation, consider a classical description of a function $f$ which takes a pair of quantum channels $A, B$ as input and outputs a quantum channel $f(A,B)$. Consider also a higher-order transformation $\map{C}$ that simulates the function $f$ deterministically and exactly---that is, such that $\map{C}(A^{\otimes k_A}, B^{\otimes k_B} ) = f(A, B)$ for all $A, B$---for some number $k_A$ and $k_B$ of black-box queries to the quantum channels $A$ and $B$, respectively.
In the case where one of the channels is fixed to being called $k_B=1$ times, we can define a simple notion of quantum query complexity that depends only on the number of calls to the other channel, $k_A$. We define the  {one-sided quantum query complexity} of a function $f$, with respect to a class of higher-order quantum transformations $\mathscr{C}$, as the minimum number of queries $k_A$ while $k_B=1$, over all higher-order quantum transformations $\map{C}\in\mathscr{C}$ such that $\map{C}$ simulates $f$. This definition can be seen as a step towards a fully quantum generalization of the notion of query complexity. While the standard notion of quantum query complexity has so far typically been defined for classical (e.g. boolean) functions~\cite{abbott2024quantum}, here we consider the query complexity of functions whose inputs and outputs are themselves quantum channels (see also~\cite{odake2024analytical}). This is similar in spirit to recent works on the complexity of preparing quantum states~\cite{rosenthal2021interactive,metger2023stateqip}.

In this context, we see that Thm.~\ref{thm::nogo_exp} implies that the one-sided quantum query complexity of the action of the quantum switch, with respect to all QC-CC transformations, is lower-bounded by $2^n$.

%%%%%%%%%%%%%%%%%
\subsection*{No-go theorems: Probabilistic simulations}\label{subsubsec::nogo_prob}
%%%%%%%%%%%%%%%%%

Following from the above discussion, one might wonder whether a simulation remains impossible if more than one call of the quantum channel $B$ is allowed. Here, we also prove a no-go theorem for the simulation of the quantum switch when $2$ calls of each input quantum channel is available (i.e., $k_A=k_B=2$). In fact, we prove an even stronger result: such a simulation is impossible even for single-qubit channels, and even in a probabilistic and restricted simulation scenario.

Let us start by defining the  {restricted simulation} scenario. Fixing the input control system to be in state $\sigma_c=\ketbra{+}{+}$ and the input target system to be in state $\rho_t=\ketbra{0}{0}$, a restricted simulation of the switch is possible if, for some finite number of calls $k_A$ and $k_B$, there exists a quantum comb or QC-CC $\map{C}$ such that
\begin{align}
\begin{split}
\label{eq::restrictedsimulation}
    \tr_{t_O}\Big(\map{C}(A^{\otimes k_A},\,B^{\otimes k_B})&[\,\ketbra{+}{+}\otimes\ketbra{0}{0}\,] \Big) \\
    = \tr_{t_O}\Big(\map{S}(A,B)&[\,\ketbra{+}{+}\otimes\ketbra{0}{0}\,]\Big) \ \ \ \forall \ A, B, 
\end{split}
\end{align}
where $A$ and $B$ are arbitrary channels. Notice that Eq.~\eqref{eq::restrictedsimulation} is an equality between two qubit quantum states, namely the output control systems. Notice that, while the impossibility of a simulation in the restricted case implies the impossibility of simulation in the general case, a possibility of simulating the quantum switch in the restricted case would not imply the existence of a more general simulation. 

Let us now define a probabilistic heralded simulation of the quantum switch. A probabilistic heralded simulation is a quantum comb or QC-CC $\map{C}$ that compared to the deterministic simulation in Eq.~\eqref{eq::simulation}, outputs an extra classical bit corresponding to either the success or failure outcome of the simulation, each with a certain probability. When the value of this bit corresponds to the success outcome, the implemented simulation is exactly $\map{S}(A,B)$. Such transformations can be implemented by either a quantum comb or as a QC-CC that additionally outputs a flag system encoding the success or failure outcome, followed by a dichotomic quantum measurement of the flag system. Mathematically, a probabilistic heralded simulation is a higher-order transformation $\map{C} = \map{C}_s + \map{C}_f$, where $\map{C}_s$ and $\map{C}_f$ are higher-order maps that completely preserve completely positive inputs---one associated to the success and the other with the failure outcome of the transformation---and where $\map{C}$ is described as either a quantum comb or as a QC-CC. In this case, $\map{C} = \map{C}_s + \map{C}_f$ is a probabilistic simulation of the quantum switch that uses $k_A$ calls of channel $A$ and $k_B$ calls of channel $B$ if it satisfies
\begin{equation}\label{eq::probsimulation}
    \map{C}_s(A^{\otimes k_A},B^{\otimes k_B}) = p\, \map{S}(A,B) \ \ \ \forall \ A, B, 
\end{equation}
where $A$ and $B$ are general quantum channels, and $p$ is the probability of a successful simulation. In the case where $\map{C}$ is a quantum comb, no normalization conditions need to be imposed on $\map{C}_s$. However, when $\map{C}$ is a QC-CC, then normalization conditions must be imposed on $\map{C}_s$ to ensure that $\map{C}$ is a proper QC-CC transformation~\cite{wechs2021quantum}. We present these constraints explicitly in App.~\ref{subapp::qcccs}.

Combining Eqs.~\eqref{eq::restrictedsimulation} and~\eqref{eq::probsimulation}, we define a probabilistic restricted simulation of the quantum switch via 
\begin{align}
\begin{split}
\label{eq::prob&restrictedsimulation}
    \tr_{t_O}\Big(\map{C}_s(A^{\otimes k_A},\,B^{\otimes k_B})&[\,\ketbra{+}{+}\otimes\ketbra{0}{0}\,] \Big) \\
    = p\,\tr_{t_O}\Big(\map{S}(A,B)&[\,\ketbra{+}{+}\otimes\ketbra{0}{0}\,]\Big) \ \ \ \forall \ A, B.
\end{split}
\end{align}

It is known that any higher-order transformation with indefinite causal order can be simulated by a quantum comb in a probabilistic heralded manner with $p>0$, even without any extra calls of the input channels~\cite{milz2018entanglement,quintino2019probabilistic}. For the particular case of the quantum switch, Ref.~\cite{chiribella2013quantum} presents a probabilistic circuit based on quantum teleportation that simulates the quantum switch with $p=1/d^2$, where $d$ is the dimension of the target system, without any extra calls. Here, we will analyze the maximal probability of success for simulating the quantum switch when extra calls are available. 

As we prove in Sec.~\ref{sec::methods}, for any fixed finite $d\coloneqq \dim(\map{H}^{t_I})=\dim(\map{H}^{t_O})=\dim(\map{H}^{A_I})=\dim(\map{H}^{A_O})=\dim(\map{H}^{B_I})=\dim(\map{H}^{B_O})$, $k_A$, and $k_B$, the maximum probability of success $p$ of simulating the quantum switch acting on all general quantum channels $A$ and $B$ can be computed with a semidefinite program (SDP). This is because if a simulation exists for a finite subset of channels that form a basis for the linear subspace spanned by $k_A$ copies of a quantum channel $A$, and equivalently for $B$, then said simulation is also valid for all channels $A$ and $B$, due to the linearity of higher-order transformations. Crucially, in the case where $\map{C}$ is a quantum comb, the maximum probability of success of a simulation of the quantum switch may depend on the order of the input channels $A$ and $B$. In the case where $\map{C}$ is a QC-CC, all possible fixed and dynamical orders are automatically optimized over.

We are now ready to state this result.

\begin{theorem}\label{thm::nogo}
    There is no quantum circuit that can deterministically simulate the action of the quantum switch on all quantum channels when $(k_A,k_B) \in \{(1,1),(2,1),(2,2),(3,1)\}$, where $k_A$ is the number of calls of input channel $A$ and $k_B$ is the number of calls of input channel $B$.

    Moreover, even for a restricted simulation---i.e., for fixed input systems and discarded output target system---the action on the quantum switch on general qubit channels, when $(k_A,k_B) \in \{(1,1),(2,1),(2,2),(3,1)\}$, can be simulated with at most a probability $p<1$, with upper bounds given in Table~\ref{tb::probs}.
\end{theorem}

We prove this theorem using a method that we develop for computer-assisted proofs that transforms the numerically imperfect solution of an SDP into a rigorous upper-bound for the probability of success that can be expressed in terms of rational numbers. The method is presented in Sec.~\ref{sec::methods} with further details provided in App.~\ref{app::compassist}.

%%%%%%%%%%%%%%%%%%%%%%%%%%%%%%%%%%%%%%%%%%%%%%%%%%%%
\begin{table}%[h!]
\begin{center}
{\renewcommand{\arraystretch}{1.7}
\begin{tabular}{| c | c | c |}
	\hline
	\hspace*{3mm} $(k_A,k_B)$ \hspace*{3mm} & \hspace*{5mm} order \hspace*{5mm} & \hspace*{8mm} probability \hspace*{8mm}\\
    \hline
    \hline
    $(1,1)$ & AB  & $p<\frac{4001}{10000}$ \\
    \hline
            & AAB  & $p<\frac{5715}{10000}$ \\
    $(2,1)$ & ABA  & $p<\frac{4919}{10000}$ \\
            & BAA  & $p<\frac{5001}{10000}$ \\ 
    \hline        
            & AABB & $p< \frac{8307}{10000}$ \\
    $(2,2)$ & ABAB & $p< \frac{8484}{10000}$ \\
            & ABBA & $p< \frac{8695}{10000}$ \\	
    \hline
            & AAAB & $p< \frac{8373}{10000} $\\
    $(3,1)$ & AABA & $p< \frac{6909}{10000}$ \\ 
            & ABAA & $p<\frac{7597}{10000}$ \\
            & BAAA & $p<\frac{6845}{10000}$ \\
    \hline
\end{tabular}
}
\end{center}
\caption{\textbf{Maximum probability of a restricted simulation of the qubit quantum switch.} {Rigorous} upper bounds for the maximum probability {$p$} of a restricted simulation of the quantum switch acting on a pair of arbitrary qubit quantum channels $A,B$, using open-slot quantum circuits that take $k_{A(B)}$ calls of an arbitrary quantum channel $A(B)$ in a certain order. {All bounds were derived via computer-assisted proofs.}}
\label{tb::probs}
\end{table}
%%%%%%%%%%%%%%%%%%%%%%%%%%%%%%%%%%%%%%%%%%%%%%%%%%%%

Still in the restricted simulation scenario, we found three different particular cases where a deterministic simulation  {is possible} for qubit channels, but nevertheless impossible in the more general scenario. In the following we discuss these three cases, with more details being provided in App.~\ref{app::gorestricted}.

The first case of this kind is when the quantum switch acts on arbitrary, yet identical, qubit quantum channels, i.e., $A=B$. Although the input channels in this case are identical, it is straightforward to see from Eq.~\eqref{eq::switch} that the action of the switch is non-trivial, namely, that the output channel $\map{S}(A,A)$ is not equivalent to $A$ applied twice on the target system when $A$ is not a unitary channel. A remarkable example of this point is the case where $A$ is the depolarizing channel~\cite{ebler2018enhanced}. We find that a restricted deterministic quantum comb simulation exists for qubit channels in the ``AAAA'' scenario, i.e., with $4$ identical calls of the arbitrary qubit quantum channel $A$. Since in this case the maximum probability of success is $p=1$, one cannot certify this result with a computer-assisted proof, which can yield upper and lower bounds for $p$ with arbitrary yet only finite precision. Instead, we obtain this result by numerically evaluating an SDP with very high precision, ensuring that all positivity constraints are strictly satisfied and that all equality constraints are satisfied up to an error of at most $10^{-9}$ in the operator norm. Furthermore, $4$ identical calls of the arbitrary qubit quantum channel $A$ are not only sufficient, but necessary for a deterministic restricted simulation: if only $2$ or $3$ calls are allowed, we prove upper bounds for the maximum probability of simulation that are always strictly less than $1$ (see Table~\ref{tb::probs2}). However, this only holds in the restricted simulation case. When the output target system is not discarded, we find numerically that a deterministic simulation of the AAAA scenario is no longer possible (see App.~\ref{app::gorestricted} for more details). This shows how highly nontrivial the action of the quantum switch is even when acting on identical quantum channels. 

%%%%%%%%%%%%%%%%%%%%%%%%%%%%%%%%%%%%%%%%%%%%%%%%%%%%
\begin{table}%[h!]
\begin{center}
{\renewcommand{\arraystretch}{1.7}
\begin{tabular}{| c | c | c |}
	\hline
	\hspace*{8mm} $k$ \hspace*{8mm} & \hspace*{5mm} order \hspace*{5mm} & \hspace*{8mm} probability \hspace*{8mm}\\
    \hline
    \hline
    $2$ & AA  & $p<\frac{4001}{10000}$ \\
	\hline
    $3$ & AAA  & $p<\frac{6534}{10000}$ \\
	\hline
    $4$ & AAAA & $p = 1 \ ({}^*)$ \\
	\hline
\end{tabular}
}
\end{center}
\caption{\textbf{Bounds for the maximum probability of a restricted simulation of the qubit quantum switch acting on identical channels.} Rigorous upper bounds for the maximum probability $p$ of a restricted simulation of the quantum switch that acts on a pair of identical qubit quantum channels, using open-slot quantum circuits that take $k\in\{2,3\}$ calls of an arbitrary quantum channel $A$. {All bounds were derived via computer-assisted proofs.} (*) For the case of $k=4$, we numerically find that a deterministic simulation is possible, i.e., that $p=1$ up to an numerical precision of $10^{-9}$.}
\label{tb::probs2}
\end{table}
%%%%%%%%%%%%%%%%%%%%%%%%%%%%%%%%%%%%%%%%%%%%%%%%%%%%

The other two restricted cases where a deterministic restricted simulation is possible concern qubit unitary channels applied in the order AABB and BAAA, two cases where trivial extensions of the circuit in Eq.~\eqref{cc::chiribella} fail. Similarly to the case of identical qubit channels, such simulations no longer exist if the output target system is not discarded. More details in App.~\ref{app::gorestricted}.

%%%%%%%%%%%%%%%%%
\subsection*{No-go theorems: Approximate simulations}\label{subsec::robustness}
%%%%%%%%%%%%%%%%%

By exhibiting concrete upper bounds for the probability of successful simulation of the quantum switch that are significantly below $1$, we demonstrate the extent of the robustness of our results with respect to a figure of merit related to a probabilistic simulation. Another pertinent notion of robustness in this context is that of an approximate simulation---although the quantum switch may not be able to be exactly simulated, it could still be the case that other higher-order transformations that are $\epsilon$-close to the quantum switch can be simulated. This question is particularly relevant for experimental scenarios, where noise and imprecision hinder the ability of preparing any specific transformation exactly. Here, we prove that even in the approximate case a simulation of the quantum switch is not possible, for some particular number of calls $(k_A,k_B)$. We do so by exhibiting values of $\epsilon$ that are significantly above zero for which even a probabilistic simulation, for some $p<1$ that we also exhibit, is not possible. 

%%%%%%%%%%%%%%%%%%%%%%%%%%%%%%%%%%%%%%%%%%%
\begin{figure}%[h!]
\begin{center}
	\includegraphics[width=\columnwidth]{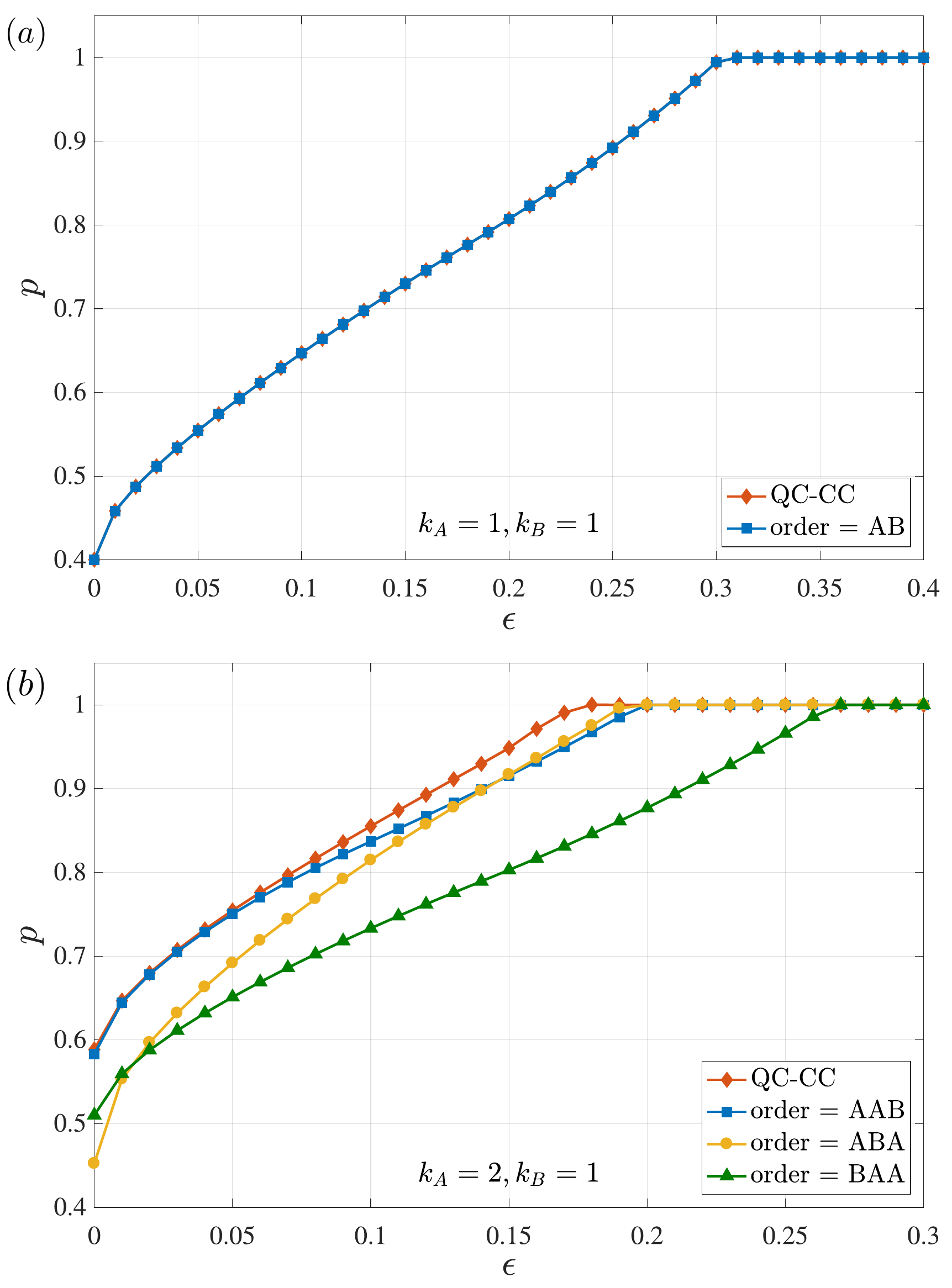}
	\caption{\textbf{Robustness of the impossibility of simulation.} Maximum probability of success {$p$} of simulating any higher-order transformation that is $\epsilon$-close to the quantum switch using quantum combs of different orders or QC-CCs, as a function of $\epsilon$. In plot $(a)$, we show the case where $k_A=k_B=1$. {Note} how the QC-CC simulation curve numerically coincides with the comb simulation curve. In plot $(b)$, we show the case where $k_A=2$ and $k_B=1$. In this case, for some values of $\epsilon$, QC-CCs show an advantage with respect to quantum combs.}
\label{fig::plot_epsilon}
\end{center}
\end{figure}
%%%%%%%%%%%%%%%%%%%%%%%%%%%%%%%%%%%%%%%%%%%

For an approximate simulation of the quantum switch, it is useful to consider a partly restricted scenario that is required to hold for fixed input control and target systems, but without discarding the output target system. Similarly to the restricted simulation case, proving the impossibility of a simulation in this partly restricted case also implies the impossibility of a simulation in more general scenarios. Let $\sigma_c=\ketbra{+}{+}$ be the state of the input control system, $\rho_t=\ketbra{0}{0}$ be the state of the input target system, and $\widetilde{\map{S}}$ be a higher-order transformation that acts on the same space as the quantum switch. We define a simulation $\map{C}$ to be $\epsilon$-close to the quantum switch $\map{S}$ in this scenario if there exists an $\widetilde{\map{S}}$ such that
\begin{align}\label{eq::approxsimulation}
\begin{split}
    \map{C}(A^{\otimes k_A},B^{\otimes k_B})&[\,\ketbra{+}{+}\otimes\ketbra{0}{0}\,] \\
    = \widetilde{\map{S}}(A,B)&[\,\ketbra{+}{+}\otimes\ketbra{0}{0}\,] \ \ \ \forall \ A, B,
\end{split}
\end{align}
where $A$ and $B$ are arbitrary quantum channels and, for $\map{S}_{+0}(\cdot,\cdot)\coloneqq \map{S}(\cdot,\cdot)[\ketbra{+}{+}\otimes\ketbra{0}{0}]$ and $\widetilde{\map{S}}_{+0}(\cdot,\cdot)\coloneqq \widetilde{\map{S}}(\cdot,\cdot)[\ketbra{+}{+}\otimes\ketbra{0}{0}]$, it holds that
\begin{equation}\label{eq::epsilon}
    F(\map{S}_{+0},\widetilde{\map{S}}_{+0}) \geq 1 - \epsilon,
\end{equation}
where $F(\map{M},\map{N})$ is the normalized fidelity between the Choi operators associated to the higher-order transformations $\map{M}$ and $\map{N}$ (see App.~\ref{subapp::switchchoi}).

Combining probabilistic-heralded and approximate protocols of simulation, for a fixed $\epsilon$, $(k_A,k_B)$, and $d$, the maximum probability of successful simulation can also be computed via an SDP. We show that, for a range of values of $\epsilon>0$, the probability of simulating the quantum switch in this partly restricted scenario is $p<1$ for the cases where $(k_A,k_B)\in\{(1,1),(2,1)\}$, considering simulations using quantum combs of all possible orders as well as QC-CCs. We present our numerical findings in Fig.~\ref{fig::plot_epsilon} for the case where $d=2$. In the first plot [Fig.~\ref{fig::plot_epsilon}$(a)$], where $(k_A,k_B)=(1,1)$, the QC-CC simulation curve numerically coincides with the quantum comb simulation curve. Here, $p<1$ for $0\leq\epsilon\lesssim0.3$. In the second plot [Fig.~\ref{fig::plot_epsilon}$(b)$], where $(k_A,k_B)=(2,1)$, each of the three possible quantum comb orders and QC-CC yield different values for the maximum probability of success for different values of $\epsilon$. Here, $p<1$ for $0\leq\epsilon\lesssim0.18$, after which point a QC-CC simulation exists. Notice that for an exact simulation, i.e., when $\epsilon=0$, a QC-CC simulation yields a probability of success that coincides with the highest among all possible orders, but as $\epsilon$ increases, it shows an advantage in the probability of success as compared to quantum combs. 

%%%%%%%%%%%%%%%%%
\subsection*{A conjecture}\label{subsec::conjecuture}
%%%%%%%%%%%%%%%%%

Taken together, our results indicate the hardness of simulating the quantum switch. Considering all the evidence we have gathered here supporting a high cost, if not altogether the impossibility, of simulating the quantum switch deterministically, we are motivated to propose the following conjecture:

\begin{conjecture} 
    There is no $(k_A+k_B)$-slot higher-order transformation $\map{C}$, described by a quantum circuit with fixed or classically-controlled causal order, that can simulate the quantum switch, i.e., that satisfies
	\begin{align}
		\map{C}(A^{\otimes k_A},B^{\otimes k_B}) &= \map{S}(A,B)\label{eq:switch_simulation_1_maintext_conj}
	\end{align}
    for all $n$-qubit quantum channels $A$ and $B$, if $\max(k_A,k_B) \leq  g(n)$, for some $g(n) = \Theta(2^n)$.
\end{conjecture}
Our rationale behind conjecturing that no simulation is possible even with multiple (albeit a sub-exponential number of) calls to  {both} input channels is the following. As mentioned above, there exists a deterministic and exact simulation of the quantum switch with a single query to a general channel $B$ and two queries to a unitary channel $A$. However, simulating the quantum switch for general channels $A$ and $B$ requires correlating each Kraus operator $A_k$ on the $\ket{0}$ branch---which can be obtained by querying $A$ before $B$---with the same $A_k$ on the $\ket{1}$ branch---which can be obtained by querying $A$ after $B$ [see Eq.~\eqref{eq::switchkraus}]. In this view, $B$ can be considered as a fixed channel~\cite{guerin2018observer}, and therefore the intuition is that querying it multiple times is no better than querying it once. The rational for the bound to be $\Theta(2^n)$ is that all the main steps in the proof of Thm.~\ref{thm::nogo_exp} except one hold for a bound of $\Theta(2^n)$ with $(k_A+k_B)$-slot higher-order transformations, and only for Lemma~\ref{lem:indep_M1} were we only able to prove the $(k_A+1)$-slot case. It remains to be seen whether the action of the quantum switch can be simulated with any finite number of queries to one or both input channels.

%%%%%%%%%%%%%%%%%%%%%%%%%%%%%%%%%%%%%%%%%%%%
\section{Discussion}\label{sec::discussion}
%%%%%%%%%%%%%%%%%%%%%%%%%%%%%%%%%%%%%%%%%%%%

In this work, we have concretely formulated the problem of simulating higher-order transformations with indefinite causal order using higher-order transformations that obey causal constraints and have access to more calls of the input operations. Our case study for this problem was the quantum switch. We defined the problem in its full generality, considering also cases where the quantum switch acts only on part of its input quantum channels.

We showed that the (one-sided) quantum query complexity of the action of the quantum switch on $n$-qubit channels, with respect to all quantum circuits with fixed and classically-controlled causal order, is lower bounded by $2^n$ (Thm.~\ref{thm::nogo_exp}). Notably, the exponential separation that we prove is formulated with respect to a computational task where the inputs and outputs of the computation are given by black-box quantum channels~\cite{chiribella2008quantum,chiribella2008transforming,chiribella2013quantum,chiribella2018indefinite}. This is in contrast to previous works on the query complexity of the quantum switch, where the output of the computation is a bit representing the evaluation of a classical function, in which case no such exponential separation has been found~\cite{chiribella2012perfect,araujo2014computational,renner2022computational,abbott2024quantum,taddei2021computational}.

We additionally proved that an extra copy of each input quantum channel, i.e., $(k_A,k_B)=(2,2)$, is not sufficient to deterministically simulate the quantum switch with a quantum circuit. Considering probabilistic simulations, we proved a range of rigorous strict upper bounds on the probability of successful simulation in various different scenarios (Thm.~\ref{thm::nogo}). In order to do so, we developed and applied a general method of computer-assisted proofs based on semidefinite programming. We also explored approximate simulations in some cases, showing that the switch cannot be simulated even approximately for some $\epsilon>0$ with probability $p=1$.

Finally, we showed that for some particular kinds of input, non-universal simulations exist. The most general such case is when $A$ is a bipartite unitary channel, $B$ is a bipartite general channel, and the quantum switch acts only on part of the input channels. Such a simulation exists a requires only an extra call of $A$ (Thm.~\ref{thm::go}). A simulation also exists for the particular case where the input pair of channels of the switch are identical qubit quantum channels, $A=B$. Here, $4$ calls of the input channel are necessary and sufficient for a restricted deterministic simulation, where the input states are fixed and the target output system is discarded. However, this result does not hold for more general simulation scenarios, where the output target system is not discarded. This possibility of simulation in the restricted case, but not in more general cases, is also true for the simulation of the action of the switch on qubit unitary channels with the orders AABB and BAAA. 

Our results have implications for the analysis of the experiments based on the quantum switch~\cite{rozema2024experimental}, particularly the ones that make use of non-unitary channels, such as those reported in Refs.~\cite{rubino2021experimental,cao2023semideviceindependent,antesberger2024higherorder,goswami2018increasing,guo2018experimental}. A long-standing debate in the community is whether these experiments are a genuine realization or some form of simulation of the quantum switch. One important criticism is that several experimental setups leave room for the interpretation that there are two independent calls of each input quantum channel available in the transformation. Here, we have proven that such access still does not allow for a causally ordered explanation for the observed data. On the other hand, an alternative causal model that can reproduce the action of the quantum switch is one that, instead of considering the inputs to be two independent calls of a general channel, considers inputs that can be described as bipartite channels with memory, which can be interpreted as two correlated uses of a general quantum channel~\cite{chiribella2019quantum,kristjansson2021witnessing}. As a corollary of our result in Thm.~\ref{thm::nogo}, we observe that there does not exist a quantum circuit that can take two independent uses of an arbitrary channel as input and output a bipartite channel with memory that corresponds to two correlated uses of the input quantum channels. Therefore, a causal model based on correlated inputs is not compatible with a simulation scenario that considers black-box inputs, as considered here. However, this point alone is not sufficient to rule out this model is a fair description of the experiments. It is our hope that our results will encourage further rigorous analysis of potential loopholes in the quantum switch experiments.

Our work opens up the study of query complexity in the context of higher-order quantum computation where the inputs and outputs of the computation are general quantum channels. The question of whether any finite number of calls suffices to perform a simulation of the switch is of high relevance and remains open. Further investigation of this topic will be crucial to determine, for example, whether advantages in specific tasks that have been previously demonstrated can persist in the asymptotic limit of the available number of calls. Should it ever be shown that the quantum switch can be simulated with a finite number of calls, a relevant follow-up question is related to the scaling of the necessary number of calls, and to whether or not the switch can be efficiently simulated in a deterministic setting. 

The existence of efficient deterministic and exact simulation of quantum processes with indefinite causal order may pose itself as an interesting physical guideline to help us understand on the one hand which kinds of higher-order transformations can be realistically implemented, and on the other hand, what advantages they may provide.

%%%%%%%%%%%%%%%%%%%%%%%%%%%%%%%%%%%%%%%%%%%%
\section{Methods}\label{sec::methods}
%%%%%%%%%%%%%%%%%%%%%%%%%%%%%%%%%%%%%%%%%%%%

%%%%%%%%%%%%%%%%%%%%%
\noindent\textbf{Preliminaries.} In our technical calculations, we make use of the Choi-Jamio\l{}kowski isomorphism~\cite{jamiolkowski1972linear,choi1975completely} and the link product~\cite{chiribella2008quantum}. 

Using this isomorphism, any linear map $M:\map{L}(\map{H}^{I})\to\map{L}(\map{H}^{O})$ that maps linear operators acting on an input space to linear operators acting on an output space can be represented by a linear operator $J^M\in\map{L}(\map{H}^{I}\otimes\map{H}^{O})$ that acts on the joint input and output space, called the  {Choi operator}, which is given by 
\begin{equation}\label{eq::choi}
    J^M \coloneqq \sum_{ij} \ketbra{i}{j}\otimes M[\ketbra{i}{j}], 
\end{equation}
where $\{\ket{i}\}_i$ is the computational basis. A linear map $M$ is completely positive (CP) if and only if its Choi operator $J^M$ is positive semidefinite, and is trace-preserving (TP) if and only if $J^M$ satisfies $\tr_O(J^M)=\id^I$, where $\id^{I}$ is the identity operator in $\map{L}(\map{H}^{I})$. Hence, the Choi operator of general quantum channels satisfy these conditions.

Any linear operator $\ope{U}:\map{H}^{I} \to \map{H}^{O}$ can be represented by its  {Choi vector} $\dket{\ope{U}}\in\map{H}^{I}\otimes\map{H}^{O}$, where
\begin{equation}
    \dket{\ope{U}}\coloneqq \sum_{i} \ket{i} \otimes \ope{U} \ket{i}.
\end{equation}
Hence, a unitary channel $U$ such that $U[\rho]=\ope{U}\rho\ope{U}^\dagger$, where $\ope{U}$ is a unitary operator that corresponds to the single Kraus operator of the channel $U$, has Choi operator $J^U = \dketbra{\ope{U}}$.

Using this representation, the composition $G \circ F$ of two maps $F:\map{L}(\map{H}^{1})\to\map{L}(\map{H}^{2})$ and $G:\map{L}(\map{H}^{2})\to\map{L}(\map{H}^{3})$, with respective Choi operators $J^F\in\map{L}(\map{H}^{1}\otimes\map{H}^{2})$ and $J^G\in\map{L}(\map{H}^{2}\otimes\map{H}^{3})$, is given by the  {link product} $J^F*J^G\in\map{L}(\map{H}^{1}\otimes\map{H}^{3})$, according to
\begin{equation}\label{eq::linkproduct}
    J^F*J^G \coloneqq \tr_2[(J^F\otimes\id^3)(\id^1\otimes {(J^G)}^{T_2})],
\end{equation}
where $(\cdot)^{T_X}$ denotes partial transposition on the space $\map{L}(\map{H}^{X})$. 

The link product for Choi vectors $\dket{\ope{Q}}\in\map{H}^{1}\otimes\map{H}^{2}$ and $\dket{\ope{R}}\in\map{H}^{2}\otimes\map{H}^{3}$ is given by $\dket{\ope{Q}}\ast\dket{\ope{R}}\in\map{H}^{1}\otimes\map{H}^{3}$, defined as
\begin{equation}
    \dket{\ope{Q}} \ast \dket{\ope{R}} \coloneqq \sum_i (\id^1 \otimes \bra{i})\dket{\ope{Q}} \otimes (\bra{i} \otimes \id^3)\dket{\ope{R}}.
\end{equation}
\\

%%%%%%%%%%%%%%%%%%%%%
\noindent{\textbf{Sketch of the proof of Theorem~\ref{thm::nogo_exp}.}} 
While we provide the full proof of Thm.~\ref{thm::nogo_exp} in App.~\ref{app::fullproof_nogoexp}, here we give a sketch of the proof for the case where $k_A=2$.

Here we adopt a slightly different notation from the main text, and write the higher-order transformation $\map{C}$ as a function $\map{C}(A,B,C)$ that takes three arguments, each corresponding to a quantum channel that can be plugged into one of its three slots. Hence, instead of writing, e.g., $\map{C}(A^{\otimes 2},B)$ we will use $\map{C}(A,A,B)$.

Let $\map{C}(A,A,B)$ be a simulation of the action of the quantum switch $\map{S}(A,B)$ on all mixed unitary quantum channels $A$ and unitary channels $B$. Then, for arbitrary unitary channels $U_1, U_2, V$, the simulation $\map{C}$ necessarily respects
\begin{equation}\label{eq:C_linear_unitary}   
    \map{C}\left(\frac{U_1+U_2}{2}, \frac{U_1+U_2}{2}, V \right) = \map{S} \left(\frac{U_1+U_2}{2}, V \right).
\end{equation}
By linearity, % Eqs.~\eqref{eq:C_linear_unitary_2} and 
Eq.~\eqref{eq:C_linear_unitary} implies that
\begin{align}\label{eq:C_linear_unitary_2}
    \sum_{i,j=1}^2\map{C}(U_i,U_j,V) = 2 \sum_{k=1}^2\map{S}(U_k,V).
\end{align}

Let $C$ be the Choi operator of the simulation $\map{C}$, $S=\dketbra{S}$ be the Choi operator of the quantum switch transformation $\map{S}$, defined by Eq.~\eqref{eq::switch} and written explicitly in App.~\ref{subapp::switchchoi}, and $\dketbra{\ope{U}_l}$ denote the Choi operator of a unitary channel $U_l$. Then, Eq.~\eqref{eq:C_linear_unitary_2} can be expressed as
\begin{align}
\begin{split}
    \sum_{i,j=1}^2 C &\ast \Big(\dketbra{\ope{U}_i}\otimes \dketbra{\ope{U}_j}\otimes\dketbra{\ope{V}}\Big)\\
    &= 2 \sum_{k=1}^2 \dketbra{S}\ast \Big(\dketbra{\ope{U}_k}\otimes \dketbra{\ope{V}}\Big).
\end{split}
\end{align}

Assuming that $C$ is positive semidefinite, it accepts a decomposition $C = \sum_a \dketbra{C^{(a)}}$ where $\dketbra{C^{(a)}}\geq0$ for all $a$. From the positivity of $\dketbra{\ope{U}_l}$ and $\dketbra{\ope{V}}$, it follows that
\begin{align}\label{eq:inequality_Cijka}
\begin{split}
    &\dketbra{C^{(a)}}\ast\Big(\dketbra{\ope{U}_1}\otimes \dketbra{\ope{U}_2}\otimes\dketbra{\ope{V}}\Big)\\
    &\leq 2 \dketbra{S}\ast \Big[\big(\dketbra{\ope{U}_1}+\dketbra{\ope{U}_2}\big)\otimes \dketbra{\ope{V}}\Big],
\end{split} 
\end{align}
for every $a$.

Following the definition of the link product for Choi vectors, $\dketbra{\ope{Q}} \ast \dketbra{\ope{R}}$ is given by $(\dket{\ope{Q}}\ast \dket{\ope{R}}) (\dket{\ope{Q}}\ast \dket{\ope{R}})^\dagger$ (see, e.g., Lemma~1 of Ref.~\cite{yokojima2021consequences}).
Thus, the support of the right-hand side of Eq.~\eqref{eq:inequality_Cijka} is given by $\mathrm{span}\{\dket{S} \ast (\dket{\ope{U}_k} \otimes \dket{\ope{V}})\}_{k=1}^{2}$ and that of the left-hand side of Eq.~\eqref{eq:inequality_Cijka} is the one-dimensional subspace spanned by $\dket{C^{(a)}}\ast (\dket{\ope{U}_1}\otimes \dket{\ope{U}_2}\otimes \dket{\ope{V}})$.
Therefore, one can write
\begin{align}\label{eq:span}
\begin{split}
    &\dket{C^{(a)}}\ast \Big(\dket{\ope{U}_1}\otimes \dket{\ope{U}_2}\otimes \dket{\ope{V}}\Big) \\
    &= \sum_{k=1}^{2} \xi_{k}^{(a)} (\ope{U}_1, \ope{U}_2, \ope{V}) \dket{S}\ast \Big(\dket{\ope{U}_k}\otimes \dket{\ope{V}}\Big),
\end{split}    
\end{align}
for some coefficients $\xi_{k}^{(a)} (\ope{U}_1, \ope{U}_2, \ope{V}) \in \mathbb{C}$. The proof of this fact uses Lemma~\ref{lem:span} from App.~\ref{subapp::lem:span} and Lemma~\ref{lem:linearity_MN} from App.~\ref{subapp::lem:linearity_MN}. 

We now invoke Lemma~\ref{lem:p_as_vector_MN} from App.~\ref{subapp::lem:p_as_vector_MN} and Lemma~\ref{lem:indep_M1} from App.~\ref{subapp::lem:indep_M1} to ensure that, when Eq.~\eqref{eq:span} is satisfied, there exist vectors $\dket{\xi_{1}^{(a)}}\in \map{H}^{I_2}\otimes \map{H}^{O_2}$ and $\dket{\xi_{2}^{(a)}}\in \map{H}^{I_1}\otimes \map{H}^{O_1}$, such that
\begin{align}
\label{eq:Ca}
    \dket{C^{(a)}} &= \sum_{k=1}^{2}\dket{S} \otimes \dket{\xi_{k}^{(a)}},
\end{align}
for all $a$, where $\dket{\xi_{1}^{(a)}}$ and $\dket{\xi_{2}^{(a)}}$ are independent of $\ope{U}_1$, $\ope{U}_2$, and $\ope{V}$.

Next, we argue why this is the case.

The basic idea for this part of the proof is based on differentiation with respect to a parametrization of the input unitary operators, a technique introduced in Ref.~\cite{odake2024analytical} by some of the present authors. Suppose that $\ope{U}_1$, $\ope{U}_2$, and $\ope{V}$ are taken from the set $\{I,X,Y,Z\}$ of Pauli operators. 
If $\ope{U}_1 \neq \ope{U}_2$, then $\dket{S}*(\dket{\ope{U}_1}\otimes\dket{\ope{V}})$ and $\dket{S}*(\dket{\ope{U}_2}\otimes\dket{\ope{V}})$ are linearly independent. In this case, we can show using linearity that $\xi_{1}^{(a)}(\ope{U}_1,\ope{U}_2,\ope{V})$ is independent of $\ope{U}_1,\ope{V}$ and $\xi_{2}^{(a)}(\ope{U}_1,\ope{U}_2,\ope{V})$ is independent of $\ope{U}_2,\ope{V}$.  If on the other hand $\ope{U}_1=\ope{U}_2=\sigma$, then $\dket{S}*(\dket{\ope{U}_1}\otimes \dket{\ope{V}})$ and $\dket{S}*(\dket{\ope{U}_2}\otimes \dket{\ope{V}})$ are not linearly independent.

In such cases, it turns out that $\xi_{1}^{(a)}(\sigma,\sigma,\ope{V})$ and $\xi_{2}^{(a)}(\sigma,\sigma,\ope{V})$ can be suitably chosen as $\xi_{1}^{(a)}(\sigma',\sigma,\ope{V})$ and $\xi_{2}^{(a)}(\sigma,\sigma',\ope{V})$, respectively, where $\sigma'\neq\sigma$ is a Pauli operator.
Note that $\xi_{1}^{(a)}(\sigma',\sigma,\ope{V})$ and $\xi_{2}^{(a)}(\sigma, \sigma',\ope{V})$ do not depend on the choice of $\sigma'$ as long as $\sigma'\neq\sigma$ holds.
The fact that such a redefinition is consistent with Eq.~\eqref{eq:span} can be proven by differentiating the expression $\xi_{i}^{(a)}(\tilde{\sigma}(\theta), \tilde{\sigma}(\theta),\ope{V})$, where $\tilde{\sigma}(\theta)$ is a parameterized unitary operator satisfying $\tilde{\sigma}(0)=\sigma$ and $\frac{\rm d}{{\rm d}\theta}\!\mid_{\theta=0}\tilde{\sigma}(\theta)\propto\sigma'$. 
This redefinition implies that that $\xi_{1}^{(a)}(\ope{U}_1,\ope{U}_2,\ope{V})$ and $\xi_{2}^{(a)}(\ope{U}_1,\ope{U}_2,\ope{V})$ are independent of $\ope{U}_1$ and $\ope{U}_2$, respectively. By linearity, we can show that for $i\in 
\{1,2\}$, $\xi_{k}^{(a)}(\ope{U}_1,\ope{U}_2,\ope{V})$ is independent of $\ope{V}$. 

The independence relations above imply that we can write $\xi_{1}^{(a)} (\ope{U}_1,\ope{U}_2,\ope{V}) = \dket{\xi_{1}^{(a)}} \ast \dket{\ope{U}_2}$ and  $\xi_{2}^{(a)} (\ope{U}_1,\ope{U}_2,V) = \dket{\xi_{2}^{(a)}} \ast \dket{\ope{U}_1}$ for some vectors $\dket{\xi_{1}^{(a)}}, \dket{\xi_{2}^{(a)}}$. Substituting this into Eq.~\eqref{eq:span} gives
\begin{align}
\begin{split}
    &\dket{C^{(a)}}\ast\Big(\dket{\ope{U}_1}\otimes \dket{\ope{U}_2}\otimes\dket{\ope{V}}\Big) \\
    &=\sum_{k=1}^{2}\dket{S}\otimes\dket{\xi_{k}^{(a)}}\ast\Big(\dket{\ope{U}_1}\otimes\dket{\ope{U}_2}\otimes\dket{\ope{V}}\Big).
\end{split}
\end{align}
Since this holds for all combinations of Pauli operators $\ope{U}_1, \ope{U}_2, \ope{V}$, we obtain Eq.~\eqref{eq:Ca}. 

Hence, we have shown that the assumption of a simulation of the quantum switch given by a higher-order transformation with Choi operator $C=\sum_a\dketbra{C^{(a)}}\geq0$ implies the existence of vectors $\dket{\xi_{1}^{(a)}}$ and $\dket{\xi_{2}^{(a)}}$ that satisfy Eq.~\eqref{eq:Ca}. Finally, we invoke Lemma~\ref{lem:qccc_invalid} from App.~\ref{subapp::lem:qccc_invalid}, which states that such a higher-order transformation does not satisfy the QC-CC conditions.
\\

%%%%%%%%%%%%%%%%%%%%%
\noindent{\textbf{Semidefinite programming.}} Here we show that the problem of simulating the quantum switch can be phrased as an SDP.

Let $S$ be the Choi operator of the quantum switch transformation $\map{S}$, defined in Eq.~\eqref{eq::switch}, written explicitly in App.~\ref{subapp::switchchoi}. Let $J^A$ and $J^B$ be the Choi operators of the quantum channels $A$ and $B$. Finally, let $C$ be the Choi operator of the deterministic higher-order transformation $\map{C}$, which corresponds to a quantum comb, and $C_s$ be the Choi operator of the higher-order transformation $\map{C}_s$ associated with the success outcome of a probabilistic  transformation. Then, the maximum probability of simulating the action of the quantum switch $S$ on a set of $N_A$ channels $\{J^A_i\}_{i=1}^{N_A}$ and a set of $N_B$ channels $\{J^B_j\}_{j=1}^{N_B}$ using a quantum comb $C$ that acts on $k_A$ calls of $J^A_i$ and $k_B$ calls of $J^B_j$ is given by the following SDP:
\begin{flalign}\label{sdp::primal}
\begin{aligned}
    \textbf{given}\ \ & \{J^A_i\}_i,\{J^B_j\}_j, k_A,k_B \\
    \textbf{max}\ \   & p \\
    \textbf{s.t.}\ \  & C_s*[{(J^A_i)}^{\otimes k_A}\otimes {(J^B_j)}^{\otimes k_B}] = p\,S*(J^A_i\otimes J^B_j) \ \ \forall\,i,j \\
    & C_s \geq 0, \ \ C - C_s \geq 0, \\ 
    & \mathbb{P}(C) = C, \ \ \tr(C)= d_{c_I}d_{t_I}d_{A_O}^{k_A}d_{B_O}^{k_B},
\end{aligned}&&
\end{flalign}
where $\mathbb{P}$ is the projector onto the linear subspace spanned by valid quantum combs~\cite{chiribella2008quantum,araujo2015witnessing,milz2024characterising} (explicitly written in App.~\ref{subapp::combs}), and $d_X\coloneqq \text{dim}(\map{H}^X)$. For a simulation given by QC-CC transformations, the constraint $\mathbb{P}(C) = C$ should be substituted by the appropriate linear constraints that define a proper QC-CC transformations, as well as additional normalization constraints on $C_s$~\cite{wechs2021quantum}. These constraints are explicitly written in App.~\ref{subapp::qcccs} for the cases of $2$ and $3$ slots. Notice that, following results from Sec. III of Ref.~\cite{quintino2019probabilistic}, the variables $C$ and $C_s$ in this SDP can be restricted to the field of real numbers without loss of generality, a feature that improves numerical performance.

The dual problem associated to this SDP can be found by combining standard methods~\cite{skrzypczyk2023semidefinite} with dual affine techniques (see, e.g., the derivation of a similar dual problem in Appendix B of Ref.~\cite{bavaresco2021strict}, inspired by Ref.~\cite{chiribella2016optimal}). It is given by
\begin{flalign}\label{sdp::dual}
\begin{aligned}
    \textbf{given}\ \ & \{J^A_i\}_i,\{J^B_j\}_j, k_A,k_B \\
    \textbf{min}\ \   & \frac{1}{d_{A_I}^{k_A}d_{B_I}^{k_B}d_{c_O}d_{t_O}} \tr(\Gamma) \\
    \textbf{s.t.}\ \  & \sum_{i,j} \tr[R_{ij}\,(S*(J^A_i\otimes J^B_j))] = 1 \\
    & \Gamma - \sum_{i,j} R_{ij}\otimes[{(J^A_i)}^{\otimes k_A}\otimes {(J^B_j)}^{\otimes k_B}]^T \geq 0 \\
    & \Gamma \geq 0, \ \ \overline{\mathbb{P}}(\Gamma) = \Gamma,
\end{aligned}&&
\end{flalign}
where $\overline{\mathbb{P}}$ is the projector onto the linear subspace spanned by the dual affine of valid combs~\cite{chiribella2016optimal,bavaresco2021strict,milz2024characterising} (also explicitly written in App.~\ref{subapp::combs}). In the case of this dual SDP problem, the variable $\Gamma$ may also be restricted to the field of real numbers, since it is the Lagrange multiplier associated to the constraint $\mathbb{P}(C)=C$ (an equality between real matrices); however, the same is not true for the variables $\{R_{ij}\}$, which are the Lagrange multipliers associated to the constraints $C_s*[{(J^A_i)}^{\otimes k_A}\otimes {(J^B_j)}^{\otimes k_B}] = p\,S*(J^A_i\otimes J^B_j)$, an equality between complex matrices. Nonetheless, restricting $\{R_{ij}\}$ to real values still allows for feasible points that yield a valid upper bound for the optimal solution of this problem.

Finally, the SDPs~\eqref{sdp::primal} and~\eqref{sdp::dual} satisfy the condition of strong duality, which is implied by the fact that a strictly feasible point for the primal problem can be created from the probabilistic simulation of Ref.~\cite{chiribella2013quantum} that requires no extra calls.
\\

%%%%%%%%%%%%%%%%%%%%%
\noindent{\textbf{Basis design.}} 
For any input set of channels, the solution of SDP~\eqref{sdp::primal}, or equivalently of SDP~\eqref{sdp::dual}, corresponds to the maximum probability of success of a simulation of the action of the quantum switch  {on the specific given inputs}, with a simulation that has access to the specified number of calls of the input channels and that satisfies the specified causal constraints. Then, the maximum probability of success of a universal switch simulation---one that works for all possible input channels and not only for the given inputs---can be obtained by setting $\{{(J^A_i)}\}_i$ to be a set of operators that forms a basis for the $\text{span}(\{(J^A_i)^{\otimes k_A}\}_i)$, i.e., the subspace spanned by $k_A$ copies of an arbitrary quantum channel, and equivalently for $\{{(J^B_i)}\}$. In this case, if the constraint $C_s*[{(J^A_i)}^{\otimes k_A}\otimes {(J^B_j)}^{\otimes k_B}] = p\,S*(J^A_i\otimes J^B_j)$ holds for all $i,j$, then it also holds for arbitrary quantum channels. In other words, it implies that $C_s*[{(J^A)}^{\otimes k_A}\otimes {(J^B)}^{\otimes k_B}] = p\,S*(J^A\otimes J^B)$ holds for  {all} channels $J^A$ and $J^B$, thereby implying the existence of a universal probabilistic simulator with success probability $p$.

There are a few properties that the elements $J^A_i$ and $J^B_j$ of these bases must satisfy in order to be valid inputs of the SDPs~\eqref{sdp::primal} and~\eqref{sdp::dual}. The first is that it is necessary to ensure that the operators $J^A_i$ and $J^B_j$ individually correspond to TP maps, in order for the total trace of both sides of the constraint $C_s*[{(J^A_i)}^{\otimes k_A}\otimes {(J^B_i)}^{\otimes k_B}] = p\,S*(J^A_i\otimes J^B_i)$ to match. However, they do not need to correspond to CP maps, i.e., to be positive semidefinite operators. That is because all elements of the $\text{span}(\{J^A_i\}_i)$ can be written as linear combinations of the elements of a set $\{{J'}^A_i\}_i$ which correspond to TP maps (but not necessarily CP maps) and form a basis for the space spanned by quantum channels. Finally, it is also necessary that these operators can be themselves expressed as a tensor power of a TP map, because they play the role of the inputs of the quantum switch, which takes only a single call of each input. That is, they appear on both sides of the constraint $C_s*[{(J'^A_i)}^{\otimes k_A}\otimes {(J'^B_j)}^{\otimes k_B}] = p\,S*(J'^A_i\otimes J'^B_j)$.

We now construct a convenient basis for the linear space spanned by the set of $k$ copies of any quantum channel, which will be used in the computer-assisted proofs presented subsequently.

First, note that an arbitrary self-adjoint two-qubit operator $M\in\mathcal{L}(\mathcal{H}^I\otimes \mathcal{H}^O)\cong\mathcal{L}(\mathbb{C}^2\otimes \mathbb{C}^2)$ can always be written as
\begin{align}
\begin{split}
    M =\, &\lambda\, \id\otimes\id \\
    &+ \sum_i \alpha_i\, \sigma_i\otimes \id + \sum_j \beta_j\, \id \otimes \sigma_j  + \sum_{ij} \gamma_{ij} \sigma_i\otimes\sigma_j,
\end{split}
\end{align}
where all $\sigma_i\in\{X,Y,Z\}$ are Pauli operators, and $\lambda,\alpha_i,\beta_j,\gamma_{ij}\in\mathbb{R}$. Since operators $M$ that correspond to a TP map satisfy $\tr_O(M)=\id^I$, in this case one has that $\lambda=1/2$ and $\alpha_i=0$. This implies that the dimension of the linear space spanned by the set of qubit quantum channels is $3+9+1=13$. One can then construct a convenient basis for the subspace spanned by qubit quantum channels, given by
\begin{align}\label{eq::basis1}
\begin{split}
    \mathcal{B}_1\coloneqq \Big\{&\id\otimes\frac{\id}{2}, \\
    &\id\otimes\frac{\id}{2} + \id\otimes\sigma_i, \\
    &\id\otimes\frac{\id}{2} + \sigma_j\otimes\sigma_k \Big\}_{i,j,k}.
\end{split}
\end{align}
Notice that all elements of the basis set $\mathcal{B}_1$ correspond to TP maps, even if they are not necessarily positive semidefinite.

We now consider the linear space spanned by the set of $k$ copies of any arbitrary qubit channel, i.e., the span of the set of all ${J}^{\otimes k}$ such that $\tr_O(J)=\id^I$. If a linear subspace $\mathcal{V}$ has dimension $d_\mathcal{V}$, then the dimension of the space $\text{span}(\{{J}^{\otimes k}|\,\text{$J$}\in\mathcal{V}\})$ is given by $\binom{d_\mathcal{V}-1+k}{k}$~\cite{watrous2018theory}. By setting $d_\mathcal{V}=13$ and $k=2$, we see that the space spanned by the set of two identical copies of qubit channels has dimension $91$. Hence, in order to find a basis for two identical copies of qubit channels, it is enough to exhibit a set of $91$ operators $J_i$, all respecting $\tr_O(J_i)=\id^I$, such that the set $\{{J_i}^{\otimes 2}\}_{i=1}^{91}$ is composed of linearly independent operators.

Let $\rho_i\in\{\id\otimes\frac{\id}{2},2 \ketbra{\phi^+}{\phi^+},\id\otimes\ketbra{0}{0},\id\otimes\ketbra{+}{+},\id\otimes\ketbra{+_Y}{+_Y}\}$, where $\ket{\phi^{\pm}}\coloneqq \frac{1}{\sqrt{2}}(\ket{00}\pm\ket{11})$ are maximally entangled two-qubit states and $\ket{\pm_Y}\coloneqq \frac{1}{\sqrt{2}}(\ket{0}\pm i\ket{1})$. Then, let $\sigma^I_i\in\{\id,X,Y,Z\}$ and $\sigma^O_i\in\{X,Y,Z\}$. We define the set of operators 
\begin{align}\label{eq::basis2}
\begin{split}
    \mathcal{B}_2\coloneqq 
    \Big\{& \Big(\id\otimes\frac{\id}{2}\Big)^{\otimes 2}, \\
    & \big(\rho_i \pm \sigma^I_j\otimes\sigma^O_k\big)^{\otimes 2}, \\ 
    & \big(\rho_l + \sigma^I_m\otimes\sigma^O_n + \sigma^I_p\otimes\sigma^O_q\big)^{\otimes 2} \Big\}_{\substack{i,j,k,l,\\m,n,p,q}},
\end{split}
\end{align}
which can be shown to contain a subset of $91$ linearly independent operators by standard computational methods. Hence, the set $\mathcal{B}_2$ forms an overcomplete basis, from which one can obtain a standard basis (containing only $91$ operators) by discarding any operators that are not linearly independent. This set of $91$ operators forms a basis for the linear subspace spanned by two identical copies of any arbitrary qubit channel.

For the case of $k=3$, we begin by calculating the dimension of the space spanned by three identical copies of arbitrary qubit channels, by setting $d_\map{V}=13$ and $k=3$, obtaining that the dimension of the relevant space is $455$. Again, in order to find a basis for three identical copies of qubit channels, it is enough to exhibit $455$ operators $J_i$ respecting $\tr_O(J_i)=\id^I$, such that such that the set $\{{J_i}^{\otimes 3}\}_{i=1}^{455}$ is composed of linearly independent operators.

Let $\rho_i\in\{\id\otimes\frac{\id}{2}, 2\ketbra{\phi^+}{\phi^+}, 2\ketbra{\phi^-}{\phi^-}, 2\ketbra{\psi^+}{\psi^+},$ $2\ketbra{\psi^-}{\psi^-}, \id\otimes\ketbra{0}{0}, \id\otimes\ketbra{1}{1}, \id\otimes\ketbra{+}{+}, \id\otimes\ketbra{-}{-}, \id\otimes\ketbra{+_Y}{+_Y}\}, \id\otimes\ketbra{-_i}{-_i}\}$, where $\ket{\psi^{\pm}}\coloneqq \frac{1}{\sqrt{2}}(\ket{01}\pm\ket{10})$ are maximally entangled two-qubit states and $\ket{-}\coloneqq \frac{1}{\sqrt{2}}(\ket{0}-\ket{1})$. Then, let $\sigma^I_i\in\{\id,X,Y,Z,X+Z,X+Y\}$ and $\sigma^O_i\in\{X,Y,Z,X+Z,X+Y\}$. We define the set of operators 
\begin{align}\label{eq::basis3}
\begin{split}
    \mathcal{B}_3\coloneqq 
    \Big\{& \big(\rho_i \pm \sigma^I_j\otimes\sigma^O_k + \sigma^I_l\otimes\sigma^O_m\big)^{\otimes 3} \Big\}_{\substack{i,j,k,\\l,m}},
\end{split}
\end{align}
which can be shown to contain a subset of $455$ linearly independent operators by standard computational methods. This set of $455$ operators forms a basis for the linear subspace spanned by three identical copies of arbitrary qubit channels.

In general, one can always construct a basis for the space spanned by $k\in\mathbb{N}$ copies of arbitrary qubit channels by finding coefficients $\alpha_{i|l},\gamma_{ij|l}\in\mathbb{R}$, such that the set
\begin{align}\label{eq::basisk}
    \left\{ \left(\id\otimes \frac{\id}{2} + \sum_i \alpha_{i|l} \, \id\otimes\sigma_i + \sum_{ij}\gamma_{ij|l} \, \sigma_i\otimes \sigma_j\right)^{\otimes k}\right\}_{l}
\end{align}
contains $\binom{13-1+k}{k}$ linearly independent operators. One simple way to find such coefficients is simply to choose them at random.

To evaluate the SDPs~\eqref{sdp::primal} and~\eqref{sdp::dual} when the inputs form a basis for the space spanned by $k_A$ copies of channel $A$ and $k_B$ copies of channel $B$, the overall number of input pairs of channels are: for $(k_A,k_B)=(1,1)$, $N_A N_B = 13\cdot13 = 169$; for $(k_A,k_B)=(2,1)$, $N_A N_B = 91\cdot13 = 1183$; for $(k_A,k_B)=(3,1)$, $N_A N_B = 455\cdot13 = 5915$; and finally for $(k_A,k_B)=(2,2)$, $N_A N_B = 91\cdot91 = 8281$, making these SDPs very computationally demanding.
\\

%%%%%%%%%%%%%%%%%%%%%
\noindent{\textbf{Computer-assisted proofs.}} While floating-point arithmetic provides an efficient and powerful numerical method to treat real numbers, it suffers from some fundamentally unavoidable issues. For instance, addition and multiplication of floats is not associative and equality and inequality constraints are not satisfied exactly, but only up to some numerical precision. We now show how efficient numerical solvers that make use of floating-point arithmetic can be used to obtain a rigorous upper bound on the maximal success probability of simulating the quantum switch, hence leading to a bona-fide computer-assisted proof. Our methods are based on Ref.~\cite{bavaresco2021strict}.

The duality aspects of semidefinite programming~\cite{skrzypczyk2023semidefinite} ensure that any feasible point of the dual problem, presented in SDP~\eqref{sdp::dual}, yields an upper bound for the solution of the maximisation problem in the primal SDP~\eqref{sdp::primal}. That is, any set of operators $\{R_{ij}\}_{ij}$ and $\Gamma$ that respects
$\sum_{i,j} \tr[R_{ij}\,(S*(J^A_i\otimes J^B_j))] = 1$, $\Gamma - \sum_{i,j} R_{ij}\otimes[{(J^A_i)}^{\otimes k_A}\otimes {(J^B_j)}^{\otimes k_B}]^T \geq 0$, $\Gamma \geq 0$, and $\overline{\mathbb{P}}(\Gamma) = \Gamma$, implies that the probability of simulating the quantum switch is necessarily $p\leq\tr(\Gamma)/d_{c_I}d_{t_I}d_{A_O}^{k_A}d_{B_O}^{k_B}$. Standard numerical SDP solvers can be used to find a floating-point solution for the dual problem in SDP~\eqref{sdp::dual}, and to provide explicit operators $\{R_{ij}^\texttt{float}\}_{ij}$ and $\Gamma^\texttt{float}$ that approximately satisfy the SDP constraints. We now show how the operators $\{R_{ij}^\texttt{float}\}_{ij}$ and $\Gamma^\texttt{float}$ can be used as a good initial ansatz to construct symbolic operators $R_{ij}^\texttt{OK}$ and $\Gamma^\texttt{OK}$, which are not stored as floating-point variables and which satisfy the SDP constraints exactly, ensuring that $\tr(\Gamma^\texttt{OK})/d_{c_I}d_{t_I}d_{A_O}^{k_A}d_{B_O}^{k_B}$ is a legitimate upper bound for the probability of success of simulating the quantum switch. In order to do so, we use the bases presented in Eqs.~\eqref{eq::basis1},~\eqref{eq::basis2},~\eqref{eq::basis3}, and~\eqref{eq::basisk}, which only contain small rational numbers.  
In the following, we present an algorithm to extract a computer-assisted proof from numerical solvers.
\\

\paragraph*{\textbf{\emph{Algorithm:}}}
\begin{enumerate}
    \item 
    \texttt{Construct symbolic non-floating-point operators 
    $\Gamma^{\texttt{sym}}$ and $R_{ij}^{\texttt{sym}}$
    by truncating $\Gamma^\texttt{float}$ and $R_{ij}^\texttt{float}$ to obtain symbolic operators expressed only in terms of rational numbers.} \\ 
    This allows us to work with fractions and to avoid numerical imprecision.
    \item
    \texttt{Force the operators $\Gamma^{\texttt{sym}}$ and  $R_{ij}^{\texttt{sym}}$ to be self-adjoint by making use of the expression ${(M+M^\dagger)/2}$, which is self-adjoint for any $M$.} \\ 
    This ensures that we are dealing with self-adjoint operators.
    \item \texttt{Evaluate ${t^\texttt{sym}\coloneqq \sum_{i,j} \tr[R^\texttt{sym}_{ij}\,(S*(J^A_i\otimes J^B_j))]}$, where $S,J^A_i,$ and $J_j^B$ are also symbolic operators. Define $R_{ij}^\texttt{OK}\coloneqq R^\texttt{sym}_{ij}/t^{\texttt{sym}}$ for all $i,j$.} \\ 
    Here, we use the basis with rational coefficients constructed earlier to ensure that the operators $R_{ij}^\texttt{OK}$ satisfy the SDP equality constraint exactly. 
    \item \texttt{Project $\Gamma^\texttt{sym}$ onto the appropriate subspace to obtain $\overline{\mathbb{P}}(\Gamma^\texttt{sym})$.} \\ This ensures that the resulting operator $\overline{\mathbb{P}}(\Gamma^\texttt{sym})$ is in the correct linear subspace.
    \item  \texttt{Find $\eta\in\mathbb{R}$ such that $\Gamma^\texttt{OK}\coloneqq \overline{\mathbb{P}}(\Gamma^\texttt{sym}) + \eta \id\geq0$ and $\Gamma^\texttt{OK} - \sum_{i,j} R_{ij}^\texttt{OK}\otimes({J^A_i}^{\otimes k_A}\otimes {J^B_j}^{\otimes k_B})^T \geq 0$} \\
    This ensures the positivity constraints in the dual SDP \eqref{sdp::dual} are satisfied.
    \item  \texttt{Output the quantity $\tr(\Gamma^\texttt{OK})/d_{c_I}d_{t_I}d_{A_O}^{k_A}d_{B_O}^{k_B}$, which is a rigorous upper bound of the primal problem.}\\
    All equality and inequality constraints hold without relying on floating point precision, and the operators $\Gamma^\texttt{OK}$ and $R_{ij}^\texttt{OK}$ yield a proof certificate that $p\leq\tr(\Gamma^\texttt{OK})/d_{c_I}d_{t_I}d_{A_O}^{k_A}d_{B_O}^{k_B}$.
\end{enumerate}
The algorithm described above deserves two clarifications. First, since the projector $\overline{\mathbb{P}}$ is unital, i.e., $\overline{\mathbb{P}}(\id)=\id$, for any $\eta\in\mathbb{R}$ and any operator $M$, we have that $\overline{\mathbb{P}}(M) + \eta\id = \overline{\mathbb{P}}(M+\eta\id)$. Second, checking if an operator is positive semidefinite can be done efficiently through the Cholesky decomposition---more details are provided in App.~\ref{app::compassist}.

%%%%%%%%%%%%%%
\section*{Data availability}

No data sets were generated or analyzed during the current study.

%%%%%%%%%%%%%%
\section*{Code availability}

\noindent All code developed for this work is available at our online repository in Ref.~\cite{github_switchsimulation}. For numerically evaluating our SDPs, we used the Splitting Conic Solver (SCS)~\cite{odonoghue2016conic,scs} and MOSEK~\cite{mosek}.

%%%%%%%%%%%%%%
%\bibliography{refs}


%%%%%%%%%%%%%%

\hspace*{1em}

%%%%%%%%%%%%%%
\section*{Acknowledgements}

We would like to thank Alastair Abbott, Emanuel-Cristian Boghiu, Cyril Branciard, Anne Broadbent, Giulio Chiribella, Arthur Mehta, Simon Milz, Pierre Pocreau, Louis Salvail, Mykola Semenyakin, Jacopo Surace and Matt Wilson for helpful discussions. We are also grateful to Denis Rosset and Siegfried M. Rump for helpful discussions regarding computer-assisted proofs.
JB acknowledges funding from the Swiss National Science Foundation (SNSF) through the funding schemes Swiss Postdoctoral Fellowship (project~216979), NCCR SwissMAP (project~182902), and project~192244;
PT acknowledges funding from the Japan Society for the Promotion of Science (JSPS) Postdoctoral Fellowships for Research in Japan.
SY acknowledges support by Japan Society for the Promotion of Science (JSPS) KAKENHI Grant Number 23KJ0734, FoPM, WINGS Program, the University of Tokyo, and DAIKIN Fellowship Program, the University of Tokyo. 
This work was supported by the MEXT Quantum Leap Flagship Program (MEXT QLEAP)  JPMXS0118069605, JPMXS0120351339, the Japan Society for the Promotion of Science (JSPS) KAKENHI Grant Number 21H03394 and 23K21643, and IBM Quantum.
We acknowledge the Japanese-French Laboratory for Informatics (JFLI) for the support on organizing the Japanese-French Quantum Information 2023 workshop.
Research at Perimeter Institute is supported in part by the Government of Canada through the Department of Innovation, Science and Economic Development and by the Province of Ontario through the Ministry of Colleges and Universities.

%%%%%%%%%%%%%%
\section*{Author contributions}

MM conceived the main idea and initiated the project. HK, MM, TO, PT, and SY developed the results related to deterministic exact simulations. These authors contributed equally: HK, TO, and SY. JB and MTQ developed the results related to probabilistic and approximate simulations. Code: JB developed the SDP-related code; MTQ developed the code for the computer-assisted proofs. All authors contributed to discussions, refining of the proofs, and writing the paper.
%%%%%%%%%%%%%%
\section*{Competing interests}

\noindent The authors declare no competing interests.

\clearpage
\onecolumngrid
\setcounter{theorem}{0}
\newpage 

\appendix

\renewcommand\thefigure{\thesection\arabic{figure}}   

\renewcommand\thesubsection{\thesection\arabic{subsection}}
\makeatletter
\renewcommand{\p@subsection}{}  % remove extra 'A' prefix from subsection
\makeatother

\section*{APPENDIX}

The Appendix is organized in the following way:

\begin{itemize}
    \item Appendix~\ref{app::definitions}: Definitions and Choi operators 
    \dotfill \hyperref[app::definitions]{\pageref*{app::definitions}}
        \begin{itemize}
            \item[--] \ref{subapp::switchchoi}. The quantum switch Choi operator and Choi operator fidelity 
            \dotfill \hyperref[subapp::switchchoi]{\pageref*{subapp::switchchoi}}
            \item[--] \ref{subapp::combs}. Quantum combs, probabilistic quantum combs, and projectors
            \dotfill \hyperref[subapp::combs]{\pageref*{subapp::combs}}
            \item[--] \ref{subapp::qcccs}. QC-CCs and probabilistic QC-CCs
            \dotfill \hyperref[subapp::qcccs]{\pageref*{subapp::qcccs}}
        \end{itemize}
    \item Appendix~\ref{app::generalsimulation}: General simulation scenarios
    \dotfill \hyperref[app::generalsimulation]{\pageref*{app::generalsimulation}}
        \begin{itemize}
            \item[--] \ref{subapp::switchonextendedchannels}. The action of the quantum switch on bipartite quantum channels
            \dotfill \hyperref[subapp::switchonextendedchannels]{\pageref*{subapp::switchonextendedchannels}}
            \item[--] \ref{subapp::switchoninstruments}. The action of the quantum switch on quantum instruments
            \dotfill \hyperref[subapp::switchoninstruments]{\pageref*{subapp::switchoninstruments}}
            \item[--] \ref{subapp::unitarydilation}. The relationship with Stinespring dilation of quantum channels
            \dotfill \hyperref[subapp::unitarydilation]{\pageref*{subapp::unitarydilation}}
            \item[--] \ref{subapp::proofoursimulation}. Theorem~\ref{thm::go}: A simulation for bipartite unitary channels $U_A$ and bipartite general channels $B$
            \dotfill \hyperref[subapp::proofoursimulation]{\pageref*{subapp::proofoursimulation}}
        \end{itemize}
    \item Appendix~\ref{app::fullproof_nogoexp}: Full proof of Theorem~\ref{thm::nogo_exp}
    \dotfill \hyperref[app::fullproof_nogoexp]{\pageref*{app::fullproof_nogoexp}}
        \begin{itemize}
            \item[--] \ref{subapp::lem:span}. Lemma~\ref{lem:span}
            \dotfill \hyperref[subapp::lem:span]{\pageref*{subapp::lem:span}}
            \item[--] \ref{subapp::lem:linearity_MN}. Lemma~\ref{lem:linearity_MN} (for $M,N\in\mathbb{N}^+$)
            \dotfill \hyperref[subapp::lem:linearity_MN]{\pageref*{subapp::lem:linearity_MN}}
            \item[--] \ref{subapp::lem:p_as_vector_MN}. Lemma~\ref{lem:p_as_vector_MN} (for $M,N\in\mathbb{N}^+$)
            \dotfill \hyperref[subapp::lem:p_as_vector_MN]{\pageref*{subapp::lem:p_as_vector_MN}}
            \item[--] \ref{subapp::lem:indep_M1}. Lemma~\ref{lem:indep_M1} (for $M < 4^{n}/2 +2$ and $N=1$)
            \dotfill \hyperref[subapp::lem:indep_M1]{\pageref*{subapp::lem:indep_M1}}
            \item[--] \ref{subapp::lem:qccc_invalid}. Lemma~\ref{lem:qccc_invalid} (for $\max\{M, N\} \leq  \max\{2, d-1\}$)
            \dotfill \hyperref[subapp::lem:qccc_invalid]{\pageref*{subapp::lem:qccc_invalid}}
            \item[--] \ref{subapp::lem:linear_independence}. Lemma~\ref{lem:linear_independence}
            \dotfill \hyperref[subapp::lem:linear_independence]{\pageref*{subapp::lem:linear_independence}}
            \item[--] \ref{subapp::lem:linear_independence2}. Lemma~\ref{lem:linear_independence2}
            \dotfill \hyperref[subapp::lem:linear_independence2]{\pageref*{subapp::lem:linear_independence2}}
        \end{itemize}    
    \item Appendix~\ref{app::gorestricted}: Possible restricted simulations for qubit channels
    \dotfill \hyperref[app::gorestricted]{\pageref*{app::gorestricted}}
    \item Appendix~\ref{app::nogounitary}: No-go results for unitary channels
    \dotfill \hyperref[app::nogounitary]{\pageref*{app::nogounitary}}
    \item Appendix~\ref{app::compassist}: Efficient certification that a matrix is positive semidefinite
    \dotfill \hyperref[app::compassist]{\pageref*{app::compassist}}
\end{itemize}

%%%%%%%%%%%%%%%%%%%%%%%%%%%%%%%%%%%%%%%%%%%%
\section{Definitions and Choi operators}\label{app::definitions}
\setcounter{figure}{0}
%%%%%%%%%%%%%%%%%%%%%%%%%%%%%%%%%%%%%%%%%%%%

In this section, we explicitly define the key higher-order transformations used in the main text in terms of their Choi operators, which are crucial for the SDP implementation of our methods.

%%%%%%%%%%%%%%%%%
\subsection{The quantum switch Choi operator and Choi operator fidelity}\label{subapp::switchchoi} 
%%%%%%%%%%%%%%%%%

We begin with the Choi operator of the quantum switch transformation~\cite{araujo2015witnessing}. The quantum switch higher-order transformation $\map{S}$ defined in the main text in terms of its Kraus decomposition [Eqs.~\eqref{eq::switch} and~\eqref{eq::switchkraus}], can be equivalently expressed by its associated Choi operator $S\in\map{L}(\map{H}^{c_I}\otimes\map{H}^{t_I}\otimes\map{H}^{A_I}\otimes\map{H}^{A_O}\otimes\map{H}^{B_I}\otimes\map{H}^{B_O}\otimes\map{H}^{t_O}\otimes\map{H}^{c_O})$. Let the vector $\dket{S}$ be defined as
\begin{equation}\label{eq::switch_choivec}
    \dket{S} \coloneqq  \ket{0}^{c_I}\dket{\id}^{t_IA_I}\dket{\id}^{A_OB_I}\dket{\id}^{B_Ot_O}\ket{0}^{c_O} + \ket{1}^{c_I}\dket{\id}^{t_IB_I}\dket{\id}^{B_OA_I}\dket{\id}^{A_Ot_O}\ket{1}^{c_O}, 
\end{equation}
where $\dket{\id}^{XY}\coloneqq \sum_{i}\ket{i}^X\ket{i}^Y$ is the vector related to the Choi operator $\dketbra{\id}{\id}$ of an identity channel from space $X$ to $Y$. Then, the Choi operator of the quantum switch is defined as
\begin{equation}
    S \coloneqq  \dketbra{S}{S}.
\end{equation}
This is the operator $S$ that appears in the first constraint of the SDPs associated to the primal problem [SDP~\eqref{sdp::primal}] and dual problem [SDP~\eqref{sdp::dual}] in the main text.

The Choi operator of the quantum channel $\map{S}(A,B)$ that results from the action of the quantum switch transformation $\map{S}$ on input channels $A$ and $B$ can be expressed in terms of their respective Choi operators $S$, $J^A$, and $J^B$, as $S*(J^A\otimes J^B)\in\map{L}(\map{H}^{c_I}\otimes\map{H}^{t_I}\otimes\map{H}^{t_O}\otimes\map{H}^{c_O})$, where $*$ is the link product defined in the Methods section of the main text [Eq.~\eqref{eq::linkproduct}].

In the case considered in the main text where the quantum switch has a fixed input state for its control and target systems, respectively given by $\sigma_{c}=\ketbra{+}{+}\in\map{L}(\map{H}^{c_I})$ and $\rho_{t}=\ketbra{0}{0}\in\map{L}(\map{H}^{t_I})$, we defined the resulting transformation as $\map{S}_{+0}(\cdot,\cdot)\coloneqq \map{S}(\cdot,\cdot)[\ketbra{+}{+}\otimes\ketbra{0}{0}]$. This higher-order operation has an associated Choi operator $S_{+0}\in\map{L}(\map{H}^{A_I}\otimes\map{H}^{A_O}\otimes\map{H}^{B_I}\otimes\map{H}^{B_O}\otimes\map{H}^{t_O}\otimes\map{H}^{c_O})$ given by
\begin{align}\label{eq::switchchoi_+0}
    S_{+0} \coloneqq & \Big(\ketbra{+}{+}\otimes\ketbra{0}{0}\Big) * S \\
    =& \dketbra{S_{+0}}{S_{+0}},
\end{align}
where
\begin{equation}
   \dket{S_{+0}} \coloneqq  \Big(\bra{+}^{c_I}\bra{0}^{t_I}\otimes\id\Big)\dket{S} =  \frac{1}{\sqrt{2}}\Big(\ket{0}^{A_I}\dket{\id}^{A_OB_I}\dket{\id}^{B_Ot_O}\ket{0}^{c_O} + \ket{1}^{B_I}\dket{\id}^{B_OA_I}\dket{\id}^{A_Ot_O}\ket{1}^{c_O}\Big).
\end{equation}

We now define the Choi operator fidelity. Let $\widetilde{\map{S}}$ be a higher-order transformation that acts on the same spaces as the quantum switch $\map{S}$, and let $\widetilde{\map{S}}_{+0}\coloneqq \widetilde{\map{S}}(\cdot,\cdot)[\ketbra{+}{+}\otimes\ketbra{0}{0}]$ be $\widetilde{\map{S}}$ with fixed input states for its control and target systems. The Choi operator fidelity between $\map{S}_{+0}$ and any higher-order operation $\widetilde{\map{S}}_{+0}$, with Choi operator $\widetilde{S}_{+0}$ is given by
\begin{equation}
    F(\map{S}_{+0},\widetilde{\map{S}}_{+0}) \coloneqq  \frac{1}{(d_{A_O}d_{B_O})^2} \tr (S_{+0}\, \widetilde{S}_{+0}) = \frac{1}{(d_{A_O}d_{B_O})^2} \Big|\bra{s_{+0}}\widetilde{\map{S}}_{+0}\ket{s_{+0}}\Big|,
\end{equation}
where the factor $d_{A_O}d_{B_O}=\tr(S_{+0})=\tr(\widetilde{S}_{+0})$ ensures that $F(\map{S}_{+0},\widetilde{\map{S}}_{+0})\in[0,1]$.

%%%%%%%%%%%%%%%%%
\subsection{Quantum combs, probabilistic quantum combs, and projectors}\label{subapp::combs} 
%%%%%%%%%%%%%%%%%

A probabilistic quantum comb is a quantum comb that yields a classical output with a certain probability. In our case, we consider probabilistic quantum combs that have two possible classical outcomes, success or failure. We recall that such transformations can be implemented by a quantum comb that additionally output a flag system that encodes the success or failure outcome, followed by a dichotomic quantum measurement of the flag system. In the case of a quantum comb $\map{C}=\map{C}_s+\map{C}_f$ that performs a probabilistic simulation of the quantum switch, its associated Choi operator is given by $C=C_s+C_f$, where $C,C_s,C_f\in\map{L}(\map{H}^{c_I}\otimes\map{H}^{t_I}\otimes(\map{H}^{A_I}\otimes\map{H}^{A_O})^{\otimes k_A}\otimes(\map{H}^{B_I}\otimes\map{H}^{B_O})^{\otimes k_B}\otimes\map{H}^{t_O}\otimes\map{H}^{c_O})$. These operators are characterised by
\begin{align}
    C_s&\geq 0 \\
    C_f = C - C_s &\geq 0 \\
    \tr(C) &= d_{c_I}d_{t_I}d_{A_O}^{k_A}d_{B_O}^{k_B} \\
    C &= \mathbb{P}_k(C),
\end{align}
where $k=k_A+k_B$ is the total number of slots in the quantum comb $C$, and $\mathbb{P}_k$ is the projector onto the subspace spanned by $k$-slot quantum combs. 

The projector $\mathbb{P}_k$ is defined in full generality in Ref.~\cite{milz2024characterising}. For sake of completeness, we explicitly write $\mathbb{P}_k$ here in the cases where $k\in\{2,3,4\}$, which are the ones involved in our numerical calculations. To simplify the notation in the definition of the projector, let us define a $k$-slot quantum comb by its Choi operator $C\in\map{L}(\map{H}^{P}\otimes\map{H}^{I_1}\otimes\map{H}^{O_1}\otimes\ldots\otimes\map{H}^{I_k}\otimes\map{H}^{O_k}\otimes\map{H}^{F})$. Then, let us define the trace-and-replace operation acting on the subspace $\map{H}^X$ of an operator $C$ as
\begin{equation}
    _{X} C \coloneqq  \tr_X(C)\otimes\frac{\id^C}{d_C}.
\end{equation}
We are now ready to explicitly write the projector $\mathbb{P}_k$ for  $k\in\{2,3,4\}$. For the case where $k=2$,
\begin{equation}
    \mathbb{P}_2(C) =\, C -\, _{F}C +\, _{O_2F}C -\, _{I_2O_2F}C +\, _{O_1I_2O_2F}C -\, _{I_1O_1I_2O_2F}C +\, _{PI_1O_1I_2O_2F}C.
\end{equation}
For the case where $k=3$,
\begin{align}
\begin{split}
    \mathbb{P}_3(C) =&\ C -\, _{F}C +\, _{O_3F}C -\, _{I_3O_3F}C +\, _{O_2I_3O_3F}C -\, _{I_2O_2I_3O_3F}C +\, _{O_1I_2O_2I_3O_3F}C \\ 
    &-\, _{I_1O_1I_2O_2I_3O_3F}C +\, _{PI_1O_1I_2O_2I_3O_3F}C.
\end{split}
\end{align}
Finally, for the case where $k=4$,
\begin{align}
\begin{split}
    \mathbb{P}_4(C) =&\ C -\, _{F}C +\, _{O_4F}C -\, _{I_4O_4F}C +\, _{O_3I_4O_4F}C -\, _{I_3O_3I_4O_4F}C +\, _{O_2I_3O_3I_4O_4F}C -\, _{I_2O_2I_3O_3I_4O_4F}C \\
    &+\, _{O_1I_2O_2I_3O_3I_4O_4F}C -\, _{I_1O_1I_2O_2I_3O_3I_4O_4F}C +\, _{PI_1O_1I_2O_2I_3O_3I_4O_4F}C. 
\end{split}
\end{align}

For convenience, we also define the dual affine projector used in the dual formulation of the SDP in the main text [Eq.~\eqref{sdp::dual}]. Given a set of operators $\map{A}$, its dual affine set $\map{B}$ is the set of all operators $B$ such that $\tr(A^\dagger B)=1$ for all $A\in\map{A}$. Hence, the dual affine set of the set of Choi operators of quantum combs $C\in\map{L}(\map{H}^{P}\otimes\map{H}^{I_1}\otimes\map{H}^{O_1}\otimes\ldots\otimes\map{H}^{I_k}\otimes\map{H}^{O_k}\otimes\map{H}^{F})$ is the set of all operators $\Gamma\in\map{L}(\map{H}^{P}\otimes\map{H}^{I_1}\otimes\map{H}^{O_1}\otimes\ldots\otimes\map{H}^{I_k}\otimes\map{H}^{O_k}\otimes\map{H}^{F})$ such that $\tr(C\,\Gamma)=1$. The operators in the dual affine set of quantum combs are themselves a particular case of quantum combs, and can be characterized by projectors $\overline{\mathbb{P}}_k$ given by~\cite{milz2024characterising}
\begin{equation}
    \overline{\mathbb{P}}_k(\Gamma) = \Gamma - \mathbb{P}_k(\Gamma) +\, _{PI_1I_2\ldots I_kO_kF} \Gamma. 
\end{equation}
In our numerical calculations, we used the code available in the repository of Ref.~\cite{milz2024characterising} to generate the above projector constraints.

%%%%%%%%%%%%%%%%%
\subsection{QC-CCs and probabilistic QC-CCs}\label{subapp::qcccs}
%%%%%%%%%%%%%%%%%

We now present the explicit constraints for QC-CC transformations. QC-CCs and probabilistic QC-CCs have been defined in full generality and for any number of slots in Ref.~\cite{wechs2021quantum}. Here, we write these definitions explicitly for the cases of $k=2$ and $k=3$ slots, which were used in our numerical calculations. Since we only evaluated the maximal probability of simulating the quantum switch with a QC-CC in the restricted simulation scenario, where the input control and target systems are fixed, we write the explicit constraints for a probabilistic QC-CC with fixed input systems (compared to the definition of quantum combs in this section, this is the equivalent of a scenario where $d_P=1$).

Unlike the case of quantum combs, to define the Choi operator $W = W_s + W_f$ associated to a probabilistic QC-CC, where $W,W_s,W_f\in\map{L}(\map{H}^{I_1}\otimes\map{H}^{O_1}\otimes\ldots\otimes\map{H}^{I_k}\otimes\map{H}^{O_k}\otimes\map{H}^{F})$, it is necessary but not sufficient to say that $W_s\geq 0$, $W_f=W-W_s\geq 0$ and $W$ is a QC-CC, as individual constraints must be applied to $W_s$ and $W_f$ as well to ensure validity of the overall transformation. 

In the case where $k=2$, we have that 
\begin{align}
    W &=  W_{s} + W_{f} \\
    \tr(W) &= d_{O_1}d_{O_2},
\end{align}
where
\begin{equation}
    W_{s} =  W_{s}^{12F} + W_{s}^{21F}, \hspace*{1cm} W_{f} =  W_{f}^{12F} + W_{f}^{21F},
\end{equation}
such that $W_{s}^{12F}\geq 0$, $W_{s}^{21F}\geq 0$, $ W_{f}^{12F}\geq 0$, and $ W_{f}^{21F}\geq 0$, and additionally $W^{12F} \coloneqq  W_{s}^{12F} + W_{f}^{12F}$ must be a quantum comb with the order $I_1O_1I_2O_2F$, and $W^{21F}\coloneqq  W_{s}^{21F} + W_{f}^{21F}$ must be a quantum comb with the order $I_2O_2I_1O_1F$. 

In the case where $k=3$, we have that 
\begin{align}
    W &=  W_{s} + W_{f} \\
    \tr(W) &= d_{O_1}d_{O_2}d_{O_3}.
\end{align}
where
\begin{align}
    W_{s} &=  W_{s}^{123F} + W_{s}^{132F} + W_{s}^{213F} + W_{s}^{231F} + W_{s}^{312F}+ W_{s}^{321F} \\
    W_{f} &=  W_{f}^{123F} + W_{f}^{132F} + W_{f}^{213F} + W_{f}^{231F} + W_{f}^{312F}+ W_{f}^{321F},
\end{align}
where $W_{s}^{xyzF}\geq0$ and $W_{f}^{xyzF}\geq0$ for all $xyz \in \mathrm{Perm}(1,2,3)$. Moreover, we define $W^{xyzF}\coloneqq  W_{s}^{xyzF} + W_{f}^{xyzF}$ for all $xyz\in \mathrm{Perm}(1,2,3)$ and impose that the operators $W^{123F}$ and $W^{132F}$ must satisfy
\begin{align}
    _{F}W^{123F} =\, _{O_3F}W^{123F}, \hspace*{6mm} & \hspace*{6mm} \hspace*{6mm} _{F}W^{132F} =\, _{O_2F}W^{132F} \\
    _{I_3O_3F}W^{123F} =\, _{O_2I_3O_3F}W^{123F}, & \hspace*{6mm} _{I_2O_2F}W^{132F} =\, _{O_3I_2O_2F}W^{132F} \\
    _{I_2O_2I_3O_3F}W^{123F} + _{I_3O_3I_2O_2F}W^{132F} &=\, _{O_1}(_{I_2O_2I_3O_3F}W^{123F} + _{I_3O_3I_2O_2F}W^{132F}).
\end{align}
Similarly, $W^{213F}$ and $W^{231F}$ must satisfy
\begin{align}
    _{F}W^{213F} =\, _{O_3F}W^{213F}, \hspace*{6mm} & \hspace*{6mm} \hspace*{6mm} _{F}W^{231F} =\, _{O_1F}W^{231F} \\
    _{I_3O_3F}W^{213F} =\, _{O_1I_3O_3F}W^{213F}, & \hspace*{6mm} _{I_1O_1F}W^{231F} =\, _{O_3I_1O_1F}W^{231F} \\
    _{I_1O_1I_3O_3F}W^{213F} + _{I_3O_3I_1O_1F}W^{231F} &=\, _{O_2}(_{I_1O_1I_3O_3F}W^{213F} + _{I_3O_3I_1O_1F}W^{231F}). 
\end{align}
Finally, $W^{312F}$ and $W^{321F}$ must satisfy
\begin{align}
    _{F}W^{312F} =\, _{O_2F}W^{312F}, \hspace*{6mm} & \hspace*{6mm} \hspace*{6mm} _{F}W^{321F} =\, _{O_1F}W^{321F} \\
    _{I_2O_2F}W^{312F} =\, _{O_1I_2O_2F}W^{312F}, & \hspace*{6mm} _{I_1O_1F}W^{321F} =\, _{O_2I_1O_1F}W^{321F} \\
    _{I_1O_1I_2O_2F}W^{312F} + _{I_2O_2I_1O_1F}W^{321F} &=\, _{O_3}(_{I_1O_1I_2O_2F}W^{312F} + _{I_2O_2I_1O_1F}W^{321F}). 
\end{align}

%%%%%%%%%%%%%%%%%%%%%%%%%%%%%%%%%%%%%%%%%%%%
\section{General simulation scenarios}\label{app::generalsimulation}
\setcounter{figure}{0}
%%%%%%%%%%%%%%%%%%%%%%%%%%%%%%%%%%%%%%%%%%%%

In this section, we discuss the differences between simulation scenarios where the quantum switch acts on general quantum channels, on only part of general quantum channels, or on quantum instruments, highlighting how they relate to the action of the quantum switch on unitary channels. We also demonstrate how the switch simulation that we introduced in the circuit in Eq.~\eqref{cc::ourcircuit} related to Thm.~\ref{thm::go} fits into this context.

%%%%%%%%%%%%%%%%%
\subsection{The action of the quantum switch on bipartite quantum channels}\label{subapp::switchonextendedchannels}
%%%%%%%%%%%%%%%%%

%%%%%%%%%%%%%%%%%%%%%%%%%%%%%%%%
\begin{figure}%[h!]
\begin{center}
	\includegraphics[width=\columnwidth]{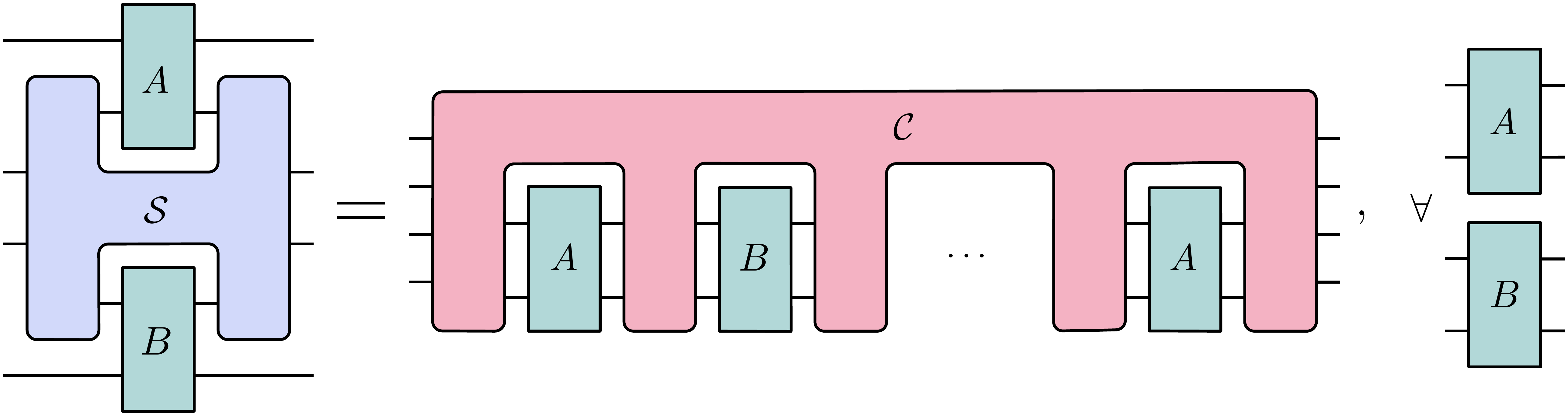}
	\caption{\textbf{General simulation of the quantum switch.} A higher-order transformation $\map{C}$, which can be a quantum comb or a QC-CC, that acts on part of several copies of input bipartite quantum channels $A$ and $B$ is a simulation of the quantum switch $\map{S}$ if it reproduces the action of the quantum switch {on} all arbitrary pairs of bipartite channels $A$ and $B$.}
\label{fig::simulation_general}
\end{center}
\end{figure}
%%%%%%%%%%%%%%%%%%%%%%%%%%%%%%%%

The first simulation scenario discussed in the main text is the one where the quantum switch $\map{S}$ acts on the ``entire'' quantum channels $A$ and $B$, as represented in Fig.~2, in the main text. Formally, this is the case where the quantum switch $\mathcal{S}$ acts on a pair of single-party (single input system, single output system) channels
$A: \map{L}(\map{H}^{A_I}) \to \map{L}(\map{H}^{A_O})$ 
and 
$B: \map{L}(\map{H}^{B_I}) \to \map{L}(\map{H}^{B_O})$,
and outputs another channel 
$\map{S}(A,B): 
\map{L}(\map{H}^{c_I}\otimes\map{H}^{t_I}) \to
\map{L}(\map{H}^{c_O}\otimes\map{H}^{t_O})$.
However, quantum theory also allows one to apply the quantum switch on \textit{parts} of bipartite (two input systems, two output systems) channels $A: \map{L}(\map{H}^{A_I}\otimes\map{H}^{A'_I}) \to \map{L}(\map{H}^{A_O}\otimes\map{H}^{A'_O})$ and $B: \map{L}(\map{H}^{B_I}\otimes\map{H}^{B'_I}) \to \map{L}(\map{H}^{B_O}\otimes\map{H}^{B'_O})$, where the dimension of the primed spaces is arbitrary. This results in the quantum channel $\map{S}\otimes\map{I}(A,B): \map{L}(\map{H}^{A'_I}\otimes\map{H}^{c_I}\otimes\map{H}^{t_I}\otimes\map{H}^{B'_I}) \to \map{L}(\map{H}^{A'_O}\otimes\map{H}^{c_O}\otimes\map{H}^{t_O}\otimes\map{H}^{B'_O})$, where $\map{I}$ is the identity higher-order transformation. Since the dimension of the primed spaces is arbitrary, all multi-partite channels can be described in this context as bipartite channels, which are the most general kinds of deterministic transformations between quantum states one {can} consider. This case is illustrated in Fig.~3 of the main text.

As mentioned in the main text, the impossibility of simulating the action of the quantum switch on all single-party quantum channels, for some number of calls $k_A$ and $k_B$, implies the impossibility of simulating the action of the quantum switch on all bipartite channels with the same number of calls. Hence, when focusing on no-go simulation proofs, we can restrict ourselves to considering only single-party channels. However, conversely, should a simulation of the action of the quantum switch on all single-party channels exist for some number of calls $k_A$ and $k_B$, this does not necessarily imply that a simulation would also exists for the action of the quantum switch on all bipartite channels. 

For this reason, in order to have a fully general simulation of the quantum switch, one must consider the case where the quantum switch acts on only part of a bipartite channel, as illustrated in Fig.~\ref{fig::simulation_general}. That is, a full simulation of the quantum switch is only obtained when, for every pair of bipartite channels $A: \map{L}(\map{H}^{A_I}\otimes\map{H}^{A'_I}) \to \map{L}(\map{H}^{A_O}\otimes\map{H}^{A'_O})$ 
and 
$B: \map{L}(\map{H}^{B_I}\otimes\map{H}^{B'_I}) \to \map{L}(\map{H}^{B_O}\otimes\map{H}^{B'_O})$, there exists $k_A, k_B\in\mathbb{N}$ such that,
\begin{align}
    \map{C}(A^{\otimes k_A},B^{\otimes k_B}) = \map{S}\otimes\map{I}(A,B)
\end{align}
where $\mathcal{C}$ is a quantum comb, or a QC-CC. This most general simulation scenario is depicted in Fig.~\ref{fig::simulation_general}. 

In the next section, we show how a simulation of the action of the quantum switch on all pairs of bipartite channels indeed covers all possible quantum operations the quantum switch could take as input, including the probabilistic ones, such as quantum instruments.

%%%%%%%%%%%%%%%%%
\subsection{The action of the quantum switch on quantum instruments}\label{subapp::switchoninstruments}
%%%%%%%%%%%%%%%%%

Higher-order transformations like the quantum switch can also be applied to quantum instruments, which describe the most general probabilistic transformation between quantum states. 

\begin{figure}%[h!]
\begin{center}
	\includegraphics[width=0.9\columnwidth]{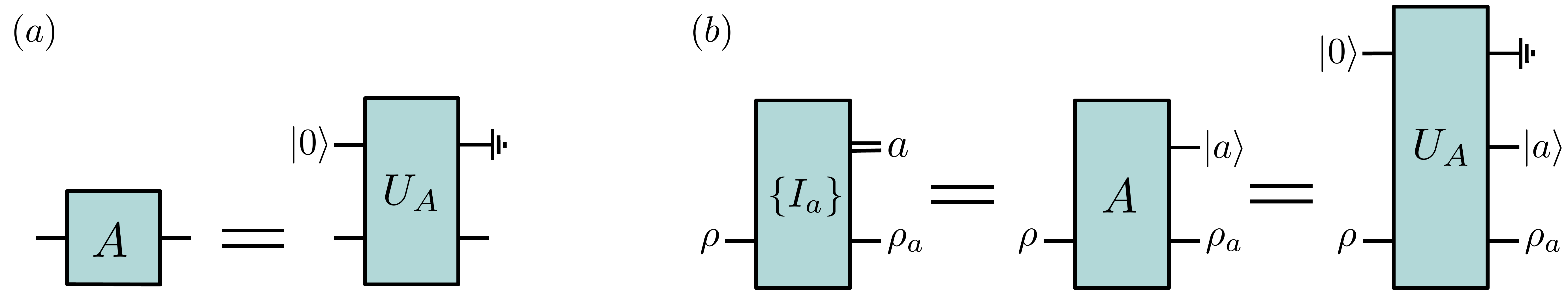}
	\caption{\textbf{General quantum channels and quantum instruments.} $(a)$ Quantum circuit representation of the Stinespring dilation $U_A$ of the general quantum channel $A$ presented in Eq.~\eqref{eq::Stinespring}. $(b)$ Quantum circuit representation of a quantum instrument $\{I_a\}_a$ by a quantum channel $A[\rho] \coloneqq  \sum_a I_a[\rho] \otimes \ketbra{a}$, which is then represented by its Stinespring dilation $U_A$.}
\label{fig::stinespring+instrument}
\end{center}
\end{figure}

Quantum instruments are transformations that map a quantum state to another quantum state with a certain probability, also outputting a classical outcome. Physically, this describes, for instance, a measurement process where both a {classical} outcome and a post-measurement {quantum} state are produced. They can be described by a collection of CP maps $\{I_a\}_a$ which add to a CPTP map, i.e., such that $\sum_a I_a$ is CPTP. A quantum instrument takes a quantum state $\rho$ as input, and outputs a classical outcome $a$, together with the quantum state $\rho_a\coloneqq  I_a[\rho]/\tr(I_a[\rho])$, with probability $p_a\coloneqq \tr(I_a[\rho])$. Quantum instruments can be equivalently represented by deterministic quantum channels, which instead of yielding a classical outcome, prepare an additional {output} pure quantum state $\ket{a}$ that encodes the value of the instrument's classical outcome $a$ in a perfectly discriminable manner~\cite{ozawa1984quantum,wilde2017from,buscemi2023unifying}. In other words, a quantum instrument with $N\in\mathbb{N}$ outcomes $\{I_a\}_{a=1}^N$ with $I_a:\mathcal{L}(\map{H}^I)\to \map{H}(\map{H}^O)$ is equivalent to a quantum channel $A:\mathcal{L}(\map{H}^I)\to \map{H}(\map{H}^O\otimes \mathbb{C}^N)$ given by
\begin{align}\label{eq::instrument2channel}
    A[\,\rho\,] \coloneqq  \sum_{a=1}^N I_a[\,\rho\,] \otimes \ketbra{a},
\end{align}
and illustrated in Fig.~\ref{fig::stinespring+instrument}$(b)$. By representing a quantum instrument as a particular case of a bipartite quantum channel, namely one with a single input system and two output systems, it is straightforward to see that the action of the quantum switch on a pair of instruments $\{I^A_a\}_a$ and $\{I^B_b\}_b$ constitutes a special case of its action on arbitrary bipartite channels $A$ and $B$. Hence, a simulation of the quantum switch for all bipartite channels can be used for a simulation of the quantum switch for all instruments.

Lastly, one might also consider the scenario where the quantum switch acts on {a part} of bipartite instruments. Notice, however, that from the channel representation of instruments, it is clear that this scenario is included in that of arbitrary bipartite channels discussed in the previous section.

%%%%%%%%%%%%%%%%%%%%%%%%%%%%%%%%
%%%%%%%%%%%%%%%%%%%%%%%%%%%%%%%%

%%%%%%%%%%%%%%%%%
\subsection{The relationship with Stinespring dilation of quantum channels}\label{subapp::unitarydilation}
%%%%%%%%%%%%%%%%%

%%%%%%%%%%%%%%%%%%%%%%%%%%%%%%%%
\begin{figure}%[h!]
\begin{center}
    \includegraphics[height=0.925\textheight]{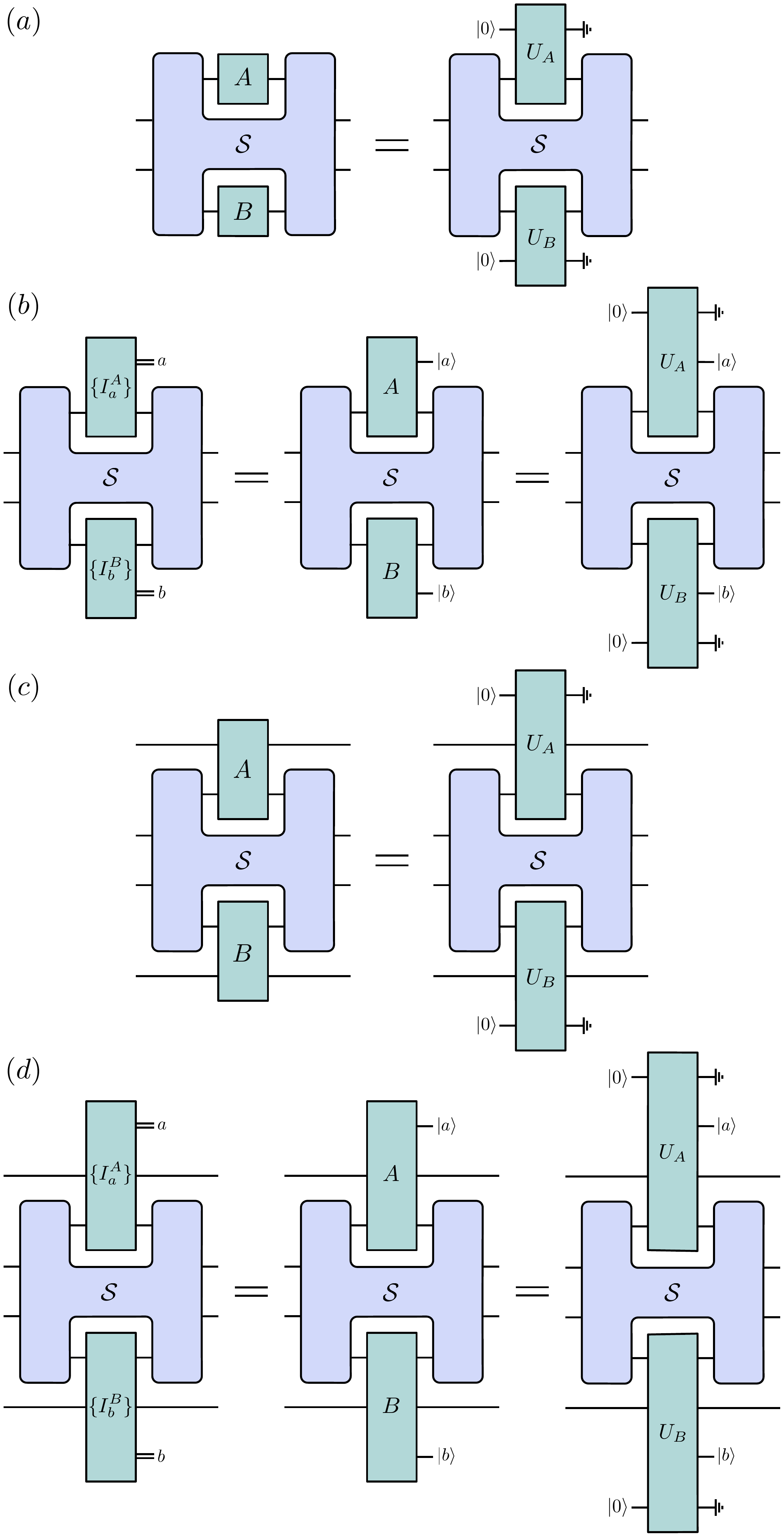}
	\caption{\textbf{The action of the quantum switch.} Depiction of the action of the quantum switch on $(a)$ general single-party quantum channels; $(b)$ general single-party quantum instruments; $(c)$ part of general bipartite quantum channels, and; $(d)$ part of general bipartite quantum instruments; as a function of the action of the quantum switch on the Stinespring dilation of the input channels.}
    \label{fig::switchaction}
\end{center}
\end{figure}
%%%%%%%%%%%%%%%%%%%%%%%%%%%%%%%%

The Stinespring dilation theorem~\cite{watrous2018theory} states that the action of every general quantum channel, i.e, CPTP map, $A:\mathcal{L}(\mathcal{H}^I)\to\mathcal{L}(\mathcal{H}^O)$ can be written as
\begin{align} \label{eq::Stinespring}
    A[\,\rho\,] = \tr_{\text{aux}_O} \Big(U_A[\,\rho^I\otimes \ketbra{0}{0}\,]\Big),
\end{align}
for some unitary channel $U_A:\map{L}(\mathcal{H}^I\otimes\mathcal{H}^{\text{aux}_I})\to\map{L}(\mathcal{H}^O\otimes\mathcal{H}^{\text{aux}_O})$ that acts jointly on the input system and an auxiliary system (of sufficiently large dimension), which can be initialized to the state $\ket{0}$ without loss of generality. We emphasize that the discarding of the output auxiliary system is necessary for the equivalence between non-unitary channels and their unitary channel dilation. This equivalence between arbitrary quantum channels and their unitary channel dilation is depicted in Fig.~\ref{fig::stinespring+instrument}$(a)$. Additionally, note that the Stinespring dilation can be combined with the channel representation of a quantum instrument presented in Eq.~\eqref{eq::instrument2channel}, so that all instruments can also be viewed as a unitary channel that makes use of an auxiliary system which is later discarded, as depicted in Fig.~\ref{fig::stinespring+instrument}$(b)$.

Exploiting the Stinespring dilation of general quantum instruments and general quantum channels into bipartite unitary channels that discard their auxiliary system, one can express the action---and consequently the simulation---of the quantum switch solely in terms of unitary channels. However, in order to do so, it is necessary to keep track of exactly which systems are acted upon, preserved, and discarded. In Fig.~\ref{fig::switchaction}, we show precisely how the action of the quantum switch on the different kinds of inputs discussed so far can be expressed in terms of its action on unitary channels. 

In Fig.~\ref{fig::switchaction}$(a)$, we depict the action of the quantum switch on general single-party channels. In Fig.~\ref{fig::switchaction}$(b)$, we represent the action of the quantum switch on single-party quantum instruments. In Fig.~\ref{fig::switchaction}$(c)$, we show the action of the switch in its most general case: when it acts on only part of bipartite general channels. Finally, in Fig.~\ref{fig::switchaction}$(d)$, the action of the switch on part of bipartite instruments is pictured. Trivially, the case in Fig.~\ref{fig::switchaction}$(a)$ is a particular case of Fig.~\ref{fig::switchaction}$(c)$. Notice, moreover, how both the cases of Fig.~\ref{fig::switchaction}$(b)$ and $(d)$, concerning quantum instruments, are also particular cases of Fig.~\ref{fig::switchaction}$(c)$. In the latter case in particular, this is because the dimension of the primed spaces of the input bipartite channels---the ones that the switch does not act upon---are arbitrary, and hence, the quantum outputs $\ket{a}$ and $\ket{b}$ of the instruments can be absorbed into the other output system of the unitary channels that is not acted upon by the quantum switch.

%%%%%%%%%%%%%%%%%
\subsection{Theorem~\ref{thm::go}: A simulation for bipartite unitary channels $U_A$ and bipartite general channels $B$}\label{subapp::proofoursimulation}
%%%%%%%%%%%%%%%%%

In this section, we prove Thm.~\ref{thm::go} from the main text and discuss how this particular-case simulation compares to more general simulation scenarios. We start by reproducing Eq.~\eqref{eq::simulation_general} of the main text which defines a general simulation of the quantum switch. Namely, a simulation $\map{C}$ that is able to prepare, with some finite number of calls $k_A$ and $k_B$, a channel $\map{C}(A^{\otimes k_A},B^{\otimes k_B})$ such that 
\begin{equation}\label{eq::simulation_general_app}
    \map{C}(A^{\otimes k_A},B^{\otimes k_B}) = \map{S}\otimes\map{I}(A,B) \ \ \ \forall \ A,B,
\end{equation}
where $A,B$ are arbitrary quantum channels, is a higher-order transformation that can simulate the action of the quantum switch in its most general form. 

We are now ready to restate and prove Theorem~\ref{thm::go}: 

\begin{theorem}\label{thm::go_app}
    The action of the quantum switch on part of bipartite quantum channels can be deterministically simulated by a quantum circuit that has access to an extra call to one the input channels, as long as that channel is restricted to being unitary.

    In other words, if $A$ is a bipartite unitary channel and $B$ is a bipartite general channel, there exists a quantum circuit described by a higher-order transformation $\map{C}$ that satisfies Eq.~\eqref{eq::simulation_general_app} for $k_A=2$ and $k_B=1$.
\end{theorem}

The proof is based on the explicit construction of the following quantum circuit, presented as Eq.~\eqref{cc::ourcircuit} in the main text and repeated here for convenience:

\begin{equation}\label{cc::ourcircuit_app}
	%\begin{tikzpicture}	\node[scale=0.8]{
	\begin{quantikz}[column sep=3.8mm]
\lstick{$c_I$}		&\ctrl{3}&\ctrl{4}&           &\ctrl{3}&\ctrl{4}&   &\ctrl[open]{3}&\ctrl[open]{4}&        &\ctrl[open]{3}&\ctrl[open]{4}& \rstick{$c_O$}	\\
\lstick{${\text{aux}_1}_I$}		&\targX{}&		  &			  &\targX{}&		&	        &\targX{}&		  &	   &\targX{}&	     & \rstick{${\text{aux}_1}_O$}\\
\lstick{${\text{aux}_2}_I$}		& 		 &\targX{}&		      &	   	   &\targX{}&	        &		 &\targX{}&		   &		&\targX{}& \rstick{${\text{aux}_2}_O$}\\
\lstick{$A'_I$}		&\targX{}&        &\gate[2,style={fill=JessicaColour!30},label style=black]{A}&\targX{}&        &		    &\targX{}&        &\gate[2,style={fill=JessicaColour!30},label style=black]{A}&\targX{}&	 & \rstick{$A'_O$} \\
\lstick{$t_I$}		& 		 &\targX{}& 	      &	       &\targX{}&\gate[2,style={fill=JessicaColour!30},label style=black]{B}&        &\targX{}&		   &		&\targX{}& \rstick{$t_O$} \\
\lstick{$B'_I$}		& 		 &		  & 	      &	       &	    &		    &        &		  &		   &		&	     &  \rstick{$B'_O$}
	\end{quantikz}
	%};
	%\end{tikzpicture}
\end{equation}

\begin{proof}
Let $A: \map{L}(\map{H}^{A_I}\otimes\map{H}^{A'_I}) \to \map{L}(\map{H}^{A_O}\otimes\map{H}^{A'_O})$ be a bipartite channel with Kraus operators $\{\ope{A}_i\}$ where $\ope{A}_i:\map{H}^{A_I}\otimes\map{H}^{A'_I}\to \map{H}^{A_O}\otimes\map{H}^{A'_O}$. Similarly, let $B:\map{L}(\map{H}^{B_I}\otimes\map{H}^{B'_I}) \to \map{L}(\map{H}^{B_O}\otimes\map{H}^{B'_O})$ be a bipartite channel with Kraus operators $\{\ope{B}_j\}$ where $\ope{B}_j:\map{H}^{B_I}\otimes\map{H}^{B'_I}\to \map{H}^{B_O}\otimes\map{H}^{B'_O}$. The action of the quantum switch $\map{S}$ on part of these bipartite channels results in the quantum channel $\map{S}\otimes\map{I}(A,B): 
\map{L}(\map{H}^{c_I}\otimes\map{H}^{t_I}\otimes\map{H}^{A'_I}\otimes\map{H}^{B'_I}) \to
\map{L}(\map{H}^{c_O}\otimes\map{H}^{t_O}\otimes\map{H}^{A'_O}\otimes\map{H}^{B'_O})$, whose definition is implied by the action of the quantum switch on single-party channels and linearity. It is given by
\begin{equation}\label{eq::switchBipartite}
    \map{S}\otimes\mathcal{I}(A,B)[\,\sigma_c\otimes\rho_t\otimes \omega_{A'B'} \,] = \sum_{i,j} \ope{S}_{ij} (\sigma_c\otimes\rho_t\otimes \omega_{A'B'}) \ope{S}_{ij}^\dagger,
\end{equation}
where $\omega_{A'B'}\in\mathcal{L}(\map{H}^{A'_I}\otimes\map{H}^{B'_I})$ is an arbitrary state of the quantum system in the primed spaces of Alice and Bob, which are not acted upon by the quantum switch. Before explicitly writing the $\ope{S}_{ij}$, it is convenient to decompose the Kraus operators $\ope{A}_i$ and $\ope{B}_i$ into linear combinations of operators that factorize between the primed and nonprimed spaces, which can always be done without loss of generality. It follows from linearity that one can always write
\begin{align}
    \ope{A}_i = \sum_a \alpha_{a|i}\, \ope{A}(a)\otimes \ope{A}'(a)
\end{align}
for some $\alpha_{a|i}\in\mathbb{C}$, $\ope{A}(a):\map{H}^{A_I}\to \map{H}^{A_O}$, and $\ope{A}'(a):\map{H}^{A'_I}\to \map{H}^{A'_O}$. Similarly, one can write 
\begin{align}
    \ope{B}_j = \sum_b  \beta_{b|j}\, \ope{B}(b)\otimes \ope{B}'(b)
\end{align}
for some $\beta_{b|j}\in\mathbb{C}$, $\ope{B}(b):\map{H}^{B_I}\to \map{H}^{B_O}$, and $\ope{B}'(b):\map{H}^{B'_I}\to \map{H}^{B'_O}$.
Then, the Kraus operators $\{\ope{S}_{ij}\}$ in Eq.~\eqref{eq::switchBipartite} take the form
\begin{align}\label{eq::switchBipartiteKrausGeneral}
    \ope{S}_{ij} = \ketbra{0}\otimes  \sum_{ab}  \alpha_{a|i}\, \beta_{b|j}\, \ope{B}(b)\ope{A}(a) \otimes \ope{A}'(a)\otimes \ope{B}'(b) + \ketbra{1}\otimes  \sum_{ab}  \alpha_{a|i}\, \beta_{b|j}\, \ope{A}(a)\ope{B}(b) \otimes \ope{A}'(a)\otimes \ope{B}'(b),
\end{align}
where the order of the spaces from left to right in each term is control, target, and Alice and Bob's primed systems.

The transformation of the quantum circuit in Eq.~\eqref{cc::ourcircuit_app} that uses two calls of the quantum channel $A$ and a single call of the quantum channel $B$ in the order ABA results in a quantum channel $\mathcal{C}(A^{\otimes 2},B)$ which has Kraus operators 
\begin{align}\label{eq::switchBipartiteKrausABA}
    \ope{C}_{ijk} =  \ketbra{0}\otimes \ope{A}_k \otimes \sum_{ab}  \alpha_{a|i}\, \beta_{b|j}\, \ope{A}'(a) \otimes \ope{B}(b)\ope{A}(a) \otimes \ope{B}'(b)+  \ketbra{1}\otimes \ope{A}_i \otimes \sum_{ab} \alpha_{a|k}\, \beta_{b|j} \ope{A}'(a) \otimes  \ope{A}(a)\ope{B}(b) \otimes \ope{B}'(b),
\end{align}
where the order of the spaces from left to right in each term coincides with the order of the wires from top to bottom in the quantum circuit in Eq.~\eqref{cc::ourcircuit_app} (i.e., control, first auxiliary, second auxiliary, Alice's primed, target, and Bob's primed systems).

Whenever the bipartite channel $A=U_A$ is unitary, it has a single Kraus operator, i.e., $A[\rho]=\ope{A}_0 \rho \ope{A}_0^\dagger$. Hence, following Eq.~\eqref{eq::switchBipartiteKrausGeneral}, in this case the quantum channel resulting from action of the quantum switch $\map{S}\otimes\map{I}(U_A,B)$ has Kraus operators
\begin{align}\label{eq::switchBipartiteKrausAUnitary}
    \ope{S}_{j} = \ketbra{0}\otimes  \sum_{ab}  \alpha_{a|0}\, \beta_{b|j}\, \ope{B}(b)\ope{A}(a) \otimes \ope{A}'(a)\otimes \ope{B}'(b) + \ketbra{1}\otimes  \sum_{ab}  \alpha_{a|0}\, \beta_{b|j}\, \ope{A}(a)\ope{B}(b) \otimes \ope{A}'(a)\otimes \ope{B}'(b).
\end{align}
Also in this case, following Eq.~\eqref{eq::switchBipartiteKrausABA}, the quantum channel resulting from action of the quantum circuit in Eq.~\eqref{cc::ourcircuit_app} is described by the Kraus operators
\begin{align}\label{eq::ourcircuir_kraus_unitary}
    \ope{C}_{j} &= \ketbra{0}\otimes \ope{A}_0 \otimes \sum_{ab}  \alpha_{a|0}\, \beta_{b|j}\, \ope{A}'(a) \otimes \ope{B}(b)\ope{A}(a) \otimes \ope{B}'(b)+  \ketbra{1}\otimes \ope{A}_0 \otimes \sum_{ab} \alpha_{a|0}\, \beta_{b|j}\, \ope{A}'(a) \otimes  \ope{A}(a)\ope{B}(b) \otimes \ope{B}'(b) \\
    &= \left( \ketbra{0} \otimes \sum_{ab}  \alpha_{a|0} \,\beta_{b|j}\, \ope{B}(b)\ope{A}(a) \otimes \ope{A}'(a) \otimes \ope{B}'(b)+  \ketbra{1}\otimes  \sum_{ab} \alpha_{a|0}\, \beta_{b|j}\,  \ope{A}(a)\ope{B}(b) \otimes \ope{A}'(a) \otimes \ope{B}'(b)\right) \otimes \ope{A}_0.
\end{align}
where, in the last line, we reordered the spaces from control, first auxiliary, second auxiliary, Alice's primed, target, and Bob's primed systems to control, target, and Alice and Bob's primed, first auxiliary, and second auxiliary systems. 

Note that the term in parenthesis is equal to that of Eq.~\eqref{eq::switchBipartiteKrausAUnitary}. Hence, the action of the quantum switch on part of a bipartite unitary channel $A=U_A$ and a bipartite general channel $B$ is equivalent to the action of the quantum circuit in Eq.~\eqref{cc::ourcircuit_app} that uses two calls of bipartite unitary channel $A=U_A$ and a single call of bipartite general channel $B$ in the order ABA. Moreover, note that one call of $A=U_A$ in the quantum circuit in Eq.~\eqref{cc::ourcircuit_app} is recovered, since $A_0$ factorizes in Eq.~\eqref{eq::ourcircuir_kraus_unitary}. This phenomenon is referred to as a catalytic higher-order transformation in the literature~\cite{yoshida2023reversing}, since the extra use of $A=U_A$ is recovered after the completion of the process. 
\end{proof}

We now contrast the different scenarios of the action of the quantum switch presented in Secs.~\ref{subapp::switchonextendedchannels}--\ref{subapp::unitarydilation} with the particular-case simulation of Theorem~\ref{thm::go}.

\begin{figure}%[h!]
\begin{center}
	\includegraphics[width=0.9\columnwidth]{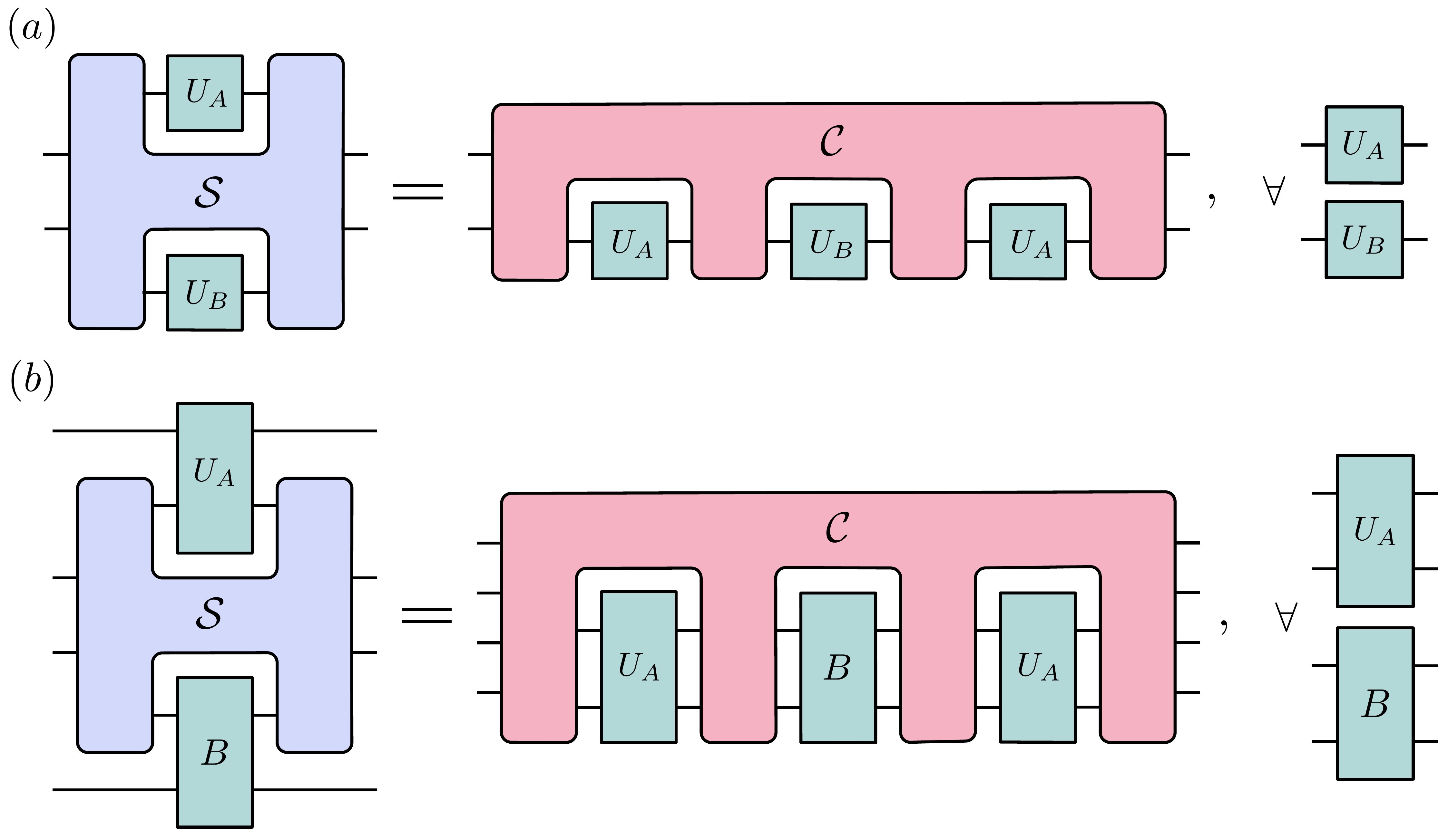}
	\caption{\textbf{Quantum comb representation of possible quantum switch simulations.} $(a)$ Representation of the simulation of the action of the quantum switch on single-party unitary channels as a quantum comb, presented in Ref.~\cite{chiribella2013quantum} and reproduced in the main text as the quantum circuit in Eq.~\eqref{cc::chiribella} $(b)$ Representation of the simulation of the action of the quantum switch on part of a bipartite unitary channel and part of a bipartite general quantum channel by a quantum comb, which we have shown to be possible by explicitly constructing the quantum circuit in Eq.~\eqref{cc::ourcircuit_app} {in Thm.}~\ref{thm::go}.}
\label{fig::oursimulation}
\end{center}
\end{figure}

Let us begin with the circuit from Ref.~\cite{chiribella2013quantum}, reproduced in Eq.~\eqref{cc::chiribella} of the main text, and depicted here as the quantum comb in Fig.~\ref{fig::oursimulation}$(a)$. This circuit simulates the action of the quantum switch on unitary channels by making use of one extra call to one of the input channels. Although every general quantum channel can be dilated into a unitary channel acting on a larger space, the simulation of the quantum switch on unitary channels in the case depicted in Fig.~\ref{fig::oursimulation}$(a)$ is not the most general case, because here the \textit{entire} unitary channel is ``plugged into'' the open slots of the quantum switch and of the quantum comb that performs its simulation.

The simulation presented in Ref.~\cite{chiribella2013quantum}, here in Fig.~\ref{fig::oursimulation}$(a)$, is a particular case of the simulation scenario we proved possible via the explicit construction of the quantum circuit in Eq.~\eqref{cc::ourcircuit_app}. Our generalization addresses the case where the quantum switch acts on only part of a unitary channel and part of a general quantum channel, with the simulation also requiring only one extra copy of the input unitary channel. We depict this scenario and our quantum circuit simulation as a quantum comb in Fig.~\ref{fig::oursimulation}$(b)$. Although more general than the previous result from Ref.~\cite{chiribella2013quantum}, this simulation result still does not hold in the most general case, depicted in Fig.~\ref{fig::simulation_general}$(c)$, because the input quantum channel $U_A$ is required to be a unitary channel. 

%%%%%%%%%%%%%%%%%%%%%%%%%%%%%%%%
%%%%%%%%%%%%%%%%%%%%%%%%%%%%%%%%

The quantum switch simulation presented in Fig.~\ref{fig::oursimulation}$(b)$ is also useful to illustrate the crucial aspect of the partial trace involved in the Stinespring dilation. Notice that this simulation covers the scenario where $B$ is an arbitrary bipartite channel, and $U_A$ is an arbitrary bipartite unitary channel. Na{\" i}vely, one could expect that---since our simulation covers all bipartite unitary channels $U_A$---due to the Stinespring dilation theorem, this simulation would also apply to the case where $A$ is a single-party arbitrary quantum channel, as in Fig.~\ref{fig::switchaction}$(a)$. However, this line of argumentation is false, as it is not possible to simulate the quantum switch for a pair of single-party arbitrary quantum channels, as proven in the main text. The logical gap in the argument is a misuse of the Stinespring dilation theorem, which requires the auxiliary system to be discarded, whereas in the simulation presented in Fig.~\ref{fig::oursimulation}$(b)$, there is no partial trace after the unitary operation $U_A$. When the auxiliary system of the Stinespring dilation is not discarded, the corresponding operation is not equivalent to the  quantum channel in question. From this circuit simulation perspective, keeping track of the auxiliary systems may be viewed as a ``loophole'', since the auxiliary system may carry additional information that is not provided by the channel $A$. To see this intuitively, consider a quantum channel $A$ with a Kraus decomposition given by $A[\rho]=\sum_i \ope{K}_i \rho \ope{K}_i^\dagger$ and which is dilated by the isometry $\ope{V}=\sum_i \ope{K}_i \otimes \ket{i}$, where $\ket{i}$ is a quantum state on the auxiliary system. The quantum state $\ket{i}$ may be viewed as a flag that indicates which Kraus operator was applied to $\rho$. If the flag state is not discarded, one could use this information to correlate Kraus elements between a first and a second call of the channel $A$, which cannot be done when the auxiliary system is traced out.

%%%%%%%%%%%%%%%%%%%%%%%%%%%%%%%%%%%%%%%%%%%%
\section{Full proof of Theorem~\ref{thm::nogo_exp}}\label{app::fullproof_nogoexp}
\setcounter{figure}{0}
%%%%%%%%%%%%%%%%%%%%%%%%%%%%%%%%%%%%%%%%%%%%

This section alone will adopt a slightly different notation from the main text and the remainder of the Appendix. This is due to the fact that the proof technique of this particular theorem is different from those applied to prove and present the other results, and it benefits from a different notation. Below is a summary of the changes, to facilitate the transition into this next section:

\begin{itemize}

    \item 
    We will denote a Hilbert space simply by the label of the system it describes, according to $X\coloneqq\map{H}^X$. In particular, the input and output Hilbert spaces of the control and target systems in the quantum switch higher-order transformation will be denoted instead as ``past'' and ``future'', using the notation $P_C\coloneqq\map{H}^{c_I}$, $P_T\coloneqq\map{H}^{t_I}$, $F_C\coloneqq\map{H}^{c_O}$, $F_T\coloneqq\map{H}^{t_O}$, and collectively as $P\coloneqq P_C\otimes P_T$ and $F\coloneqq F_C\otimes F_T$. 

    \item 
    Quantum channels will be denoted by cursive letters $\map{A}$ and $\map{B}$ instead of $A$ and $B$. Kraus operators and unitary operators will be denoted by $A_i$, $B_j$ and $U$, $V$ instead of $\ope{A}_i$, $\ope{B}_j$ and $\ope{U}$, $\ope{V}$. Higher-order transformations will continue to be denoted with cursive letters, such as $\map{S}$ for the quantum switch and $\map{C}$ for quantum combs or QC-CC simulators, as in the remainder of the Appendix.
    
    \item 
    Higher-order transformations that correspond to the simulator, be it a quantum comb or a QC-CC, will be denoted as a function $\map{C}$ that has as many arguments as there are slots in the transformation (e.g., $\map{C}=\map{C}(\map{A},\map{B},\map{C})$ for a $3$-slot higher-order transformation). The number of available calls to channel $\map{A}$ and $\map{B}$ will be denoted as $M$ (instead of $k_A$) and $N$ (instead of $k_B$). Hence, what would be written as $\map{C}(A^{\otimes{k_A}},B^{\otimes{k_B}})$ in the remainder of the Appendix, in this section becomes instead $\map{C}(\underbrace{\map{A},\dots,\map{A}}_M,\underbrace{\map{B},\dots,\map{B}}_N)$.

    \item 
    Choi operators, Choi vectors, and the link product will be used as defined in Sec. IV Methods of the main text.
    
\end{itemize}

We are now ready to restate and prove Thm.~\ref{thm::nogo_exp}.

\begin{theorem}[expanded]\label{thm::nogo_exp_app}
    Let $\map{S} :  [\map{L}(I) \to \map{L}(O)] \otimes  [\map{L}(I') \to \map{L}(O')]  \to [\map{L}(P) \to \map{L}(F)]$, where $I, O, I', O'$ are $n$-qubit Hilbert spaces and $P, F$ are $(n+1)$-qubit Hilbert spaces, be the quantum switch.
    
    There is no ($M+1$)-slot quantum circuit with classical control of causal order (QC-CC) higher-order transformation $\map C : \bigotimes_{i=1}^{M}[\map{L}(I_i) \to \map{L}(O_i)] \otimes [\map{L}(I'_1) \to \map{L}(O'_1)] \to [\map{L}(P) \to \map{L}(F)]$, where $\{I_i\}_{i}, \{O_i\}_{i}, I'_1, O'_1$ are $n$-qubit Hilbert spaces, satisfying
    \begin{align}
        \supermap{C}( \underbrace{\map{A},\dots, \map{A}}_M, \map{B} ) &= \map{S}(\map{A}, \map{B})\label{eq:switch_simulation_1}
    \end{align}
    for all mixed unitary channels $\map{A}$ and unitary channels $\map{B}$, if $M \leq  \max(2, 2^{n}-1)$.
\end{theorem}

\begin{proof} 
The proof is based upon a series of lemmas, proven below. First, we assume that $\mathcal{C}$ is a multilinear higher-order transformation whose Choi operator $C$ is positive semidefinite, and hence can be written as
    \begin{align}
    	&C= \sum_a \dketbra{C^{(a)}}
\end{align}
with $\dketbra{C^{(a)}}\geq0$ for all $a$.

Let us write the mixed unitary channel $\map{A}$ in Eq.~\eqref{eq:switch_simulation_1} as $\map{A}=\frac{1}{M}\sum_{i=1}^{M}\map{A}_i$, for some $M\in\mathbb{N}^+$, where $\{\map{A}_i\}_i$ are unitary channels with associated unitary operators $\{U_i\}_i$, and let unitary channel $\map{B}$ be associated to unitary operator $V_1$. We then invoke Lemma~\ref{lem:linearity_MN} (which requires Lemma~\ref{lem:span} for its proof) for $N=1$ to say that, since it is assumed that $\map{C}$ satisfies Eq.~\eqref{eq:switch_simulation_1}, each $\dketbra{C^{(a)}}$ satisfies
\begin{align}
      \dket{C^{(a)}} \ast \Big(  \bigotimes_{k=1}^{M} \dket{U_k}
         \otimes \dket{V_1} \Big)
          = \sum_{k=1}^{M} 
          \xi^{(a)}_{k1}(\{U_i\}_i,\ V_1) \,
          \dket{S}\ast \Big(\dket{U_k}\otimes \dket{V_1}\Big) \,  \label{eq:pure_C_ar_M1}
\end{align}
for some coefficients $\xi^{(a)}_{k1}(\{U_i\}_i,\ V_1) \in \mathbb{C}$, where $\dket{S}$ is the Choi vector of the quantum switch.

From Lemma~\ref{lem:p_as_vector_MN}, we know that if there exist coefficients $\xi^{(a)}_{k1} \leftarrow \tilde{\xi}^{(a)}_{k1}$ such that, for all $k \in \{1, \dots, M\}$, $\tilde{\xi}^{(a)}_{k1}$ is simultaneously  
\begin{enumerate}
    \item  independent of $U_{k}$ and $V_1$, and 
    \item  linear in $U_{k'}$ for all $k' \neq k$,
\end{enumerate}
then, there exist vectors $\dket{\tilde{\xi}^{(a)}}^{\{I_1 O_1, \cdots , I_M O_M \} \setminus \{I_k O_k\}}$ independent of $\{U_i\}_i$ and $V_1$ such that 
\begin{align} \label{eq:pure_C_lemmapasvector_thm1}
    &\dket{C^{(a)}}^{P I_1 O_1 \cdots  I_M O_M  I'_1 O'_1 F}
    = \sum_{k=1}^M 
    \dket{S}^{P I_k O_k I'_1 O'_1 F} \otimes
    \dket{\tilde{\xi}^{(a)}_{k1}}^{\{I_1 O_1, \cdots , I_M O_M \} \setminus \{I_k O_k\} } 
\end{align}
for all $a$. Such conditions are guaranteed to be satisfied via Lemma~\ref{lem:indep_M1} when $M<4^n/2 +2$ [which is implied by $M\leq \max(2,2^n-1)$].

Finally, invoking Lemma~\ref{lem:qccc_invalid} (which requires Lemmas~\ref{lem:linear_independence} and~\ref{lem:linear_independence2} for its proof), we find that if $M \leq \max(2, 2^n-1)$, a higher-order transformation with Choi operator $C=\sum_a\dketbra{C^{(a)}}$, where $\dket{C^{(a)}}$ satisfies Eqs.~\eqref{eq:pure_C_lemmapasvector_thm1} for all $a$ does not satisfy the conditions of a QC-CC transformation.
\end{proof}

%%%%%%%%%%%%%%%%%%%%%%%%%%%%
\subsection{Lemma~\ref{lem:span}}\label{subapp::lem:span}
%%%%%%%%%%%%%%%%%%%%%%%%%%%%

\begin{lemma}\label{lem:span}
    Let $\ket{\phi}$ and $\{\ket{\psi_i}\}_i$ be vectors in $\mathbb{C}^d$. 
    \begin{align}
        \text{If }
        \ketbra{\phi}\leq \sum_i \ketbra{\psi_i}, \;
        \text {then} \ket{\phi}\in\mathrm{span}(\{\ket{\psi_i}\}).
    \end{align}    
    That is, there exist complex numbers $\alpha_i$ such that $\ket{\phi}=\sum_i \alpha_i \ket{\psi_i}$.
\end{lemma}

\begin{proof}
The proof will go by contradiction. We start by pointing out that any vector $\ket{\phi}\in\mathbb{C}^d$ can be decomposed as 
\begin{align}
    \ket{\phi} = \ket{\psi}+\ket{\psi_\perp},
\end{align}
where $\ket{\psi}\in \mathrm{span}(\{\ket{\psi_i}\})$, $\ket{\psi_\perp}\notin \mathrm{span}(\{\ket{\psi_i}\})$. Also, since $\ket{\psi_\perp}\notin \mathrm{span}(\{\ket{\psi_i}\})$, we have that $\braket{\psi_\perp}{\psi_i}=0$ for every $i$.

Now, assume that  $\ket{\phi}\notin\mathrm{span}(\{\ket{\psi_i}\})$. In this case, we necessarily have that $\ket{\psi_\perp}\neq0$.
Using this decomposition $\ket{\phi} = \ket{\psi}+\ket{\psi_\perp}$, we can write the inequality $\ketbra{\phi}\leq \sum_i \ketbra{\psi_i},$ as 
\begin{align} \label{eq:psi_psiperp}
    \ketbra{\psi} + \ketbra{\psi}{\psi_\perp} +  \ketbra{\psi_\perp}{\psi} + \ketbra{\psi_\perp} \leq \sum_i \ketbra{\psi_i}.
\end{align}
We then apply $\bra{\psi_\perp}$ and  $\ket{\psi_\perp}$ on both sides of operator inequality in Eq.~\eqref{eq:psi_psiperp} to obtain the real number inequality
\begin{align} 
    \braket{\psi_\perp} \braket{\psi_\perp} \leq 0.
\end{align}
However, since $\ket{\psi_\perp}\neq0$, $\braket{\psi_\perp} \braket{\psi_\perp}$ is strictly positive, hence we have arrived at a contradiction. Therefore, $\ket{\phi}$ must belong to the $\mathrm{span}(\{\ket{\psi_i}\})$.
\end{proof}

%%%%%%%%%%%%%%%%%%%%%%%%%%%%
\subsection{Lemma~\ref{lem:linearity_MN} (for $M,N\in\mathbb{N}^+$)}\label{subapp::lem:linearity_MN}
%%%%%%%%%%%%%%%%%%%%%%%%%%%%

\begin{lemma}\label{lem:linearity_MN}
    Let $\map{S} : [\map{L}(I) \to \map{L}(O)] \otimes  [\map{L}(I') \to \map{L}(O')]  \to [\map{L}(P) \to \map{L}(F)]$, where $I, O, I', O'$ correspond to $d$-dimensional Hilbert spaces and $P, F$ correspond to $(2\times d)$-dimensional Hilbert spaces, be a multilinear higher-order transformation that acts on two quantum channels and has a rank-$1$ Choi operator $S=\dketbra{S}$ (such as, e.g., the quantum switch). Furthermore, let $\map C : \bigotimes_{i=1}^{M}[\map{L}(I_i) \to \map{L}(O_i)] \otimes \bigotimes_{j=1}^{N}[\map{L}(I'_j) \to \map{L}(O'_j)] \to [\map{L}(P) \to \map{L}(F)]$, where $M,N\in\mathbb{N}^+$ and $\{I_i\}_{i}, \{O_i\}_{i}, \{I'_j\}_{j}, \{O'_j\}_{j}$ correspond to $d$-dimensional Hilbert spaces, be a multilinear higher-order transformation that acts on $M+N$ quantum channels with a positive semidefinite Choi operator $C$, which, therefore, can be written as 
    \begin{align}\label{eq::Cspectral}
    	C= \sum_{a}\dketbra{C^{(a)}}\, ,
    \end{align}
    where each $\dketbra{C^{(a)}}\geq0$.
    Finally, let $\map{A}$ and $\map{B}$ be mixed unitary channels, such that $\map{A}=\frac{1}{K}\sum_{i=1}^{K}\map{A}_i$ and $\map{B}=\frac{1}{L}\sum_{j=1}^{L}\map{B}_j$ for some $K,L\in\mathbb{N}^+$, where $\{\map{A}_i\}_i$ and $\{\map{B}_j\}_j$ are unitary channels with associated unitary operators $\{U_i\}_i$ and $\{V_j\}_j$, respectively. 

    If $\map{C}$ corresponds to a simulation of the action of $\map{S}$ on mixed unitary channels $\map{A}$ and $\map{B}$, i.e., if $\map{C}$ satisfies 
    \begin{align}
        \supermap{C}( \underbrace{\map{A},\dots, \map{A}}_M, \underbrace{\map{B}, \dots , \map{B}}_N ) &= \map{S}(\map{A}, \map{B})
    \label{eq:switch_simulation_2}
    \end{align}
    for all mixed unitary channels $\map{A}$ and $\map{B}$ (or, if $N=1$, for all mixed unitary channels $\map{A}$ and unitary channels $\map{B}$), then, for every $a$, there exist coefficients $\{\xi^{(a)}_{kl}(\{U_i\}_i,\ \{V_j\}_j) \}_{kl} \in \mathbb{C}$ such that
    \begin{align}
        \dket{C_{a}} \ast \left(\dket{U_1} \otimes \ldots \otimes \dket{U_M} \otimes
        \dket{V_1} \otimes \ldots \otimes \dket{V_N}\right)
        = \sum_{k=1}^{M} \sum_{l=1}^{N}
        \xi^{(a)}_{kl} (\{U_i\}_i,\ \{V_j\}_j)
        \dket{S}\ast \left(\dket{U_k}\otimes \dket{V_l}\right) \, .
    \label{eq:pure_C_ar}
    \end{align}
\end{lemma}

\begin{proof}
Since, by assumption, Eq.~\eqref{eq:switch_simulation_2} holds for all mixed unitary channels $\map A, \map B$, for any sets of unitary channels $\{\map A_1, \dots \map A_K\}$ and $\{\map B_1, \dots \map{B}_L\}$ with $K,L \ge 1 $, one has that
\begin{equation}
	\map C \left(
	 \sum_{i_1 = 1}^K  \frac{\map A_{i_1}}K , \dots,  \sum_{i_M = 1}^K  \frac{\map A_{i_M}}K ;  
	 \sum_{j_1 = 1}^L \frac{\map B_{j_1}}L , \dots, \sum_{j_N = 1}^L \frac{\map B_{j_N}}L
	\right)
	=
	\map{S}
	\left(
	\sum_{k=1}^K \frac{\map A_{k}}K ,
	\sum_{l=1}^L \frac{\map B_{l}}L
	\right) \, .
\end{equation}
Note that for $N=1$, it is sufficient to assume that Eq.~\eqref{eq:switch_simulation_2} holds for all mixed unitary channels $\map{A}$ and unitary channels $\map{B}$, in which case we take $L=1$.

From the multilinearity of $\map{C}$ and $\map{S}$, it follows that
\begin{align}
    \frac{1}{K^M L^N} \sum_{i_1, \dots, i_M = 1}^K \; \sum_{ j_1, \dots , j_N = 1}^L
		\map C \left( 
    \map A_{i_1}, \dots,   \map A_{i_M} ;  
    \map B_{j_1}, \dots, \map B_{j_N}
	\right)
	=
	\frac{1}{KL}
	\sum_{k=1}^K \sum_{l=1}^L
	\map{S}
	\left(
    \map A_{k},
    \map B_{l}
	\right) \, .
\end{align}
Rewriting this expression in the Choi representation---using the convention that $C, S, A_i, B_j$ are the Choi matrices of  $\map C, \map S, \map{A}_i, \map{B}_j$, respectively---gives
\begin{align}
	\sum_{i_1, \dots, i_M\text{$=1$}}^{K} \sum_{ j_1, \dots , j_N = 1}^L
	 C \ast \left(
	 	 A_{i_1} \otimes \dots  \otimes     A_{i_M} \otimes   
	 B_{j_1}  \otimes  \dots  \otimes   B_{j_N}
	\right)
	=
	(K^{M-1} L^{N-1})
	\sum_{k=1}^K \sum_{l=1}^L
	 S \ast
	\left(
	 A_{k} \otimes
	 B_{l}
	\right) \, .
\end{align}

Now, using the decomposition of $C$ given by Eq.~\eqref{eq::Cspectral}, and $S=\dketbra{S}$, one has that
\begin{align}
\begin{split}
    \sum_a \dketbra{C^{(a)}}
	 \ast \sum_{i_1, \dots, i_M =1 }^K  \sum_{\substack{j_1, \dots , j_N = 1}}^L
	 \left(
	\bigotimes_{k=1}^M  A_{i_k} \otimes \bigotimes_{l=1}^N  B_{j_l} 
	\right) 
	= \,
     (K^{M-1} L^{N-1})
	\dketbra{S} \ast
    \sum_{k=1}^K \sum_{l=1}^L
	\left(
	 A_{k} \otimes
	 B_{l}
	\right) \, .
\end{split}
\end{align}	

The positivity of each $\dketbra{C^{(a)}}$ implies that
\begin{align}
\begin{split}\label{eq:big_sum}
    \dketbra{C^{(a)}}  \ast 
    \sum_{i_1, \dots, i_M =1 }^K  \sum_{\substack{j_1,\dots,j_N = 1}}^L
	\left(
	\bigotimes_{k=1}^M  A_{i_k} \otimes \bigotimes_{l=1}^N B_{j_l} 
	\right)
    \leq
	(K^{M-1} L^{N-1} )
    \dketbra{S} \ast
	\sum_{k=1}^K \sum_{l=1}^L
	\left(
	 A_{k} \otimes
	 B_{l}
	\right)
    \end{split}
\end{align}
for all $a$. 
    
Consider now the case where $K=M, L=N$.
Since $A_i$ and $B_j$ are Choi operators of unitary channels, they can be expressed as $A_{i} = \dketbra{U_i}$ and $B_j = \dketbra{V_j}$. Since the left-hand side of Eq.~\eqref{eq:big_sum} is a sum of positive operators, the inequality also holds for the sum of any subset of the terms on the left-hand side. Then, by considering only the term in the sum over $i_1, \dots, i_M, j_1, \dots, j_N$ that corresponds to $i_k=k, j_l=l$ for all $k,l$, we obtain that 
\begin{align}
\begin{split}
    \dketbra{C^{(a)}}  \ast \Big(
    \bigotimes_{k=1}^M \dketbra{U_k} \otimes
    \bigotimes_{l=1}^N \dketbra{V_l} \Big) 
    \leq  
    (M^{(M-1)} N^{(N-1)})
    \dketbra{S} \ast
	\sum_{k=1}^M 	\sum_{l=1}^N
	\left(
	\dketbra{U_k} \otimes \dketbra{V_l}
	\right) \, ,
\end{split}
\end{align}
for all $a$. Following the link product rule for vectors, as in, e.g., Lemma~1 of Ref.~\cite{yokojima2021consequences}, the above equation can be rewritten as 
\begin{align}
\begin{split}
    \bigotimes_{k=1}^M \bigotimes_{l=1}^N (\dket{C^{(a)}}  \ast 
    \dket{U_k} \otimes
    \dket{V_l})
    (\dbra{C^{(a)}}  \ast 
    \dbra{U_k} \otimes
    \dbra{V_l})
    \leq  
    (M^{(M-1)} N^{(N-1)}) \sum_{k=1}^M 	\sum_{l=1}^N
    (\dket{S} \ast
	\dket{U_k} \otimes \dket{V_l})
    (\dbra{S} \ast
	\dbra{U_k} \otimes \dbra{V_l})
	 \, .
\end{split}
\end{align}

Now, invoking Lemma~\ref{lem:span}, we arrive at the conclusion that 
\begin{align}
    \dket{C^{(a)}}  \ast  \bigotimes_{k=1}^M \dket{U_k} \otimes
    \bigotimes_{j=1}^N \dket{V_j}
    \in
    \text{span}(\{\dket{S} \ast
	\left(\dket{U_k} \otimes \dket{V_l}\right)
	\}_{kl})\, ,
\end{align}
for all $a$. Therefore, for every $a$, we have that
\begin{align}
    \dket{C^{(a)}}  \ast  \Big( \bigotimes_{k=1}^M \dket{U_k} \otimes
    \bigotimes_{j=1}^N \dket{V_j} \Big)
    = 	\sum_{k=1}^M 	\sum_{l=1}^N
    \xi^{(a)}_{kl}(\{U_i\}_i,\ \{V_j\}_j)\,
    \dket{S}\ast \left(\dket{U_k}\otimes \dket{V_l}\right) \, , 
\end{align}
for some coefficients $\xi^{(a)}_{kl}(\{U_i\}_i,\ \{V_j\}_j) \in \mathbb{C}$ for all $k,l$.

\end{proof}

%%%%%%%%%%%%%%%%%%%%%%%%%%%%
\subsection{Lemma~\ref{lem:p_as_vector_MN} (for $M,N\in\mathbb{N}^+$)}\label{subapp::lem:p_as_vector_MN}
%%%%%%%%%%%%%%%%%%%%%%%%%%%%

\begin{lemma}\label{lem:p_as_vector_MN}
    Let $\dket{S} \in P \otimes I \otimes O \otimes I' \otimes O' \otimes F$, where $I,O,I',O'$ correspond to $d$-dimensional Hilbert spaces and $P,F$ correspond to $(2\times d)$-dimensional Hilbert spaces, and let $\dket{C} \in  P \otimes I_1 \otimes O_1 \otimes \cdots \otimes I_M \otimes O_M \otimes I'_1 \otimes O'_1 \otimes \cdots \otimes I'_N \otimes O'_N \otimes F$, where $M, N \in \mathbb{N}^+$ and $\{I_i\}_{i}, \{O_i\}_{i}, \{I'_j\}_{j}, \{O'_j\}_{j}$ correspond to $d$-dimensional Hilbert spaces. 
    
    If, for a given $M,N$, for all $(M+N)$-tuples of $n$-qubit unitary operators $(U_1, \dots, U_M, V_1, \dots, V_N)$,
    \begin{equation}\label{eq:indep}
        \dket{C} * \dket{U_1}^{I_1 O_1} \otimes \dots \otimes \dket{U_M}^{I_M O_M} \otimes  \dket{V_1}^{I'_1 O'_1} \otimes \dots \otimes \dket{V_N}^{I'_N O'_N} = \sum_{k=1}^M \sum_{l=1}^N  \tilde{\xi}_{kl}(\{U_i\}_i,\ \{V_j\}_j) \dket{S} * \dket{U_k}^{I_k O_k} \otimes \dket{V_l}^{I'_l O'_l}
    \end{equation}
    for some coefficients $\tilde{\xi}_{kl}(\{U_i\}_i,\ \{V_j\}_j) \in \mathbb{C}$ for all $k,l$ that are simultaneously
    \begin{enumerate}
        \item  independent of $U_{k}$ and $V_{l}$, and 
        \item  linear in $U_{i}$ and $V_{j}$ for all $i \neq k, j \neq l$,
    \end{enumerate}
    then 
    \begin{align} \label{eq:pure_C_lemmapasvector}
        &\dket{C}^{P I_1 O_1 \ldots I_M O_M I'_1 O'_1 \ldots I'_N O'_N F}
        = \sum_{k=1}^M \sum_{l=1}^N 
        \dket{S}^{P I_k O_k I'_l O'_l F} \otimes
        \dket{\tilde{\xi}_{kl}}^{\{I_1 O_1 \ldots I_M O_M I'_1 O'_1 \ldots I'_N O'_N\} \setminus \{I_k O_k I'_l O'_l\}} 
    \end{align}
    for some vectors $\dket{\tilde{\xi}_{kl}}^{\{I_1 O_1 \ldots I_M O_M I'_1 O'_1 \ldots I'_N O'_N\} \setminus \{I_k O_k I'_l O'_l\} }$ that are independent of $\{U_i\}_i$ and $\{V_j\}_j$.
\end{lemma}

\begin{proof}
The first condition (independence) implies that we can write $\tilde{\xi}_{kl}(\{U_i\}_i,\ \{V_j\}_j) = \tilde{\xi}_{kl}(\{U_i\}_{i \neq k},\ \{V_j\}_{j \neq l})$.
The second condition (linearity) combined with the Riesz representation theorem implies that we can write the linear functions $\tilde{\xi}_{kl}(\{U_i\}_{i \neq k},\ \{V_j\}_{j \neq l})$ using vectors $\dket{\tilde{\xi}_{kl}}$ according to
\begin{align}
    \tilde{\xi}_{kl}(\{U_i\}_{i \neq k},\ \{V_j\}_{j \neq l})
    &=
    \dket{\tilde{\xi}_{kl}}\ast \bigotimes_{i \neq k} \dket{U_i} \otimes
    \bigotimes_{j \neq l} \dket{V_j} \, ,
\end{align}
where the definition of the link product of two vectors $\ket{\psi},\ket{\phi}\in\mathbb{C}^d$ acting on the same space implies that $\ket{\psi}*\ket{\phi}=\overline{\bra{\psi}}\ket{\phi}$, where the bar denotes complex conjugation.

Then, by explicitly writing in the system labels, for any sets of $n$-qubit unitaries $\{U_i\}_{i=1}^M$ and $\{V_j\}_{j=1}^N$, Eq.~\eqref{eq:indep} becomes
\begin{align}
\begin{split}
    &\dket{C}^{P I_1 O_1 \ldots I_M O_M I'_1 O'_1 \ldots I'_N O'_N F} 
    \ast   
    \left[
    \bigotimes_{i=1}^M \dket{U_i}^{I_i O_i} \otimes
    \bigotimes_{j=1}^N \dket{V_j}^{I'_j O'_j}
    \right]
    \\          
    &= \left[ \sum_{k= 1}^M \sum_{l= 1}^N 
    \dket{\tilde{\xi}_{kl}}^{\{I_1 O_1 \ldots I_M O_M I'_1 O'_1 \ldots I'_N O'_N\} \setminus \{I_k O_k I'_l O'_l\}} \otimes 
    \dket{S}^{P I_k O_k I'_l O'_l F}
    \right]
    \ast 
    \left[
    \bigotimes_{i=1}^M  \dket{U_i}^{I_i O_i} \otimes
    \bigotimes_{j=1}^N  \dket{V_j}^{I'_j O'_j}
    \right] \, , 
\end{split}
\end{align}
for some vectors $\dket{\tilde{\xi}_{kl}}^{\{I_1 O_1 \ldots I_M O_M I'_1 O'_1 \ldots I'_N O'_N\} \setminus \{I_k O_k I'_l O'_l\}}$.

Since the equation is true for all unitaries $\{U_i\}$ and $\{V_j\}$, and the  $\mathrm{span}(\{\dket{U}\,|\,U\in\text{SU}(d)\})=\mathbb{C}^d\otimes\mathbb{C}^d$, it implies that
\begin{align}
    &\dket{C}^{P I_1 O_1 \ldots I_M O_M I'_1 O'_1 \ldots I'_N O'_N F}
    =  \sum_{k=1}^M \sum_{l= 1}^N 
    \dket{S}^{P I_k O_k I'_l O'_l F} \otimes
    \dket{\tilde{\xi}_{kl}}^{\{I_1 O_1 \ldots I_M O_M I'_1 O'_1 \ldots I'_N O'_N\} \setminus \{I_k O_k I'_l O'_l\}} 
\end{align}
for some vectors $\dket{\tilde{\xi}_{kl}}^{\{I_1 O_1 \ldots I_M O_M I'_1 O'_1 \ldots I'_N O'_N\} \setminus \{I_k O_k I'_l O'_l\}}$.
\end{proof}

%%%%%%%%%%%%%%%%%%%%%%%%%%%%
\subsection{Lemma~\ref{lem:indep_M1} (for $M < 4^{n}/2 +2$ and $N=1$)}\label{subapp::lem:indep_M1}
%%%%%%%%%%%%%%%%%%%%%%%%%%%%

\begin{lemma}\label{lem:indep_M1}
    Let $\dket{S} \in P \otimes I \otimes O \otimes I' \otimes O' \otimes F$, where $I,O,I',O'$ correspond to $n$-qubit Hilbert spaces for some $n \in \mathbb{N}^+$ and $P,F$ correspond to $(n+1)$-qubit Hilbert spaces, and let $\dket{C} \in  P \otimes I_1 \otimes O_1 \otimes \cdots \otimes I_M \otimes O_M \otimes I'_1 \otimes O'_1 \otimes F$, where $M \in \mathbb{N}^+$ and $\{I_i\}_{i}, \{O_i\}_{i}, I'_1, O'_1$ correspond to $n$-qubit Hilbert spaces. 

    If, for all $(M+1)$-tuples of $n$-qubit unitary operators $(U_1, \dots, U_M, V_1)$, it holds that
    \begin{equation}\label{eq:indep_M1}
        \dket{C} * \dket{U_1}^{I_1 O_1} \otimes \dots \otimes \dket{U_M}^{I_M O_M} \otimes  \dket{V_1}^{I'_1 O'_1} = \sum_{k=1}^M \xi_{k1} \dket{S} * \dket{U_k}^{I_k O_k} \otimes \dket{V_1}^{I'_1 O'_1}
    \end{equation}
    for some complex numbers $\xi_{k1} \coloneqq \xi_{k1}(\{U_i\}_i,\ V_1) \in \mathbb{C}$, then there exist complex numbers $\tilde{\xi}_{k1} \coloneqq \tilde{\xi}_{k1} (\{U_i\}_i,\ V_1) \in \mathbb{C} $ such that Eq.~\eqref{eq:indep_M1} with  $\{{\xi}_{k1}\}_{k=1}^M \leftarrow \{\tilde{\xi}_{k1}\}_{k=1}^M$ remains satisfied and, for all $k \in \{1, \dots, M\}$, $\tilde{\xi}_{k1}$ is simultaneously
    \begin{enumerate}
        \item  independent of $U_{k}$ and $V_1$ (independence condition), and 
        \item  linear in $U_{i}$ for all $i \neq k$ (linearity condition),
    \end{enumerate}
    as long as $M < 4^{n}/2 +2$. 
\end{lemma}

\begin{proof}
Assume that Eq. \eqref{eq:indep_M1} holds. Then, in particular, it holds for the choice $U_i=\sigma_{\vec{r_i}}$ for $i\in \{1,\ldots,M\}$ and $V_1=\sigma_{\vec{q}_1}$, where $\vec{r}_i,\vec{q}_1\in \{0,1,2,3\}^{\times n}$ and $\{\sigma_{\vec{r}_i}\}_i,\ \sigma_{\vec{q}_1}$ are $n$-qubit Pauli operators.
Here, the set of $n$-qubit Pauli operators is defined by
\begin{align}
    \left\{\sigma_{\vec{r}}\coloneqq \bigotimes_{i=1}^{n} \sigma_{r_i}\middle|\vec{r}\in\{0,1,2,3\}^{\times n}\right\},
\end{align}
where $\sigma_0, \sigma_1, \sigma_2, \sigma_3$ are 1-qubit Pauli operators defined by
\begin{align}
    \sigma_0 \coloneqq\left(\begin{matrix}
        1 & 0\\
        0 & 1
    \end{matrix}
    \right),\quad 
    \sigma_1\coloneqq \left(\begin{matrix}
        0 & 1\\
        1 & 0
    \end{matrix}
    \right),\quad
    \sigma_2\coloneqq
    \left(\begin{matrix}
        0 & -i\\
        i & 0
    \end{matrix}
    \right),\quad
    \sigma_3\coloneqq \left(\begin{matrix}
        1 & 0\\
        0 & -1
    \end{matrix}
    \right).
\end{align}
Thus,
\begin{equation}
    \dket{C} * \dket{\sigma_{\vec{r}_1}}^{I_1 O_1} \otimes \dots \otimes \dket{\sigma_{\vec{r}_M}}^{I_M O_M} \otimes  \dket{\sigma_{\vec{q}_1}}^{I'_1 O'_1} = \sum_{k=1}^M \xi_{k1} \dket{S} * \dket{\sigma_{\vec{r}_k}}^{I_k O_k} \otimes \dket{\sigma_{\vec{q}_1}}^{I'_1 O'_1}.
\end{equation}
Now suppose that $F \ge 1$ elements of the $\{\sigma_{\vec{r}_i} \}_{i=1}^M$ are equal to some fixed $n$-qubit Pauli operator $\sigma_{\vec{w}}$. Let the set of integers labeling those Pauli operators be denoted by $\mathbb{F}\coloneqq \{1 \leq i \leq M | \sigma_{\vec{r}_i}=\sigma_{\vec{w}} \} $, such that $|\mathbb{F}|=F$. Equation~\eqref{eq:indep_M1} then reads 
\begin{equation}\label{eq:indep_w_M1}
\begin{split}
    \dket{C} * \bigotimes_{i \in \mathbb{F}} \dket{\sigma_{\vec{w}}}^{I_i O_i} \otimes \bigotimes_{i \in \{1,\dots , M\}\backslash \mathbb{F}}
    \dket{\sigma_{\vec{r}_{i}}}^{I_{i} O_{i}} 
    \otimes \dket{\sigma_{\vec{q}_1}}^{I'_1 O'_1} 
    = \sum_{k=1}^M \xi_{k1}^{\{\vec{r}_l\}_l,\vec{q}_1}  \dket{S} * \dket{\sigma_{\vec{r}_k}}^{I_k O_k} \otimes \dket{\sigma_{\vec{q}_1}}^{I'_1 O'_1} \, ,
\end{split}
\end{equation}
where, for the input unitaries chosen as $n$-qubit Paulis, we define
\begin{align}
    \xi_{k1}^{\{\vec{r}_l\}_l,\vec{q}_1} 
    \coloneqq
    \xi_{k1}(  \{U_i=\sigma_{\vec{r}_i}\}_{i =1}^M ,V_1=\sigma_{\vec{q}_1}) \ .
\end{align} 
In the following, we will adopt the shorthand convention that any changes to the dependence of $\xi_{k1}$ from $\xi_{k1}^{\{\vec{r}_l\}_l,\vec{q}_1}$ will be specified as $\xi_{k1}^{\{\vec{r}_l\}_l,\vec{q}_1}[U_i= (\dots), V_j = (\dots)]$, with all unspecified arguments $U_i, V_1$ defined to be the same as for $\xi_{k1}^{\{\vec{r}_l\}_l,\vec{q}_1}$ defined above.
A key point that we note for later is that the value of each individual variable $\xi_{k1}^{\{\vec{r}_l\}_l,\vec{q}_1}$ for $k \in \mathbb{F}$ is not uniquely determined from Eq.~\eqref{eq:indep_M1}, so we can take a different set of variables $\tilde{\xi}_{k1}$ still satisfying Eq.~\eqref{eq:indep_M1}.

We now show that for all $k \in \{1,\dots, M\}$, the variables $\xi_{k1}$ can be 
replaced by $\tilde{\xi}_{k1}$ defined by
\begin{align}
    \tilde{\xi}_{k1}(\{U_i=\sigma_{\vec{r}_i}\}_{i=1}^M,V_1=\sigma_{\vec{q}_1})&\coloneqq\xi_{k1}^{\{\vec{r}_l\}_l,\vec{q}_1}[U_k=\sigma_{\vec{r}^*_k}, V_1=\sigma_{\vec{q}^*}] \, ,
    \label{eq:global_def_M1}
    \\
    \tilde{\xi}_{k1}\left(\left\{U_i=\sum_{\vec{r}_i}{\alpha_{\vec{r}_i}^i \sigma_{\vec{r}_i}}\right\}_{i=1}^M,\ V_1=\sum_{\vec{q}_1}\beta_{\vec{q}_1}^1\sigma_{\vec{q}_1}\right)&\coloneqq
    \sum_{\{\vec{r}_l\}_{l\neq k}}
    \left(
    \prod_{j\neq k}
    \alpha^j_{\vec{r}_j}\right)
    \tilde{\xi}_{k1}(\{U_i=\sigma_{\vec{r}_i}\}_{i=1}^M,V_1=\sigma_{\vec{q}_1}) \, ,
    \label{eq:global_def_M1_linear}
\end{align}
where $\vec{r}^*_k \in \{0,1,2,3\}^{\times n}$ is an arbitrary vector \textit{outside} of the set $\{\vec{r}_1,\ldots,\vec{r}_{k-1}, \vec{r}_{k+1},\ldots,\vec{r}_M\}$, $\vec{q}^* \in \{0,1,2,3\}^{\times n}$ is an arbitrary fixed vector, and $\alpha_{\vec{r}_i}^i, \beta_{\vec{q}_1}^1$ are complex numbers. 
In the discussion below, we pick one choice of $\vec{r}_k^*$ defined as a function of $\vec{r}_1,\ldots,\vec{r}_{k-1}, \vec{r}_{k+1},\ldots,\vec{r}_M$, i.e., $\vec{r}_k^* = \vec{r}_k^*(\vec{r}_1,\ldots,\vec{r}_{k-1}, \vec{r}_{k+1},\ldots,\vec{r}_M)$. Such an $\vec{r}_k^*$ always exists for $M < 4^n$.

By construction, if the definition in Eqs.~\eqref{eq:global_def_M1}--\eqref{eq:global_def_M1_linear} satisfies Eq.~\eqref{eq:indep_M1}, then $\tilde{\xi}_{k1}$ satisfies both the linearity and independence conditions outlined in the statement of the lemma. We now proceed to show that the definition in Eqs.~\eqref{eq:global_def_M1}--\eqref{eq:global_def_M1_linear} indeed satisfies Eq.~\eqref{eq:indep_M1}. We do this by considering the dependence of $\xi_{k1}^{\{\vec{r}_l\}_l,\vec{q}_1}$ on the unitaries $\{U_i=\sigma_{\vec{r}_i}\}_{i =1}^M $ and $V_1=\sigma_{\vec{q}_1}$  in turn.

%%%%%%%%%%%%%
\subsection*{Dependence on $\{U_i\}_{i=1}^M$}
%%%%%%%%%%%%%

We focus on the case where $\{U_i\}_i,\ V_1$ are chosen from the set of Pauli operators. 
For every $k \in \{1, \dots M\}$, we choose one $\sigma_{\vec{r}^*_k}$ and take $\sigma$ to be the unique $n$-qubit Pauli operator such that $\sigma  \sigma_{\vec{r}_k} = \beta \sigma_{\vec{r}^*_k}$, where $\beta \in \{-1,1,i,-i\}$. We then consider the following expression
\begin{equation}
    \mathbf{E}\coloneqq  
    \frac{\d}{\d \theta} \bigg |_{\theta=0} 
    \left[
    \dket{C} 
    * \bigotimes_{m \in \mathbb{F}_k}  \dket{e^{i \theta \sigma } \sigma_{\vec{r}_m}}^{I_m O_m} 
    * \bigotimes_{m=1 | m \notin \mathbb{F}_k}^M \dket{\sigma_{\vec{r}_{m}}}^{I_{m} O_{m}} 
    * \dket{\sigma_{\vec{q}_1}}^{I'_1 O'_1} 
    \right] \, ,
\end{equation}
where $\mathbb{F}_k\coloneqq\{1\leq i\leq M\mid \sigma_{\vec{r}_i}=\sigma_{\vec{r}_k}\}$. 
Due to linearity, this expression can be evaluated in two ways: either by first computing the derivative and then applying Eq.~\eqref{eq:indep_M1}, or by first applying Eq.~\eqref{eq:indep_M1} and then computing the derivative. The former method gives
\begin{align}
    \mathbf{E} &=  
    i \beta \sum_{m\in \mathbb{F}_k}
    \left[
    \dket{C} 
    * \dket{ \sigma_{\vec{r}^*_k}}^{I_m O_m} 
    * \bigotimes_{i \in \mathbb{F}_k | i \neq m}  \dket{ \sigma_{\vec{r}_{i}}}^{I_{i} O_{i}} 
    * \bigotimes_{i=1 | i \notin \mathbb{F}_k}^M \dket{\sigma_{\vec{r}_{i}}}^{I_{i} O_{i}} 
    *  \dket{\sigma_{\vec{q}_1}}^{I'_1 O'_1} 
    \right] \notag
     \\ \nonumber
    &= 
    i \beta \sum_{m\in \mathbb{F}_k} 
    \Bigg[
    \xi_{m1}^{\{\vec{r}_l\}_l,\vec{q}_1} [U_m = \sigma_{\vec{r}^*_k}]
    \dket{S} * \dket{\sigma_{\vec{r}^*_k}} * \dket{\sigma_{\vec{q}_1}} 
    \\
    &~~~~~~~~+
    \sum_{i \in \mathbb{F}_k | i \neq m} \xi_{i1}^{\{\vec{r}_l\}_l,\vec{q}_1} [U_m = \sigma_{\vec{r}^*_k}]
    \dket{S} * \dket{\sigma_{\vec{r}_k}} * \dket{\sigma_{\vec{q}_1}} \nonumber
    \\ 
    &~~~~~~~~+
    \sum_{i=1 | i \notin \mathbb{F}_k}^M \xi_{i1}^{\{\vec{r}_l\}_l,\vec{q}_1} [U_m = \sigma_{\vec{r}^*_k}]
    \dket{S} * \dket{\sigma_{\vec{r}_{i}}} * \dket{\sigma_{\vec{q}_1}}
    \Bigg] \, , 
\end{align}
while the latter gives
\begin{align}
	\mathbf{E} &=  
	\frac{\d}{\d \theta} \bigg |_{\theta=0} 
	\Bigg[
	 \sum_{m\in \mathbb{F}_k}
	\xi_{m1}^{\{\vec{r}_l\}_l,\vec{q}_1} 	 [\{ U_{i} = e^{i \theta \sigma} \sigma_{\vec{r}_k}\}_{i\in \mathbb{F}_k}]
	\dket{S} * \dket{e^{i \theta \sigma} \sigma_{\vec{r}_k}}^{I_m O_m} * \dket{\sigma_{\vec{q}_1}}^{I'_1 O'_1} 
	\notag \\
	&~~~~~~~~~~~~+
	 \sum_{m=1 | m\notin \mathbb{F}_k}^M \xi_{m1}^{\{\vec{r}_l\}_l,\vec{q}_1} [\{ U_{i} = e^{i \theta \sigma} \sigma_{\vec{r}_k}\}_{i\in \mathbb{F}_k}]
	\dket{S} * \dket{ \sigma_{\vec{r}_m}}^{I_m O_m} * \dket{\sigma_{\vec{q}_1}}^{I'_1 O'_1} 
	\Bigg] \notag
	\\
    &=
    i  \beta 
    \sum_{m\in \mathbb{F}_k} 
    \xi_{m1}^{\{\vec{r}_l\}_l,\vec{q}_1}
    \dket{S} *  \dket{\sigma_{\vec{r}^*_k}} * \dket{\sigma_{\vec{q}_l}} \notag
    \\ 
    &~~~~~~~~~~~~+
	\frac{\d}{\d \theta} \bigg |_{\theta=0} 
    \Bigg[
    \sum_{m\in \mathbb{F}_k} 
    \xi_{m1}^{\{\vec{r}_l\}_l,\vec{q}_1} [\{ U_{i} = e^{i \theta \sigma} \sigma_{\vec{r}_k}\}_{i\in \mathbb{F}_k}]
    \Bigg]
    \dket{S} * \dket{ \sigma_{\vec{r}_k}} * \dket{\sigma_{\vec{q}_1}} \notag
    \\ 
    &~~~~~~~~~~~~+
    \sum_{\vec{v}\in \{\vec{r}_1,\ldots,\vec{r}_M\}\backslash\{\vec{r}_k\}}
	\frac{\d}{\d \theta} \bigg |_{\theta=0} 
    \Bigg[
    \sum_{m\in \mathbb{F}_{\vec{v}}}
    \xi_{m1}^{\{\vec{r}_l\}_l,\vec{q}_1} [\{ U_{i} = e^{i \theta \sigma} \sigma_{\vec{r}_k}\}_{i\in \mathbb{F}_k}]
    \Bigg] 
    \dket{S} * \dket{ \sigma_{\vec{v}}}* \dket{\sigma_{\vec{q}_1}} 
    \, ,
\end{align}
where $\mathbb{F}_{\vec{v}}\coloneqq\{1\leq m\leq M\mid \vec{r}_m=\vec{v}\}$. Note that the vector
\begin{align}
    \dket{ S}*\dket{\sigma_{\vec{r}_m}}^{I_m O_m}*\dket{\sigma_{\vec{q}_1}}^{I_1' O'_1}
\end{align}
belongs to the Hilbert space $P\otimes F$, which is independent of $I_m, O_m, I_1', O'_1$, thus the superscripts $I_m, O_m, I_1', O'_1$ can be omitted. Also, note that 
\begin{align}
    \sum_{m\in \mathbb{F}_k} 
    \xi_{m1}^{\{\vec{r}_l\}_l,\vec{q}_1} [\{ U_{i} = e^{i \theta \sigma} \sigma_{\vec{r}_k}\}_{i\in \mathbb{F}_k}]
\end{align}
and
\begin{align}
    \sum_{m\in \mathbb{F}_{\vec{v}}}
    \xi_{m1}^{\{\vec{r}_l\}_l,\vec{q}_1} [\{ U_{i} = e^{i \theta \sigma} \sigma_{\vec{r}_k}\}_{i\in \mathbb{F}_k}]
\end{align}
are differentiable since their values are uniquely determined from 
\begin{align}\label{eq:diff_M1}
    &\dket{C} 
    * \bigotimes_{m \in \mathbb{F}_k}  \dket{e^{i \theta \sigma } \sigma_{\vec{r}_m}}^{I_m O_m} 
    * \bigotimes_{m=1 | m \notin \mathbb{F}_k}^M \dket{\sigma_{\vec{r}_{m}}}^{I_{m} O_{m}} 
    * \dket{\sigma_{\vec{q}_1}}^{I'_1 O'_1} 
    \nonumber\\
    =&
    \left(
    \sum_{m\in \mathbb{F}_k} 
    \xi_{m1}^{\{\vec{r}_l\}_l,\vec{q}_1} [\{ U_{i} = e^{i \theta \sigma} \sigma_{\vec{r}_k}\}_{i\in \mathbb{F}_k}]
    \right)
    \dket{ S}*\dket{e^{i\theta \sigma}\sigma_{\vec{r}_k}}*\dket{\sigma_{\vec{q}_1}}
    \nonumber\\
    +&\sum_{\vec{v}\in \{\vec{r}_1,\ldots ,\vec{r}_M\}\backslash\{\vec{r}_k\}}
    \left(
    \sum_{m \in\mathbb{F}_{\vec{v}}}
    \xi_{m1}^{\{\vec{r}_l\}_l,\vec{q}_1} [\{ U_{i} = e^{i \theta \sigma} \sigma_{\vec{r}_k}\}_{i\in \mathbb{F}_k}]
    \right)
    \dket{S} * \dket{ \sigma_{\vec{v}}}* \dket{\sigma_{\vec{q}_1}} \, ,
\end{align}
and thus can be obtained from the inner product of the vector on the left-hand side of Eq.~\eqref{eq:diff_M1} and vectors $\dket{ S}*\dket{e^{i\theta \sigma}\sigma_{\vec{r}_k}}*\dket{\sigma_{\vec{q}_1}}$ or $\dket{ S}*\dket{\sigma_{\vec{v}}}*\dket{\sigma_{\vec{q}_1}}$ (which are mutually orthogonal), which are polynomials of $e^{\pm i\theta}$.

By comparing the coefficients for $\dket{S}*\dket{\sigma_{\vec{r}^*_k}}*\dket{\sigma_{\vec{q}_1}}$, we find that
\begin{align}\label{eq:sum_equal_linear}
    \sum_{m\in \mathbb{F}_k}\xi_{m1}^{\{\vec{r}_l\}_l,\vec{q}_1}=\sum_{m\in \mathbb{F}_k}\xi_{m1}^{\{\vec{r}_l\}_l,\vec{q}_1}[U_m=\sigma_{\vec{r}_k^*}].
\end{align}

%%%%%%%%%%%%%
\subsection*{Dependence on $V_1$}
%%%%%%%%%%%%%

In this part of the proof, we adopt the following shorthand notations. For any unitary operators $U, V$,
\begin{align}
	\dket{C[V]}&\coloneqq\dket{C}*\bigotimes_{i=1}^M\dket{\sigma_{\vec{r}_i}}^{I_i O_i}*\dket{V}^{I_1'O_1'}\nonumber\\
	\dket{S(U, V)}&\coloneqq\dket{ S}*\dket{U}*\dket{V}
	\nonumber\\
	b(\sigma_{\vec{v}},V)
	&\coloneqq
	\sum_{m\in \mathbb{F}_{\vec{v}}}
	\xi^{\{\vec{r}_l\}_l,\vec{q}_1}_{m1}[V_1=V]\, ,
\end{align}
where $\mathbb{F}_{\vec{v}}\coloneqq\{1\leq i\leq M\mid \vec{r}_i=\vec{v}\}$.

For any two $n$-qubit Pauli operators $\sigma_A, \sigma_B$, either the operator $(\sigma_A+\sigma_B)/\sqrt{2}$ or  the operator  $(\sigma_A+i\sigma_B)/\sqrt{2}$ is unitary. 
When $U\coloneqq(\sigma_A+\beta \sigma_B)/\sqrt{2}$ with $\beta\in \{1, i\}$ is unitary, the following equality holds for any $\{\vec{r}_1,\ldots,\vec{r}_M\}, \sigma_A, \sigma_B$:
\begin{align}\label{eq::sq2-AB}
	0 = 
	&\dket{C[U]}-
	\frac{1}{\sqrt{2}}
	(\dket{C[\sigma_A]}+\beta \dket{C[\sigma_B]})
	\nonumber\\
	=&\frac{1}{\sqrt{2}}
	\sum_{\vec{v}\in \{\vec{r}_1,\ldots,\vec{r}_M\}}    
	\left[
	\{b(\sigma_{\vec{v}},U)-b(\sigma_{\vec{v}},\sigma_A)\}
	\dket{S(\sigma_{\vec{v}},\sigma_A)}+
	\beta\{b(\sigma_{\vec{v}},U)-b(\sigma_{\vec{v}},\sigma_B) \}
	\dket{S(\sigma_{\vec{v}},\sigma_B)}
	\right].
\end{align}

We now calculate the inner product 
\begin{equation}
    \dbraket{S(\sigma_{\vec{v}}, \sigma_A)}{S(\sigma_{\vec{v}'}, \sigma_B)} = \tr (\sigma_B \sigma_A \sigma_{\vec{v}}  \sigma_{\vec{v}'} ) + \tr ( \sigma_{\vec{v}} \sigma_A \sigma_B \sigma_{\vec{v}'} ) \, .
\end{equation}
From this, it is clear that $\dbraket{S(\sigma_{\vec{v}}, \sigma_A)}{S(\sigma_{\vec{v}'}, \sigma_B)} = 0$ if $\sigma_{\vec{v}'} \not \propto \sigma_{\vec{v}} \sigma_A\sigma_B$. Therefore, taking an inner product of Eq.~\eqref{eq::sq2-AB} with $\dket{S(\sigma_{\vec{v}}, \sigma_A)}$, we obtain
\begin{align}
	\begin{cases}
	  \{b(\sigma_{\vec{v}},U)-b(\sigma_{\vec{v}},\sigma_A)\} \dbraket{S(\sigma_{\vec{v}}, \sigma_A)} = 0 \textrm{~~~~if for all}~~ \sigma_{\vec{v}'} \in \{\sigma_{\vec{r}_1},\ldots,\sigma_{\vec{r}_M}\}: \sigma_{\vec{v}'} \not \propto \sigma_{\vec{v}} \sigma_A \sigma_B \, \\
	  \\
	 \{b(\sigma_{\vec{v}},U)-b(\sigma_{\vec{v}},\sigma_A)\} \dbraket{S(\sigma_{\vec{v}}, \sigma_A)} \\
	 ~~~+ \beta \{b(\gamma \sigma_{\vec{v}}\sigma_A \sigma_B,U)-b(\gamma \sigma_{\vec{v}}\sigma_A \sigma_B, \sigma_B) \} \dbraket{S(\sigma_{\vec{v}}, \sigma_A)}{S(\gamma\sigma_{\vec{v}}\sigma_A \sigma_B, \sigma_B)} = 0 \textrm{~~~~else},
	\end{cases}
\end{align}
where $\gamma\in \{1,-1, i,-i\}$ is defined by the unique choice of $\sigma_{\vec{v}'} \in \{\sigma_{\vec{r}_1},\ldots,\sigma_{\vec{r}_M}\}$  such that
\begin{align}\label{eq:gamma}
	\sigma_{\vec{v}'} = \gamma \sigma_{\vec{v}} \sigma_A\sigma_B \, .
\end{align}

In the first case, i.e., if for all $\sigma_{\vec{v}'} \in \{\sigma_{\vec{r}_1},\ldots,\sigma_{\vec{r}_M}\}: \sigma_{\vec{v}'} \not \propto \sigma_{\vec{v}} \sigma_A\sigma_B$, we directly obtain
\begin{align}\label{eq:UA0}
	b(\sigma_{\vec{v}},U)-b(\sigma_{\vec{v}},\sigma_A) = 0 \, .
\end{align}
In the second case, if 
\begin{equation}\label{eq:gamma_vAB}
	\dbraket{S(\sigma_{\vec{v}}, \sigma_A)}{S(\gamma \sigma_{\vec{v}}\sigma_A\sigma_B, \sigma_B)} =  \tr (\sigma_B \sigma_A \sigma_{\vec{v}} \gamma \sigma_{\vec{v}} \sigma_A \sigma_B ) + \tr ( \sigma_{\vec{v}} \sigma_A \sigma_B \gamma \sigma_{\vec{v}} \sigma_A \sigma_B ) = 0
\end{equation}
holds, we also obtain Eq.~\eqref{eq:UA0}

By a similar argument, if for all $\sigma_{\vec{v}''} \in \{\sigma_{\vec{r}_1},\ldots,\sigma_{\vec{r}_M}\}: \sigma_{\vec{v}''} \not \propto \sigma_{\vec{v}} \sigma_B \sigma_A$, we directly obtain
\begin{align}\label{eq:UB0}
	b(\sigma_{\vec{v}},U)-b(\sigma_{\vec{v}},\sigma_B) = 0 \, .
\end{align}
Alternatively, if
\begin{equation}\label{eq:delta_vAB}
	\dbraket{S(\sigma_{\vec{v}}, \sigma_B)}{S(\gamma \sigma_{\vec{v}}\sigma_B\sigma_A, \sigma_A)} =  \tr (\sigma_A \sigma_B \sigma_{\vec{v}} \delta \sigma_{\vec{v}} \sigma_B \sigma_A ) + \tr ( \sigma_{\vec{v}} \sigma_B \sigma_A \delta \sigma_{\vec{v}} \sigma_B \sigma_A ) = 0
\end{equation}
holds, where $\delta \in \{1,-1, i,-i\}$ is defined by the unique choice of $\sigma_{\vec{v}''} \in \{\sigma_{\vec{r}_1},\ldots,\sigma_{\vec{r}_M}\}$  such that
 \begin{align}\label{eq:delta}
 	\sigma_{\vec{v}''} = \delta \sigma_{\vec{v}} \sigma_B\sigma_A \, ,
 \end{align}
we also obtain Eq.~\eqref{eq:UB0}.

Consider now the conditions for Eq.~\eqref{eq:gamma_vAB} to be satisfied. The first term is given by $ \tr (\sigma_B \sigma_A \sigma_{\vec{v}} \gamma \sigma_{\vec{v}} \sigma_A \sigma_B ) = \gamma \tr \1 $. For the second term, there are four cases:
\begin{itemize}
    \item If $\gamma= \pm 1$, then $\pm \sigma_{\vec{v}} \sigma_A\sigma_B$ is an $n$-qubit Pauli operator so the second term is given by\\
    $\tr[\sigma_{\vec{v}} \sigma_A \sigma_B \gamma \sigma_{\vec{v}} \sigma_A \sigma_B ]=  \gamma \tr[(\pm \sigma_{\vec{v}} \sigma_A \sigma_B) (\pm \sigma_{\vec{v}} \sigma_A \sigma_B)]  = \gamma \tr \1 $. 
    
    \item If $\gamma=\pm i$, then $\pm i \sigma_{\vec{v}} \sigma_A\sigma_B$ is an $n$-qubit Pauli operator so the second term is given by\\
    $\tr[(\sigma_{\vec{v}} \sigma_A \sigma_B) \gamma (\sigma_{\vec{v}} \sigma_A \sigma_B)]  = - \gamma \tr[(\pm i \sigma_{\vec{v}} \sigma_A \sigma_B)  (\pm i \sigma_{\vec{v}} \sigma_A \sigma_B)] = - \gamma \tr \1 $. 
\end{itemize}
Therefore, Eq.~\eqref{eq:gamma_vAB} is satisfied if and only if $\gamma = \pm i$. By a similar argument,  Eq.~\eqref{eq:delta_vAB} is satisfied if and only if $\delta = \pm i$. 

Note that the following equivalences hold: $\forall \sigma_{\vec{v}''} \in \{\sigma_{\vec{r}_1},\ldots,\sigma_{\vec{r}_M}\}: \sigma_{\vec{v}''} \not \propto \sigma_{\vec{v}} \sigma_B \sigma_A \iff \forall \sigma_{\vec{v}'} \in \{\sigma_{\vec{r}_1},\ldots,\sigma_{\vec{r}_M}\}: \sigma_{\vec{v}'} \not \propto \sigma_{\vec{v}} \sigma_A\sigma_B$, and also $\gamma \in \{i, -i\} \iff \delta \in \{i, -i\} $. Therefore,  for all $ \{\sigma_{\vec{r}_1},\ldots,\sigma_{\vec{r}_M}\}$, for every tuple $(  \sigma_{\vec{v}} \in\{\sigma_{\vec{r}_1},\ldots,\sigma_{\vec{r}_M}\},  \sigma_A, \sigma_B)$, if one of the two following conditions is satisfied:
\begin{enumerate}
    \item  $ \forall \, \sigma_{\vec{v}'} \in \{\sigma_{\vec{r}_1},\ldots,\sigma_{\vec{r}_M}\}: \sigma_{\vec{v}'} \not \propto \sigma_{\vec{v}} \sigma_A\sigma_B$, or
    \item  $\exists \, \sigma_{\vec{v}'}$ such that $ \sigma_{\vec{v}'} = \pm i \sigma_{\vec{v}} \sigma_A \sigma_B$, 
\end{enumerate}
then 
\begin{align}\label{eq:AUB_equal}
	b(\sigma_{\vec{v}}, \sigma_A) = b(\sigma_{\vec{v}}, U) \, ,\\
  b(\sigma_{\vec{v}}, \sigma_B) = b(\sigma_{\vec{v}}, U) \, ,
\end{align}
which implies that 
\begin{align}\label{eq:AB_equal}
	b(\sigma_{\vec{v}}, \sigma_A) = b(\sigma_{\vec{v}}, \sigma_B)  \, .
\end{align}
We now consider the choices of $(\sigma_{\vec{v}}, \sigma_A,  \sigma_B)$ where neither of the above two conditions is satisfied.

\textit{The case where  $\sigma_{\vec{v}} = \sigma_A$ or $\sigma_{\vec{v}} = \sigma_B$:} First, note that if $\sigma_{\vec{v}} = \sigma_B$, and Condition 1. is not satisfied, then Eq.~\eqref{eq:gamma}  implies that $\sigma_{\vec{v}'} = \sigma_A $ and $\gamma = \pm 1$, so Condition 2. is not satisfied either. In this case, we can take another $n$-qubit Pauli operator $\sigma_C \notin \{\sigma_{\vec{r}_1},\ldots,\sigma_{\vec{r}_M}\}$ (which implies that $\sigma_C \neq \sigma_A, \sigma_B$), such that $\sigma_B  \sigma_A \sigma_C = \pm i \sigma_{BAC}$ for some $n$-qubit Pauli $\sigma_{BAC}$. The existence of such a $ \sigma_C$ is guaranteed by:
\begin{itemize}
    \item the fact that half of the total $4^n$ number of $n$-qubit Pauli operators, when multiplied after $\sigma_B \sigma_A$ (which is not equal to the identity because $\sigma_B \neq \sigma_A$ by construction), gives a Pauli operator times $\pm 1$, and the other half will give a Pauli operator times $\pm i$, 
    
    \item the fact that the set $\{\sigma_{\vec{r}_1},\ldots,\sigma_{\vec{r}_M}\}$ contains the operators $\sigma_A, \sigma_B$, which, when multiplied after $\sigma_B \sigma_A$, gives a Pauli operator times $\pm 1$,
    
    \item  the assumption that $M < 4^n/2 + 2$.
\end{itemize}
Applying the procedure in Eqs.~\eqref{eq::sq2-AB}--\eqref{eq:AB_equal} above to the unitary $U' \coloneqq (\sigma_C + \beta' \sigma_B)/\sqrt{2}$ (with $\beta' \in 
\{1,i\}$), we find that there is no $\sigma_{\vec{v}'} \in \{\sigma_{\vec{r}_1},\ldots,\sigma_{\vec{r}_M}\}$ such that $\sigma_{\vec{v}'} \propto \sigma_{\vec{v}} \sigma_C \sigma_B = \sigma_{B} \sigma_C \sigma_B \propto  \sigma_C  $. 
Therefore, Condition 1. is satisfied for $(  \sigma_{\vec{v}} \in\{\sigma_{\vec{r}_1},\ldots,\sigma_{\vec{r}_M}\},  \sigma_C, \sigma_B)$ and we have that
\begin{equation}
	b(\sigma_{\vec{v}}, \sigma_C) = b(\sigma_{\vec{v}}, U') = b(\sigma_{\vec{v}}, \sigma_B) \, .	
\end{equation}
Applying the procedure in Eqs.~\eqref{eq::sq2-AB}--\eqref{eq:AB_equal} above to the unitary $U'' \coloneqq (\sigma_A + \beta'' \sigma_C)/\sqrt{2}$ (with $\beta'' \in 
\{1,i\}$), we find that either (a) there is no $\sigma_{\vec{v}''} \in \{\sigma_{\vec{r}_1},\ldots,\sigma_{\vec{r}_M}\}$ such that $\sigma_{\vec{v}''} \propto \sigma_{\vec{v}} \sigma_A \sigma_C$, or (b) if there is, then $ \sigma_{\vec{v}} \sigma_A \sigma_C = \sigma_{B} \sigma_A \sigma_C = \pm i \sigma_{BAC}$, i.e. $\sigma_{\vec{v}''} = \sigma_{BAC}$. Therefore, for $(  \sigma_{\vec{v}} \in\{\sigma_{\vec{r}_1},\ldots,\sigma_{\vec{r}_M}\},  \sigma_A, \sigma_C)$,  either Condition 1. or 2. is satisfied and we have that
\begin{equation}
	b(\sigma_{\vec{v}}, \sigma_A) = b(\sigma_{\vec{v}}, U'') = b(\sigma_{\vec{v}}, \sigma_C) \, .	
\end{equation}
Overall, Eq.~\eqref{eq:AB_equal} is satisfied for $(  \sigma_{\vec{v}} \in\{\sigma_{\vec{r}_1},\ldots,\sigma_{\vec{r}_M}\},  \sigma_A, \sigma_B)$. An analogous argument applies if $\sigma_{\vec{v}} = \sigma_A$.

\textit{The case where  $\sigma_{\vec{v}} \neq \sigma_A, \sigma_B$:}
If neither Condition 1.\ nor Condition 2.\ are satisfied, then we can take another $n$-qubit Pauli operator 
$\sigma_C \neq \sigma_A, \sigma_B$,
such that 
\begin{align}
\sigma_{\vec{v}}  \sigma_A \sigma_C &= \pm i \sigma_{vAC} \, , \\
\sigma_{\vec{v}}  \sigma_C \sigma_B &\in \{ \pm i \sigma_{vCB} \} \iff \sigma_{\vec{v}}  \sigma_B \sigma_C \in \{ \pm i \sigma_{vCB} \} \,	,
\end{align}
for some $n$-qubit Pauli operators $\sigma_{vAC},\sigma_{vCB}$. 
The existence of such a $ \sigma_C$ is guaranteed by:
\begin{itemize}	
	\item the fact that for any two different non-identity $n$-qubit Pauli operators $\sigma_{E},\sigma_{F}$, there exists an $n$-qubit Pauli operator $\sigma_{ Q}$ such that both $\sigma_{E}\sigma_{  Q}$ and $\sigma_{F}\sigma_{\rm  Q}$ are equal to $+i$ or $-i$ times an $n$-qubit Pauli operator, 
	\item the fact that for any two different non-identity  $n$-qubit Pauli operators $\sigma_{E},\sigma_{F}$, there exists an $n$-qubit Pauli operator $\sigma_{ R}$ such that $\sigma_{E}\sigma_{  R}$ equals $\pm i$  times an $n$-qubit Pauli operator, while $\sigma_{F}\sigma_{  R}$ equals $\pm 1$ times an $n$-qubit Pauli operator.
\end{itemize}
This enables $\sigma_C$ to be chosen according to the following strategy:
\begin{itemize}
	\item if there are $\sigma_{E}, \sigma_{F}$ such that  $\sigma_{E} = \pm  \sigma_{v} \sigma_{A}$ and $\sigma_{F}= \pm \sigma_{v} \sigma_{B}$, then pick $\sigma_C = \sigma_{  Q}$ as defined above,
	\item if there are $\sigma_{E}, \sigma_{F}$ such that either $\sigma_{E} = \pm  \sigma_{v} \sigma_{A}$ and $\sigma_{F}= \pm i  \sigma_{v} \sigma_{B}$, or $\sigma_{F} = \pm i \sigma_{v} \sigma_{A}$ and $\sigma_{E}= \pm   \sigma_{v} \sigma_{B}$, then pick $\sigma_C = \sigma_{  R}$ as defined above,
	\item if there are $\sigma_{E}, \sigma_{F}$ such that $\sigma_{E} = \pm i \sigma_{v} \sigma_{A}$ and $\sigma_{F}= \pm i \sigma_{v} \sigma_{B}$, then pick $\sigma_C =  \sigma_{F}$, in which case $\sigma_{\vec{v}}  \sigma_B \sigma_C =  \sigma_{\vec{v}} \sigma_B (\pm i \sigma_{v} \sigma_{B}) = \mp i \1$ and $\sigma_{\vec{v}}  \sigma_A \sigma_C = \pm i \sigma_{\vec{v}}  \sigma_A \sigma_{\vec{v}} \sigma_B = \mp i \sigma_{\vec{v}} \sigma_{\vec{v}} \sigma_A  \sigma_B =  \mp i \sigma_{\vec{v}} \sigma_B \sigma_A  \sigma_{\vec{v}} = - \sigma_C  \sigma_A \sigma_{\vec{v}} = - (\sigma_{\vec{v}}  \sigma_A \sigma_C)^\dag$ (where in the third equality we use the assumption that Conditions 1.\ and 2.\ are not satisfied, which implies that $\sigma_{\vec{v}} \sigma_A  \sigma_B = \sigma_B \sigma_A  \sigma_{\vec{v}}$), and therefore $\sigma_{\vec{v}}  \sigma_A \sigma_C$ must be proportional to $\pm i$ times a Pauli.
\end{itemize}
Applying the procedure in Eqs.~\eqref{eq::sq2-AB}--\eqref{eq:AB_equal} above to the unitary $U' \coloneqq (\sigma_C + \beta' \sigma_B)/\sqrt{2}$ (with $\beta' \in 
\{1,i\}$), we find that Condition 1.\ or 2.\ is satisfied for $(  \sigma_{\vec{v}} \in\{\sigma_{\vec{r}_1},\ldots,\sigma_{\vec{r}_M}\},  \sigma_C, \sigma_B)$ and we have that
\begin{equation}
	b(\sigma_{\vec{v}}, \sigma_C) = b(\sigma_{\vec{v}}, U') = b(\sigma_{\vec{v}}, \sigma_B) \, .	
\end{equation}
Applying the procedure in Eqs.~\eqref{eq::sq2-AB}--\eqref{eq:AB_equal} above to the unitary $U'' \coloneqq (\sigma_A + \beta'' \sigma_C)/\sqrt{2}$ (with $\beta'' \in 
\{1,i\}$), we find that Condition 1.\ or 2.\ is satisfied for $(  \sigma_{\vec{v}} \in\{\sigma_{\vec{r}_1},\ldots,\sigma_{\vec{r}_M}\},  \sigma_A, \sigma_C)$ and we have that
\begin{equation}
	b(\sigma_{\vec{v}}, \sigma_A) = b(\sigma_{\vec{v}}, U'') = b(\sigma_{\vec{v}}, \sigma_C) \, .	
\end{equation}
Overall, Eq.~\eqref{eq:AB_equal} is satisfied for  $(  \sigma_{\vec{v}} \in\{\sigma_{\vec{r}_1},\ldots,\sigma_{\vec{r}_M}\},  \sigma_A, \sigma_B)$.

Having shown that for all $ \{\sigma_{\vec{r}_1},\ldots,\sigma_{\vec{r}_M}\}$, for every tuple $(  \sigma_{\vec{v}} \in\{\sigma_{\vec{r}_1},\ldots,\sigma_{\vec{r}_M}\},  \sigma_A, \sigma_B)$, Eq.~\eqref{eq:AB_equal} is satisfied, we conclude that $b(\sigma_{\vec{v}},\sigma_{\vec{q_1}}) \coloneqq
\sum_{m\in \mathbb{F}_{\vec{v}}}
\xi^{\{\vec{r}_l\}_l,\vec{q}_1}_{m1}$ is independent of the choice of $\sigma_{\vec{q_1}}$. This means that 
\begin{equation}\label{eq:V1_indep}
    \sum_{m\in \mathbb{F}_{\vec{v}}}
	\xi^{\{\vec{r}_l\}_l,\vec{q}_1}_{m1} = \sum_{m\in \mathbb{F}_{\vec{v}}}
	\xi^{\{\vec{r}_l\}_l,\vec{q}_1}_{m1} [V_1= \sigma_{\vec{q}^*}] \, ,
\end{equation}
for any $n$-qubit Pauli operator $\sigma_{\vec{q}^*}$.

%%%%%%%%%%%%%
\subsection*{Proving that the redefinition of Eqs.~\eqref{eq:global_def_M1} and \eqref{eq:global_def_M1_linear} satisfies Eq.~\eqref{eq:indep_M1}}
%%%%%%%%%%%%%

Equations~\eqref{eq:sum_equal_linear} and Eq.~\eqref{eq:V1_indep} together show that for all $\vec{r}_1,\ldots,\vec{r}_M, \vec{q_1} \in \{0,1,2,3\}^{\times n}$,
\begin{align}
    \dket{C}*\bigotimes_{i=1}^M \dket{\sigma_{\vec{r}_i}}^{I_iO_i}*\dket{\sigma_{\vec{q}_1}}^{I_1'O_1'} 
    &=  \sum_{k=1}^M \xi^{\{\vec{r}_l\}_l,\vec{q}_1}_{k1} \dket{ S}*\dket{\sigma_{\vec{r}_k}}*\dket{\sigma_{\vec{q}_1}} 
\nonumber\\
   &=\sum_{\vec{v}\in \{\vec{r}_1,\ldots,\vec{r}_M\}}
    \sum_{m\in \mathbb{F}_{\vec{v}}}
    \xi^{\{\vec{r}_l\}_l,\vec{q}_1}_{m1}
    \dket{ S}*\dket{\sigma_{\vec{v}}}*\dket{\sigma_{\vec{q}_1}}
\nonumber\\
   &=
   \sum_{\vec{v}\in \{\vec{r}_1,\ldots,\vec{r}_M\}}
    \sum_{m\in \mathbb{F}_{\vec{v}}}
    \xi^{\{\vec{r}_l\}_l,\vec{q}_1}_{m1} [V_1= \sigma_{\vec{q}^*}]
    \dket{ S}*\dket{\sigma_{\vec{v}}}*\dket{\sigma_{\vec{q}_1}}
\nonumber\\
    &=
    \sum_{\vec{v}\in \{\vec{r}_1,\ldots,\vec{r}_M\}}
    \sum_{m\in \mathbb{F}_{\vec{v}}}\xi^{\{\vec{r}_l\}_l,\vec{q}_1}_{m1}[U_m=\sigma_{\vec{v}^*}, V_1= \sigma_{\vec{q}^*}]
    \dket{ S}*\dket{\sigma_{\vec{v}}}*\dket{\sigma_{\vec{q}_1}}
\nonumber\\
    &=\sum_{k=1}^M\tilde{\xi}_{k1}(\{U_i=\sigma_{\vec{r}_i}\}_{i=1}^M,V_1=\sigma_{\vec{q}_1})\dket{ S}*\dket{\sigma_{\vec{r}
_k}}*\dket{\sigma_{\vec{q}_1}}
   \, ,
\end{align}
where $\mathbb{F}_{\vec{v}}\coloneqq\{1\leq i\leq M\mid \vec{r}_i=\vec{v}\}$, $\vec{v}^* \in \{0,1,2,3\}^{\times n}$ is an arbitrary vector outside of the set $\{\vec{r}_1,\ldots,\vec{r}_M\} \backslash \{\vec{v}\}$, and $\vec{q}^* \in \{0,1,2,3\}^{\times n}$ is an arbitrary fixed vector.
Therefore, $\tilde{\xi}_{k1}$ as defined in Eq.~\eqref{eq:global_def_M1} indeed satisfies Eq.~\eqref{eq:indep_M1}.

This also implies that 
\begin{align}
    &\dket{C} *  \bigotimes_{i=1}^M
    \dket{\sum_{\vec{r}_i}\alpha^i_{\vec{r}_i}\sigma_{\vec{r}_i}}^{I_{i} O_{i}} 
    * \dket{\sum_{\vec{q}_1}\beta^1_{\vec{q}_1}\sigma_{\vec{q}_1}
    }^{I'_1 O'_1} 
    =
    \sum_{\{\vec{r}_l\}_l, \vec{q}_1}\left(\prod_{j=1}^M \alpha^j_{\vec{r}_j}\right)\beta_{\vec{q}_1}^1 \dket{C}* \bigotimes_{i=1}^M
    \dket{\sigma_{\vec{r}_{i}}}^{I_{i} O_{i}} 
    * \dket{\sigma_{\vec{q}_1}}^{I'_1 O'_1} 
    \nonumber\\
    &=
    \sum_{\{\vec{r}_l\}_l, \vec{q}_1}
    \left(\prod_{j=1}^M \alpha^j_{\vec{r}_j}\right)\beta_{\vec{q}_1}^1
    \left[
    \sum_{k=1}^M \tilde{\xi}_{k1}(\{U_m=\sigma_{\vec{r}_m}\}_m,V_1=\sigma_{\vec{q}_1}) \dket{S} * \dket{\sigma_{\vec{r}_k}}^{I_k O_k} * \dket{\sigma_{\vec{q}_1}}^{I'_1 O'_1}
    \right]
    \nonumber\\
    &=
    \sum_{k=1}^M
    \left[
    \sum_{\{\vec{r}_l\}_{l\neq k}}\left(
    \prod_{j=1| j \neq k}^{M}\alpha_{\vec{r}_j}^j
    \right) \tilde{\xi}_{k1}(\{U_m=\sigma_{\vec{r}_m}\}_m,V_1=\sigma_{\vec{q}_1}) \right]
    \sum_{\vec{r}_k, \vec{q}_1}
    \alpha^{k}_{\vec{r}_k}
    \beta^1_{\vec{q}_1}
    \dket{ S}*
    \dket{
    \sigma_{\vec{r}_k}
    }^{I_kO_k}*
    \dket{\sigma_{\vec{q}_1}}^{I_1'O_1'}
    \nonumber\\
    &=\sum_{k=1}^M \tilde{\xi}_{k1}\left(\left\{U_i=\sum_{\vec{r}_i}{\alpha_{\vec{r}_i}^i \sigma_{\vec{r}_i}}\right\},\ V_1=\sum_{\vec{q}_1}\beta_{\vec{q}_1}^1\sigma_{\vec{q}_1}\right) \dket{S} * \dket{\sum_{\vec{r}_k}\alpha_{\vec{r}_k}^k\sigma_{\vec{r}_k}}^{I_k O_k} * \dket{\sum_{\vec{q}_1}\beta^1_{\vec{q}_1}\sigma_{\vec{q}_1}}^{I'_1 O'_1} \, ,
\end{align}
i.e.,\ Eq.~\eqref{eq:global_def_M1_linear} also satisfies Eq.~\eqref{eq:indep_M1}.
\end{proof}

%%%%%%%%%%%%%%%%%%%%%%%%%%%%
\subsection{Lemma~\ref{lem:qccc_invalid} (for $\max\{M, N\} \leq  \max\{2, d-1\}$)}\label{subapp::lem:qccc_invalid}
%%%%%%%%%%%%%%%%%%%%%%%%%%%%
Before proving Lemma~\ref{lem:qccc_invalid}, we present the definition of a QC-CC higher-order transformation with an arbitrary number of slots $M+N$~\cite{wechs2021quantum}. 

An operator $C\in\mathcal{L}(P \otimes I_1 \otimes O_1 \otimes \ldots \otimes I_{M+N} \otimes O_{M+N} \otimes F)$ corresponds to the Choi operator of a $(M+N)$-slot QC-CC higher-order transformation if it satisfies
\begin{align}
    &C= \sum_{\vec{r}_{M+N} \in \mathrm{Perm}(1, \ldots, M+N)} C_{P \vec{r}_{M+N} F},\\
    &\textup{such that}\quad\quad
    C_{P \vec{r}_{M+N} F} \geq 0 \quad \forall \ \vec{r}_{M+N},\\
    &\tr_{F}[C_{P \vec{r}_{M+N} F}] = C_{P \vec{r}_{M+N}} \otimes \1^{O_{r_{M+N}}} \quad \forall \ \vec{r}_{M+N},\label{eq:qc-cc_F}\\
    &\sum_{r_{m+1}}\tr_{I_{r_{m+1}}}[C_{P \vec{r}_m r_{m+1}}] = C_{P \vec{r}_m} \otimes \1^{O_{r_{m}}} \quad \forall \ m\in\{1, \ldots, M+N-1\}, \forall \ \vec{r}_m \coloneqq (r_1, \ldots, r_m),\label{eq:qc-cc_m}\\
    &\sum_{r_1} \mathrm{tr}_{I_{r_1}}[C_{P r_1}] = \1^{P},\label{eq:qc-cc_norm}
\end{align}
where $\vec{r}_m r_{m+1}$ represents a vector $(r_1, \ldots, r_m, r_{m+1})$, with each vector $\vec{r}_m$ composed of elements $r_1, \ldots , r_m$, and
the operators $C_{P\vec{r}_{m}} \in \mathcal{L}(P \otimes I_1 \otimes O_1 \otimes \cdots \otimes I_{m-1} \otimes O_{m-1} \otimes I_{m})$ for $m\in\{1, \ldots, M+N\}$ are recursively defined by
\begin{align}
    C_{P\vec{r}_{M+N}} &\coloneqq {1\over d} \tr_{O_{r_{M+N}} F}[C_{P\vec{r}_{M+N} F}],\\
    C_{P\vec{r}_m}&\coloneqq {1\over d} \sum_{r_{m+1}}\tr_{O_{r_m} I_{r_{m+1}}}[C_{P\vec{r}_m r_{m+1}}] \quad \forall \ m\in\{1, \ldots, M+N-1\}.
\end{align}
These conditions will be referred as the \textit{QC-CC conditions}. We remark that quantum combs are a subset of QC-CCs.

For convenience in the following, we rename the last $N$ input and output systems as
\begin{align}
    I'_k \coloneqq I_{M + k}, \quad O'_k \coloneqq O_{M+k} \quad \forall \ k\in\{1, \ldots, N\}. 
\end{align}
\\

\begin{lemma}\label{lem:qccc_invalid}
    Let $\dket{S}\in P \otimes I \otimes O \otimes I' \otimes O' \otimes F$, where $I, O, I', O'$ correspond to $d$-dimensional Hilbert spaces and $P, F$ correspond to $(2\times d)$-dimensional Hilbert spaces, be the Choi vector of the quantum switch and let $C\in\mathcal{L}(P \otimes I_1 \otimes O_1 \otimes \ldots \otimes I_M \otimes O_M \otimes I'_1 \otimes O'_1 \otimes \ldots \otimes I'_N \otimes O'_N \otimes F)$, where $M,N\in\mathbb{N}^+$ and $\{I_i\}_i, \{O_i\}_i, \{I'_j\}_j, \{O'_j\}_j$ correspond to $d$-dimensional Hilbert spaces, be the Choi operator of an $(M+N)$-slot QC-CC higher-order transformation, which we write as 
    \begin{align}
        C = \sum_{\vec{r}_{M+N}} C_{P \vec{r}_{M+N} F} \, ,
    \end{align}
    with $\vec{r}_{M+N} \in \mathrm{Perm}(1, \ldots, M+N)$, where all $C_{P \vec{r}_{M+N} F}\geq 0$ and, hence, can be decomposed as
    \begin{align}
        C_{P \vec{r}_{M+N} F} = \sum_a \dketbra{C^{(a)}_{P \vec{r}_{M+N} F}}\, .
    \end{align}
    
    If $\max(M, N) \leq  \max(2, d-1)$, then the set of operators $\dketbra{C^{(a)}_{P \vec{r}_{M+N} F}}$ cannot be such that
    \begin{align}
        \dket{C_{P\vec{r}_{M+N} F}^{(a)}} &= \sum_{i=1}^{M}\sum_{k=1}^{N} \dket{S}^{P I_i O_i I'_k O'_k F} \otimes \dket{\tilde{\xi}_{ik}^{(a),\vec{r}_{M+N}}},
    \end{align}
    for all $\vec{r}_{M+N}$ and $a$, where $\dket{\tilde{\xi}_{ik}^{(a),\vec{r}_{M+N}}} \in I_{\bar{i}} \otimes O_{\bar{i}} \otimes I'_{\bar{k}} \otimes O'_{\bar{k}}$ for all $i \in \{1, \ldots, M\}$ and $k \in \{1, \ldots, N\}$, and $I_{\bar{i}} \coloneqq \bigotimes_{i'\neq i} I_{i'}$, $O_{\bar{i}} \coloneqq \bigotimes_{i'\neq i} O_{i'}$, $I'_{\bar{k}} \coloneqq \bigotimes_{k'\neq k} I'_{k'}$, and $O'_{\bar{k}} \coloneqq \bigotimes_{k'\neq k} O'_{k'}$.
\end{lemma}

\begin{proof}
The proof follows by contradiction. To this end, we use Eqs.~\eqref{eq:qc-cc_F} and \eqref{eq:qc-cc_m} in the QC-CC conditions to show the following equation for $C_{P \vec{r}_m}$:
\begin{align}
\begin{split}\label{eq:C_Prm}
    \sum_{r_m}\tr_{I_{r_{m}}} [C_{P \vec{r}_m}] 
    &=\sum_{i,j\in \mathbb{A}_{\vec{r}_{m-1}}} \sum_{k,l\in \mathbb{B}_{\vec{r}_{m-1}}} \Big(\ketbra{0}^{P_C} \otimes \dket{\1}^{P_T I_i}\dbra{\1}^{P_T I_j} \otimes \dket{\1}^{O_i I'_k}\dbra{\1}^{O_j I'_l} \otimes \1^{O'_l \to O'_k} 
    \\
    &+ \ketbra{1}^{P_C} \otimes \dket{\1}^{P_T I'_k}\dbra{\1}^{P_T I'_l} \otimes \dket{\1}^{O'_k I_i}\dbra{\1}^{O'_l I_j} \otimes \1^{O_j \to O_i}\Big) \otimes C_{P\vec{r}_{m-1}}^{(ijkl)} \quad \forall m\in\{1, \ldots, M+N+1\},
\end{split}
\end{align}
where the summation over $r_m$ for $m=M+N+1$ is taken as $I_{r_{M+N+1}} \coloneqq F$, $C_{P\vec{r}_{M+N+1}}$ is defined by $C_{P\vec{r}_{M+N+1}}\coloneqq C_{P\vec{r}_{M+N}F}$, the set of indices $\mathbb{A}_{\vec{r}_{m-1}}$ and $\mathbb{B}_{\vec{r}_{m-1}}$ are defined by
\begin{align}
    \mathbb{A}_{\vec{r}_{m-1}}&\coloneqq \{r_1, \ldots, r_{m-1}\} \cap \{1, \ldots, M\},\\
    \mathbb{B}_{\vec{r}_{m-1}}&\coloneqq \{r_1-M, \ldots, r_{m-1}-M\} \cap \{1, \ldots, N\},
\end{align}
and $C_{P\vec{r}_{m-1}}^{(ijkl)}$ is an operator.
If this equation holds, since $\mathbb{A}_{\vec{r}_0}$ and $\mathbb{B}_{\vec{r}_0}$ are the empty sets, we obtain 
\begin{align} \label{eq:norm_contradict}
    \sum_{r_1} \tr_{I_{r_1}}[C_{P r_1}] = 0,
\end{align}
which contradicts with the normalization condition in Eq.~\eqref{eq:qc-cc_norm} of the QC-CC conditions.
In the rest of the proof, we show Eq.~\eqref{eq:C_Prm} by induction with respect to $m$.
\\
    
First, we show Eq.~\eqref{eq:C_Prm} for $m=M+N+1$ as follows.
Since the operator $C_{P\vec{r}_{M+N} F}$ can be written as
\begin{align}
    C_{P\vec{r}_{M+N} F} &= \sum_{i,j,k,l} \dket{S}^{P I_i O_i I'_k O'_k F}\dbra{S}^{P I_j O_j I'_l O'_l F} \otimes C_{P\vec{r}_{M+N}}^{(ijkl)},
\end{align}
where $C_{P\vec{r}_{M+N}}^{(ijkl)} \coloneqq \sum_{a} \dketbra{\tilde{\xi}_{ik}^{(a),\vec{r}_{M+N}}}{\tilde{\xi}_{jl}^{(a),\vec{r}_{M+N}}}$.
The partial trace $\tr_{F} C_{P\vec{r}_{M+N} F}$ is given by
\begin{align}
\begin{split}\label{eq:C_r}
    \tr_{F} [C_{P\vec{r}_{M+N} F}] = \sum_{i,j=1}^{M} \sum_{k,l=1}^{N} \Big(&\ketbra{0}^{P_C} \otimes \dket{\1}^{P_T I_i}\dbra{\1}^{P_T I_j} \otimes \dket{\1}^{O_i I'_k}\dbra{\1}^{O_j I'_l} \otimes \1^{O'_l \to O'_k} 
    \\
    + &\ketbra{1}^{P_C} \otimes \dket{\1}^{P_T I'_k}\dbra{\1}^{P_T I'_l} \otimes \dket{\1}^{O'_k I_i}\dbra{\1}^{O'_l I_j} \otimes \1^{O_j \to O_i}\Big) \otimes C_{P\vec{r}_{M+N}}^{(ijkl)}\, ,
\end{split}
\end{align}
i.e., Eq.~\eqref{eq:C_Prm} holds for $m=M+N+1$.

To complete the proof, we show Eq.~\eqref{eq:C_Prm} by assuming Eq.~\eqref{eq:C_Prm} for $m \gets m+1$, i.e.,
\begin{align}
\begin{split}\label{eq:induction_m+1}
    \sum_{r_{m+1}}\tr_{I_{r_{m+1}}} [C_{P \vec{r}_{m} r_{m+1}}]
    =\sum_{i,j\in \mathbb{A}_{\vec{r}_{m}}} \sum_{k,l\in \mathbb{B}_{\vec{r}_{m}}} \Big(&\ketbra{0}^{P_C} \otimes \dket{\1}^{P_T I_i}\dbra{\1}^{P_T I_j} \otimes \dket{\1}^{O_i I'_k}\dbra{\1}^{O_j I'_l} \otimes \1^{O'_l \to O'_k}
    \\
    + &\ketbra{1}^{P_C} \otimes \dket{\1}^{P_T I'_k}\dbra{\1}^{P_T I'_l} \otimes \dket{\1}^{O'_k I_i}\dbra{\1}^{O'_l I_j} \otimes \1^{O_j \to O_i}\Big) \otimes C_{P\vec{r}_{m}}^{(ijkl)}.
\end{split}
\end{align}
    
By symmetry with $(I_i, O_i)$ and $(I'_k, O'_k)$, it is sufficient to show if $r_{m} \in \{1, \ldots, M\}$ holds. 
From Eq.~\eqref{eq:qc-cc_m} [or Eq.~\eqref{eq:qc-cc_F} for $m=M+N$] in the QC-CC conditions and Eq.~\eqref{eq:induction_m+1}, we obtain
\begin{align}
\begin{split}\label{eq:apply_qccc_condition_1}
    \sum_{i,j\in \mathbb{A}_{\vec{r}_{m}}} \sum_{k,l\in \mathbb{B}_{\vec{r}_{m}}} \Big(&\ketbra{0}^{P_C} \otimes \dket{\1}^{P_T I_i}\dbra{\1}^{P_T I_j} \otimes \dket{\1}^{O_i I'_k}\dbra{\1}^{O_j I'_l} \otimes \1^{O'_l \to O'_k} 
    \\
    + &\ketbra{1}^{P_C} \otimes \dket{\1}^{P_T I'_k}\dbra{\1}^{P_T I'_l} \otimes \dket{\1}^{O'_k I_i}\dbra{\1}^{O'_l I_j} \otimes \1^{O_j \to O_i}\Big) \otimes C_{P\vec{r}_{m}}^{(ijkl)}
    \\
    =\sum_{i,j\in \mathbb{A}_{\vec{r}_{m}}} \sum_{k,l\in \mathbb{B}_{\vec{r}_{m}}} &\ketbra{0}^{P_C} \otimes \dket{\1}^{P_T I_i}\dbra{\1}^{P_T I_j} \otimes \1^{O'_l \to O'_k} \otimes A_{ijkl}
    \\
    +&\ketbra{1}^{P_C} \otimes \dket{\1}^{P_T I'_k}\dbra{\1}^{P_T I'_l} \otimes \dket{\1}^{O'_k I_i}\dbra{\1}^{O'_l I_j} \otimes B_{ijkl},
\end{split}
\end{align}
where $A_{ijkl}$ and $B_{ijkl}$ are defined by
\begin{align}
    A_{ijkl} &\coloneqq
    \begin{cases}
        \dket{\1}^{O_i I'_k}\dbra{\1}^{O_j I'_l} \otimes \tilde{C}_{P\vec{r}_{m}}^{(ijkl)} \otimes \1^{O_{r_{m}}} & (i,j\neq r_{m})
        \\
        {1\over d} C_{P\vec{r}_{m}}^{(ijkl)} \dket{\1}^{I'_k O_{r_{m}}} \dbra{\1}^{O_j I'_l} \otimes \1^{O_{r_{m}}} & (i=r_{m} \neq j)
        \\
        {1\over d}\dket{\1}^{I'_k O_i} \dbra{\1}^{O_{r_{m}} I'_l} C_{P\vec{r}_{m}}^{(ijkl)} \otimes \1^{O_{r_{m}}} & (j = r_{m} \neq i)
        \\
        {1\over d}\1^{I'_l \to I'_k} \otimes C_{P\vec{r}_{m}}^{(ijkl)} \otimes \1^{O_{r_m}}  & (i=j=r_{m})
    \end{cases},
    \label{eq:Aijkl}\\
    B_{ijkl} &\coloneqq
    \begin{cases}
        \1^{O_j \to O_i} \otimes \tilde{C}_{P\vec{r}_{m}}^{(ijkl)} \otimes \1^{O_{r_{m}}} & (i,j\neq r_{m})
        \\
        {1\over d}C_{P\vec{r}_{m}}^{(ijkl)} \1^{O_j \to O_{r_{m}}} \otimes \1^{O_{r_{m}}} & (i=r_{m} \neq j)\\
        {1\over d}\1^{O_{r_{m}} \to O_i} C_{P\vec{r}_{m}}^{(ijkl)} \otimes \1^{O_{r_{m}}} & (j = r_{m} \neq i)
        \\
        C_{P\vec{r}_{m}}^{(ijkl)} \otimes \1^{O_{r_m}} & (i=j=r_{m})
    \end{cases},
    \label{eq:Bijkl}\\
    \tilde{C}_{P\vec{r}_{m}}^{(ijkl)}&\coloneqq {1\over d}\tr_{O_{r_{m}}}C_{P\vec{r}_{m}}^{(ijkl)}.
\end{align}
Using Lemma~\ref{lem:linear_independence} for Eq.~\eqref{eq:apply_qccc_condition_1}, we obtain
\begin{align}
    \sum_{k,l\in \mathbb{B}_{\vec{r}_{m}}} \dket{\1}^{O_i I'_k}\dbra{\1}^{O_j I'_l} 
    \otimes \1^{O'_l \to O'_k}
    \otimes C_{P\vec{r}_{m}}^{(ijkl)} &= \sum_{k,l\in \mathbb{B}_{\vec{r}_{m}}} 
    \1^{O'_l \to O'_k} \otimes A_{ijkl} 
    \quad \forall i,j,\label{eq:eq_for_A}
    \\
    \1^{O_j\to O_i} \otimes C_{P\vec{r}_{m}}^{(ijkl)} &= B_{ijkl} \quad \forall i,j,k,l.\label{eq:eq_for_B}
\end{align}
From Eq.~\eqref{eq:eq_for_B}, we obtain
\begin{align}
    C_{P\vec{r}_{m}}^{(ijkl)} &= \begin{cases}
        \tilde{C}_{P\vec{r}_{m}}^{(ijkl)} \otimes \1^{O_{r_{m}}} & (i,j\neq r_{m})
        \\
        0 & (i = r_{m}\neq j \;\mathrm{or}\; j = r_{m}\neq i)
    \end{cases},
\end{align}
where the cases of $i=r_m\neq j$ and $j=r_m\neq i$ are shown as below.
If $i=r_m\neq j$ holds, from Eqs.~\eqref{eq:Bijkl} and \eqref{eq:eq_for_B}, we obtain
\begin{align}\label{i=rm_neq_j}
    \1^{O_j\to O_{r_{m}}} \otimes C_{P\vec{r}_{m}}^{(ijkl)} = {1\over d}C_{P\vec{r}_{m}}^{(ijkl)} \1^{O_j \to O_{r_m}} \otimes \1^{O_{r_{m}}}.
\end{align}
By taking the inner product of Eq.~\eqref{i=rm_neq_j} with $\1^{O_j \to O_{r_{m}}}$, we obtain
\begin{align}\label{eq:dCPrm}
    d C_{P\vec{r}_{m}}^{(ijkl)} = {1\over d} C_{P\vec{r}_{m}}^{(ijkl)},
\end{align}
i.e., $C_{P\vec{r}_{m}}^{(ijkl)} = 0$ holds for $i=r_m\neq j$.
We can similarly show that $C_{P\vec{r}_{m}}^{(ijkl)} = 0$ for $j=r_m\neq i$.
From Eq.~\eqref{eq:eq_for_A} for $i=j=r_{m}$, we obtain
\begin{align}\label{eq:i=j=rm}
    \sum_{k,l\in \mathbb{B}_{\vec{r}_{m}}} \dket{\1}^{O_{r_{m}} I'_k}\dbra{\1}^{O_{r_{m}} I'_l}
    \otimes \1^{O'_l \to O'_k} \otimes C_{P\vec{r}_{m}}^{(ijkl)} = \sum_{k,l\in \mathbb{B}_{\vec{r}_{m}}}  \1^{O'_l \to O'_k} \otimes \1^{I'_l \to I'_k} \otimes {\1^{O_{r_{m}}} \over d} \otimes C_{P\vec{r}_{m}}^{(ijkl)} \quad \mathrm{if}\; i=j=r_{m}.
\end{align}
Using Lemma~\ref{lem:linear_independence2}, we obtain
\begin{align}
    C_{P\vec{r}_{m}}^{(ijkl)} = 0 \quad \mathrm{if}\; i=j=r_{m}.
\end{align}
In conclusion, we obtain
\begin{align}
    C_{P\vec{r}_{m}}^{(ijkl)} &= \begin{cases}
        \tilde{C}_{P\vec{r}_{m}}^{(ijkl)} \otimes \1^{O_{r_{m}}} & (i,j\neq r_{m})
        \\
        0 & (\mathrm{otherwise})
    \end{cases}.
\end{align}
Thus, from Eqs.~\eqref{eq:qc-cc_m} and~\eqref{eq:induction_m+1}, we obtain
\begin{align}
    C_{P \vec{r}_m}
    =\sum_{i,j\in \mathbb{A}_{\vec{r}_{m-1}}} \sum_{k,l\in \mathbb{B}_{\vec{r}_{m-1}}} \Big(&\ketbra{0}^{P_C} \otimes \dket{\1}^{P_T I_i}\dbra{\1}^{P_T I_j} \otimes \dket{\1}^{O_i I'_k}\dbra{\1}^{O_j I'_l} \otimes \1^{O'_l \to O'_k} 
    \\
    + &\ketbra{1}^{P_C} \otimes \dket{\1}^{P_T I'_k}\dbra{\1}^{P_T I'_l} \otimes \dket{\1}^{O'_k I_i}\dbra{\1}^{O'_l I_j} \otimes \1^{O_j \to O_i}\Big) \otimes \tilde{C}_{P\vec{r}_{m}}^{(ijkl)}.\label{eq:C_r_i}
\end{align}
Thus, defining $C_{P\vec{r}_{m-1}}^{(ijkl)}$ by
\begin{align}
    C_{P\vec{r}_{m-1}}^{(ijkl)} \coloneqq \sum_{r_{m}} \tr_{I_{r_m}} [\tilde{C}_{P\vec{r}_{m}}^{(ijkl)}],
\end{align}
we obtain Eq.~\eqref{eq:C_Prm}. We finish the proof by recalling that Eq.~\eqref{eq:C_Prm} implies Eq.~\eqref{eq:norm_contradict}, which contradicts the normalization condition in Eq.~\eqref{eq:qc-cc_norm} of the QC-CC conditions.
\end{proof}

%%%%%%%%%%%%%
\subsection{Lemma~\ref{lem:linear_independence}}\label{subapp::lem:linear_independence}
%%%%%%%%%%%%%

\begin{lemma}\label{lem:linear_independence}
    The set of matrices
    \begin{align}
    \label{eq:set_A}
        \left\{\dket{\1}^{P_T I_k'} \dbra{\1}^{P_T I_l'} \otimes \dket{\1}^{O'_k I_i} \dbra{\1}^{O'_l I_j} \otimes \ket{\vec{\alpha}}^{I_{\bar{k}}'}\bra{\vec{\beta}}^{I_{\bar{l}}'} \otimes \ket{\vec{\gamma}}^{I_{\bar{i}}}\bra{\vec{\delta}}^{I_{\bar{j}}}\right\}_{\substack{i,j\in\{1, \ldots, M\}, k,l\in\{1, \ldots, N\}, \\ \vec{\alpha}, \vec{\beta}\in \{1, \ldots, d\}^{N-1}, \vec{\gamma}, \vec{\delta}\in\{1, \ldots, d\}^{M-1}}}
    \end{align}
    is linearly independent if $\max (M, N)\leq d$ holds. Similarly, the set of matrices
    \begin{align}
    \label{eq:set_B}
        \left\{\dket{\1}^{P_T I_i}\dbra{\1}^{P_T I_j} \otimes \ket{\vec{\alpha}}^{I_{\bar{i}}}\bra{\vec{\beta}}^{I_{\bar{j}}}\right\}_{\substack{i,j\in\{1, \ldots, M\}, \\ \vec{\alpha}, \vec{\beta}\in\{1, \ldots, d\}^{M-1}}}
    \end{align}
    is linearly independent if $M\leq d$ holds.
\end{lemma}

\begin{proof}
We consider the equation
\begin{align}\label{eq:linear_independence}
    \sum_{i,j,k,l,\vec{\alpha},\vec{\beta},\vec{\gamma},\vec{\delta}}A_{ijkl\vec{\alpha}\vec{\beta}\vec{\gamma}\vec{\delta}}\dket{\1}^{P_T I_k'} \dbra{\1}^{P_T I_l'} \otimes \dket{\1}^{O'_k I_i} \dbra{\1}^{O'_l I_j} \otimes \ket{\vec{\alpha}}^{I_{\bar{k}}'}\bra{\vec{\beta}}^{I_{\bar{l}}'} \otimes \ket{\vec{\gamma}}^{I_{\bar{i}}}\bra{\vec{\delta}}^{I_{\bar{j}}} = 0
\end{align}
for complex coefficients $A_{ijkl\vec{\alpha}\vec{\beta}\vec{\gamma}\vec{\delta}}$.
Since $\max (M, N)\leq d$ holds, for all $\vec{\alpha}, \vec{\beta}, \vec{\gamma}, \vec{\delta}$, there exists $\alpha^*, \beta^*, \gamma^*, \delta^* \in \{1, \ldots, d\}$ such that $\alpha^*, \beta^*, \gamma^*, \delta^*$ do not appear in $\vec{\alpha}, \vec{\beta}, \vec{\gamma}, \vec{\delta}$, respectively.
By taking an inner product of Eq.~\eqref{eq:linear_independence} with $\ket{\alpha^* \alpha^*}^{P_T I'_k}\bra{\beta^* \beta^*}^{P_T I'_l} \otimes \ket{\gamma^* \gamma^*}^{O'_k I_i} \bra{\delta^* \delta^*}^{O'_l I_j} \otimes \ket{\vec{\alpha}}^{I_{\bar{k}}'}\bra{\vec{\beta}}^{I_{\bar{l}}'} \otimes \ket{\vec{\gamma}}^{I_{\bar{i}}}\bra{\vec{\delta}}^{I_{\bar{j}}}$ for any $i,j,k,l,\vec{\alpha},\vec{\beta},\vec{\gamma},\vec{\delta}$, we obtain
\begin{align}
    A_{ijkl\vec{\alpha}\vec{\beta}\vec{\gamma}\vec{\delta}} = 0,
\end{align}
i.e., the set~\eqref{eq:set_A} is linearly independent.
We can similarly show that the set~\eqref{eq:set_B} is linearly independent.
\end{proof}

%%%%%%%%%%%%%
\subsection{Lemma~\ref{lem:linear_independence2}}\label{subapp::lem:linear_independence2}
%%%%%%%%%%%%%

\begin{lemma}\label{lem:linear_independence2}
    The set of matrices
    \begin{align}\label{eq:set_C}
        \left\{ \left(\dket{\1}^{O_{r_m} I'_k} \dbra{\1}^{O_{r_m} I'_l} - {\1^{O_{r_m}} \over d} \otimes \1^{I'_l \to I'_k} \right) \otimes \1^{O'_l \to O'_k} \otimes \ket{\vec{\alpha}}^{I'_{\bar{k}}} \bra{\vec{\beta}}^{I'_{\bar{l}}} \otimes \ket{\vec{\gamma}}^{O'_{\bar{k}}} \bra{\vec{\delta}}^{O'_{\bar{l}}} \right\}_{\substack{k,l\in\{1, \ldots, N\}, \\ \vec{\alpha}, \vec{\beta}, \vec{\gamma}, \vec{\delta}\in\{1, \ldots, d\}^{N-1}}}
    \end{align}
    is linearly independent if $N\leq \max(2, d-1)$ holds.
\end{lemma}

\begin{proof}
We numerically check the linear independence for the case $N=d=2$ (see Listing~\ref{code}).
We prove the linear independence for the case $N\leq d-1$ to complete the proof.
    
We consider the equation
\begin{align}\label{eq:linear_indeoendence2}
    \sum_{k,l,\vec{\alpha},\vec{\beta}, \vec{\gamma},\vec{\delta}} A_{kl\vec{\alpha}\vec{\beta}\vec{\gamma}\vec{\delta}} \left(\dket{\1}^{O_{r_m} I'_k} \dbra{\1}^{O_{r_m} I'_l} - {\1^{O_{r_m}} \over d} \otimes \1^{I'_l \to I'_k} \right) \otimes \1^{O'_l \to O'_k} \otimes \ket{\vec{\alpha}}^{I'_{\bar{k}}} \bra{\vec{\beta}}^{I'_{\bar{l}}} \otimes \ket{\vec{\gamma}}^{O'_{\bar{k}}} \bra{\vec{\delta}}^{O'_{\bar{l}}} = 0
\end{align}
for complex coefficients $A_{kl\vec{\alpha}\vec{\beta}\vec{\gamma}\vec{\delta}}$.
Since $N\leq d-1$ holds, for all $\vec{\alpha}, \vec{\beta}$, there exists $\alpha^*, \beta^* \in \{1, \ldots, d\}$ such that $\alpha^*\neq \beta^*$ holds and $\alpha^*, \beta^*$ do not appear in $\vec{\alpha}, \vec{\beta}$, respectively.
By taking an inner product of Eq.~\eqref{eq:linear_indeoendence2} with ${1\over d} \ket{\alpha^*\alpha^*}^{O_{r_m} I'_k} \bra{\beta^* \beta^*}^{O_{r_m} I'_l} \otimes \ket{\vec{\alpha}}^{I'_{\bar{k}}} \bra{\vec{\beta}}^{I'_{\bar{l}}} \otimes \1^{O'_l \to O'_k} \otimes \ket{\vec{\gamma}}^{O'_{\bar{k}}} \bra{\vec{\delta}}^{O'_{\bar{l}}}$ for any $k,l,\vec{\alpha},\vec{\beta},\vec{\gamma},\vec{\delta}$, we obtain
\begin{align}
    A_{kl \vec{\alpha} \vec{\beta} \vec{\gamma} \vec{\delta}} = 0,
\end{align}
i.e., the set \eqref{eq:set_C} is linearly independent.
\end{proof}

\begin{lstlisting}[caption={MATLAB \cite{matlab} code to check the linear independency of the set \eqref{eq:set_C} for the case $d=2$ and $N=2$, which uses the functions from {\sc QETLAB} \cite{qetlab}.},label=code]
clear

d=2;
N=2;

one = Tensor(IsotropicState(d, 1)*d,eye(d));
id = Tensor(eye(d)/d,eye(d),eye(d));
I = eye(d^(d-1));

for i = 1:d
    sys(i)=i;
end
PP = perms(sys);

pos=0;

% Calculate the set of matrices
for alpha = 1:d^(d-1)
    for beta=1:d^(d-1)
        for gamma = 1:d^(d-1)
            for delta=1:d^(d-1)
                for k = 1:size(PP,1)
                    for l=1:size(PP,1)
                        pos=pos+1;
                        A(:,:,pos) = Tensor(eye(d), PermutationOperator(d, PP(k,:)), PermutationOperator(d, PP(k,:))) * PermuteSystems(Tensor(one-id, I(:,alpha)*I(beta,:), I(:,gamma)*I(delta,:)), [1 2 4 3 5]) * Tensor(eye(d), PermutationOperator(d, PP(l,:)), PermutationOperator(d, PP(l,:)));
                    end
                end
            end
        end
    end
end

% Flatten the matrices to vectors
for pos = 1:size(A,3)
    B(:,pos) = reshape(A(:,:,pos), [], 1);
end

rank(B) == size(B,2)
\end{lstlisting}

%%%%%%%%%%%%%%%%%%%%%%%%%%%%%%%%%%%%%%%%%%%%
\section{Possible restricted simulations for qubit channels}\label{app::gorestricted}
\setcounter{figure}{0}
%%%%%%%%%%%%%%%%%%%%%%%%%%%%%%%%%%%%%%%%%%%%

From this section onwards, we revert to our original notation.

Using numerical methods, we have found three particular cases where a restricted simulation of the quantum switch acting on qubit channels \textit{is possible}, exactly and deterministically, using a quantum comb. These are the cases of:
\begin{itemize}
    \item Four identical calls to general qubit channels, called the order AAAA.
    \item Two calls to unitary qubit channels $A$ and two calls to unitary qubit channels $B$, in the order AABB.
    \item Three calls to unitary qubit channels $A$ and one call to unitary qubit channels $B$, in the order BAAA.
\end{itemize}
All of these results were obtained by numerically evaluating the equivalent of the primal SDP of the main text [Eq.~\eqref{sdp::primal}] in the case where the input states are fixed and the output target system is discarded, and finding that $p=1$ up to a high numerical precision. 

Effectively, this scenario amounts to an SDP that is analogous to the primal SDP~\eqref{sdp::primal} in the main text, but with the first constraint written as 
\begin{equation}\label{eq::restrictedsim_app}
    C_s*\Big[{(J^A_i)}^{\otimes k_A}\otimes {(J^B_j)}^{\otimes k_B}\Big] = p\,\tr_{t_O}(S_{+0})*(J^A_i\otimes J^B_j) \ \ \ \forall\,i,j,
\end{equation}
where $S_{+0}$ is defined as in Eq.~\eqref{eq::switchchoi_+0}. In this case, we have that $C_s,C\in\map{L}(\map{H}^{A_I}\otimes\map{H}^{A_O}\otimes\map{H}^{B_I}\otimes\map{H}^{B_O}\otimes\map{H}^{c_O})$ and $d_{c_I}=d_{t_I}=1$.

In the case of $4$ identical copies of general qubit channels, the input channels were given as a basis constructed in the form of Eq.~\eqref{eq::basisk} of the main text. The dimension of the space spanned by $k=4$ identical copies of a general qubit channel is $d_\map{I}=1820$, a value that can be obtained from the expression presented in the Methods section of the main text.

In the case of $k$ identical copies of qubit unitary channels, a basis can be constructed numerically by randomly sampling a set of qubit unitaries according to the Haar measure, guaranteeing that they are linearly independent. It is only necessary to know the dimension of this subspace beforehand to determine how many unitaries must be sampled. The dimension of the linear space spanned by $k$ copies of $d$-dimension unitary channels is the quantity $D(d,r,s)$ of Ref.~\cite{roy2008unitary} for the case $r=s=k$. For $d=2$, the dimension of the subspace spanned by $k$ identical copies of a unitary channel is given by $d_\mathcal{U} = D(d=2,k,k) = \binom{2k+3}{3}$, which implies $d_\mathcal{U}=10$ for $k=1$, $d_\mathcal{U}=35$ for $k=2$, and $d_\mathcal{U}=84$ for $k=3$.

By numerically evaluating the appropriate SDP, we obtain that the maximum probability of success is $p=1$ with very high precision in all aforementioned cases. The numerical precision was evaluated in the following way: In all three cases, all inequality constrains are strictly satisfied; as for the equality constraints, they are satisfied up to an error of at most $10^{-9}$ in the operators norm for the case of identical general qubit channels AAAA, and at most $10^{-7}$ in the cases of qubit unitary channels AABB and BAAA.

\begin{figure}%[h!]
\begin{center}
	\includegraphics[width=0.9\columnwidth]{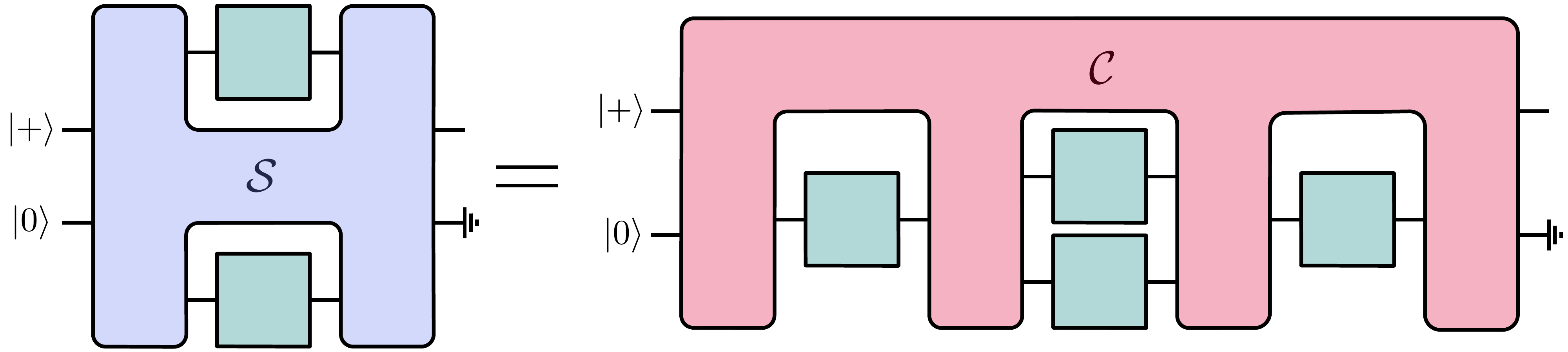}
	\caption{\textbf{Semi-parallelized strategy for restricted qubit simulations.} The quantum switch acting on qubit channels can be deterministically simulated by a $4$-slot quantum comb that has its second and third slots parallelized in a restricted scenario where the input control and target systems are fixed and the output target system is discarded in three different cases: for identical general quantum channels {(i.e. when $A=B$) called order AAAA, and for unitary channels (i.e. when $A=U_A$ and $B=U_B$) in the orders AABB and BAAA.}}
\label{fig::midpar}
\end{center}
\end{figure}

We also found that the maximum probability of success remains $p=1$, with the same precision, in all three cases when the second and third slots of the quantum comb $\map{C}$ are parallelized, as depicted in Fig.~\ref{fig::midpar}. 
\\

%%%%%%%%%%%%%%%%%%%%%%%%%%%%%%%%
%%%%%%%%%%%%%%%%%%%%%%%%%%%%%%%%

\textbf{Impossible partly restricted simulation.}
As mentioned in the main text, we also found that when the output target system is not discarded---i.e., the partly restricted simulation scenario---then a deterministic simulation is no longer possible in all 3 aforementioned cases. 

More concretely, this is the scenario where the first constraint in the primal SDP~\eqref{sdp::primal} in the main text is written as 
\begin{equation}\label{eq::openfuturesim_app}
    C_s*\Big[{(J^A_i)}^{\otimes k_A}\otimes {(J^B_j)}^{\otimes k_B}\Big] = p\, S_{+0}*(J^A_i\otimes J^B_j) \ \ \ \forall\,i,j,
\end{equation}
where $S_{+0}$ is defined as in Eq.~\eqref{eq::switchchoi_+0}, $C_s,C\in\map{L}(\map{H}^{A_I}\otimes\map{H}^{A_O}\otimes\map{H}^{B_I}\otimes\map{H}^{B_O}\otimes\map{H}^{t_O}\otimes\map{H}^{c_O})$, and $d_{c_I}=d_{t_I}=1$. 

Evaluating the above SDP, for the partly restricted scenario, we found that $p<1$ in the AAAA case of identical general qubit channels, the AABB case of qubit unitary channels, as well as the BAAA case of qubit unitary channels. Due to limitations in computational power, the partly restricted simulation SDP was evaluated in all three cases with an input that corresponds to a subset of the basis that was given as input to the restricted simulation SDP. Therefore, the solutions for the maximal probability of success correspond to upper bounds, and are given by

\begin{align}
    \text{order = AAAA (qubit, general),\phantom{.}} \ \ \  & p  \text{{ $\lesssim$ }} 0.942 \\
    \text{order = AABB (qubit, unitary)}, \ \ \  & p  \text{{ $\lesssim$ }} 0.822 \\
    \text{order = BAAA (qubit, unitary)}, \ \ \ & p \text{{ $\lesssim$ }} 0.667.
\end{align}

The difference between the two simulation scenarios discussed in this section is whether or not the output target state is discarded. Here we can see that this distinction in the requirements of the simulation is sufficient to transform cases where a deterministic simulation is possible (with a discarded output target system) into cases where a deterministic simulation is no longer possible (without discarding the output target system).

%%%%%%%%%%%%%%%%%%%%%%%%%%%%%%%%%%%%%%%%%%%%
\section{No-go results for unitary channels}\label{app::nogounitary}
\setcounter{figure}{0}
%%%%%%%%%%%%%%%%%%%%%%%%%%%%%%%%%%%%%%%%%%%%

As previously discussed, the simulation of the action of the quantum switch on unitary channels is possible when an extra call of one of the input channels is available. This was first shown by Ref.~\cite{chiribella2013quantum} for the action of the quantum switch on entire single-party unitary channels, and here we have shown this result to extend to a more general scenario where the quantum switch acts only on part of the input {bipartite} unitary channels (see Theorem~\ref{thm::go} in Sec.~\ref{subapp::proofoursimulation}). These results hold in the general simulation scenario, where the input systems are not fixed and the output systems are not discarded, and hold as well for any dimension.

However, this result crucially depends not only on the number of extra calls available, but also on the order in which the quantum unitary channels are applied in the simulation. The cases where a simulation is possible for either single-party or for bipartite unitary channels with $(k_A,k_B)=(2,1)$ use the order ABA. If the order of the applied channels is instead either AAB or BAA, we find that a deterministic simulation is no longer possible, even in the restricted qubit scenario. 

Moreover, the existence of a simulation for unitary channels in the order ABA implies that, when $(k_A,k_B)=(2,2)$, a simulation with the orders ABAB and ABBA also trivially exists. In Sec.~\ref{app::gorestricted}, we showed that a simulation using the order AABB also exists for unitary channels, albeit \textit{only} in the restricted qubit case where the output target system is discarded. For the case where $(k_A,k_B)=(3,1)$, a simulation in the orders AABA and ABAA exists, following trivially from the result for ABA. We showed in Sec.~\ref{app::gorestricted} that a simulation for unitary channels in the order BAAA exists as well, but once again \textit{only} in the restricted qubit case. The one case left to study then is a simulation of the quantum switch acting on unitary channels in the order AAAB. We show such a simulation is not possible.

Hence, our three no-go results for simulations of the quantum switch acting exclusively on unitary channels concern the cases of:
\begin{itemize}
    \item Two calls of the unitary channels $A$ and one call of unitary channels $B$, in the order AAB.
    \item Two calls of the unitary channels $A$ and one call of unitary channels $B$, in the order BAA.
    \item Three calls of the unitary channels $A$ and one call of unitary channels $B$, in the order AAAB.
\end{itemize}

In order to show these results, we numerically evaluate the maximum probability of success of simulating the quantum switch when the input channels are unitary. We do so in the case where the input channels are acting on qubit systems, i.e., when $d=2$, and in the restricted simulation scenario where the input control and target systems are fixed, and the output target system is discarded [see Eq.~\eqref{eq::restrictedsim_app}]. We remark once again that the impossibility of a deterministic simulation in the restricted, fixed-dimension case implies the impossibility of a deterministic simulation in general.

In this case, we do not construct an explicit basis analytically for the subspace spanned by $k$ identical copies of a qubit unitary channel, but instead, we randomly sample a set of $d_\map{I}$ linearly independent qubit unitary channels $\{U_i^{\otimes k}\}_{i=1}^{d_\map{I}}$ to form our basis. We repeat here that for the case where $k=1$, $d_\mathcal{U}=10$, for $k=2$, $d_\mathcal{U}=35$, and for $k=3$, and $d_\mathcal{U}=84$. This implies that in the $(k_A,k_B)=(2,1)$ case, one needs $35\cdot10=350$ pairs of qubit unitary channels and for $(k_A,k_B)=(3,1)$, one needs $84\cdot10=840$ pairs of qubit unitary channels as input for the primal and dual SDPs [Eqs.~\eqref{sdp::primal} and~\eqref{sdp::dual} of the main text]. We numerically obtain the values of 
\begin{align}
    \text{order = AAB\phantom{A} (qubit, unitary)}, \ \ \  & p \approx 0.600 \\
    \text{order = BAA\phantom{A} (qubit, unitary)}, \ \ \  & p \approx 0.851 \\
    \text{order = AAAB (qubit, unitary)}, \ \ \ & p \approx 0.708.
\end{align}

%%%%%%%%%%%%%%%%%%%%%%%%%%%%%%%%%%%%%%%%%%%%
\section{Efficient certification that a matrix is positive semidefinite}\label{app::compassist}
\setcounter{figure}{0}
%%%%%%%%%%%%%%%%%%%%%%%%%%%%%%%%%%%%%%%%%%%%

A self-adjoint linear operator $A\in\mathcal{L}(\mathbb{C}^d)$ is positive semidefinite if and only if there exists an operator $L\in\mathcal{L}(\mathbb{C}^d)$ such that $A=L L^\dagger$. Moreover, the operator $L$ can be taken to be a lower triangular matrix, i.e., a matrix in which all entries above the main diagonal are zero. The decomposition of a positive semidefinite matrix as $A=LL^\dagger$ is referred to as the Cholesky decomposition, and finding such a decomposition can be done efficiently~\cite{cholesky}.
 
In the final step of the algorithm used for computer-assisted proofs presented in the Methods section of the main text, one is required to certify that an operator $A$ is positive semidefinite. When $A$ is stored as a symbolic matrix, due to the way computers manipulate symbolic variables, if the matrix $A$ is not sparse enough, ensuring that $A\geq0$ may be a prohibitively time-consuming task even when using the Cholesky decomposition. This was the case, for instance, when trying to ensure that the matrix $\overline{\mathbb{P}}(\Gamma^\texttt{sym}) + \eta \id - \sum_{i,j} R_{ij}^\texttt{OK}\otimes({J^A_i}^{\otimes k_A}\otimes {J^B_j}^{\otimes k_B})^T$ in our computer-assisted proofs algorithm involving $k_A+k_B=4$-slot quantum combs. Below, we describe an algorithm that can be used to ensure that a symbolic matrix $A$ is positive semidefinite, which is considerably faster than performing Cholesky decomposition on a symbolic matrix $A$.

The algorithm we present below is based on three key ideas.
\begin{enumerate}
    \item It is possible to rigorously certify that a matrix is positive definite using floating-point arithmetic quickly. One way to attain this goal is to use the methods presented in Ref.~\cite{rump2006verification}. It shows how to efficiently certify that a matrix $A$ is positive definite using a rigorous algorithm that accounts for all possible computational and rounding errors and remains valid in the presence of underflow.
    \item If a matrix $A$ is ``close'' to another matrix $A'$, and $A'$  is ``far'' from the set of non-positive semidefinite matrices, then $A$ has to be positive semidefinite.
    \item If $A$ is matrix with symbolic entries, we can obtain a floating-point variable matrix $A'$ that is guaranteed to be close to $A$. This can be done via arbitrary-precision arithmetic~\cite{wikiarbitrary}.
\end{enumerate}

\paragraph*{\textbf{\textit{\uline{Algorithm to prove that a symbolic self-adjoint matrix $A$ is positive semidefinite:}}}}
\begin{enumerate}
    \item 
    \texttt{Construct a self-adjoint matrix $A'$ that is equal to $A$ up to $n$ decimal digits} \\ 
     This step can be accomplished using arbitrary-precision arithmetic~\cite{wikiarbitrary} with a precision of $n$ decimal digits. In this way, all matrix elements of $A'-A$ are between $-10^{-n+1}$ and $10^{-n+1}$, ensuring that      
    \begin{align}
        -J \cdot 10^{-n+1} \;\leq \; A'-A \; \leq\;  J \cdot 10^{-n+1},    \label{eq:J_trick}
    \end{align}
    where $J\geq0$ is a matrix in which all entries are the number one.
    \item
    \texttt{Prove that $A'-J\geq0$ using the algorithm presented in Ref.~\cite{rump2006verification}.} \\ 
    Since $A'- J \cdot 10^{-n+1}\leq A$ holds, if we ensure that $A'-J\geq0$, then by transitivity it follows that $A\geq0$. 
\end{enumerate}

\end{document}